\documentclass[aos,preprint]{imsart}

\RequirePackage{amsthm,amsmath,amsfonts,amssymb}
\RequirePackage{natbib}
\usepackage{mathrsfs}
\usepackage{amsfonts}
\usepackage{makeidx}         
\RequirePackage{graphicx}        
\usepackage{multicol}        
\usepackage[bottom]{footmisc}
\usepackage{epsfig}
\usepackage{bm}
\usepackage{enumitem}
\usepackage{epstopdf}
\usepackage{multirow}
\usepackage{booktabs}
\usepackage{mathtools}
\usepackage{wrapfig}
\usepackage[pagewise, mathlines]{lineno}
\usepackage{caption}
\usepackage{subcaption}
\usepackage{lscape}
\usepackage[colorinlistoftodos]{todonotes}
\usepackage{float}
\usepackage{titlesec}
\usepackage{verbatim}
\usepackage{rotating}
\usepackage{amsbsy}

\usepackage[allcolors=blue,colorlinks=true]{hyperref}

\let\clearpage\relax

\newtheorem{theorem}{Theorem}[section]

\newtheorem{prop}[theorem]{Proposition}

\theoremstyle{remark}
\newtheorem{defn}[theorem]{Definition}

\newlength{\tempheight}
\newlength{\tempwidth}

\newcommand{\rowname}[1]
{\rotatebox{90}{\makebox[\tempheight][c]{\textbf{#1}}}}

\newcommand{\columnname}[1]
{\makebox[\tempwidth][c]{\textbf{#1}}}

\mathtoolsset{showonlyrefs}
\startlocaldefs
\theoremstyle{remark}
\newtheorem{algorithm}{Algorithm}
\newtheorem{remark}{Remark}

\endlocaldefs

\begin{document}

\begin{frontmatter}
\title{Optimal Relevant Subset Designs in Nonlinear Models}
\runtitle{Relevant Subset Designs}

\begin{aug}
\author{\fnms{Adam} \snm{Lane}\ead[label=e1]{adam.lane@cchmc.org}},
\address{University of Cincinnati}
\end{aug}

\begin{abstract}
\cite{Fish:TwoN:1934} argued that certain ancillary statistics form a \emph{relevant subset}, a subset of the sample space on which inference should be restricted, and showed that conditioning on their observed value reduces the dimension of the data without a loss of information. The use of ancillary statistics in post-data inference has received significant attention; however, their role in the design of the experiment has not been well characterized. Ancillary statistics are unknown prior to data collection and as a result cannot be incorporated into the design \emph{a priori}. However, if the data are observed sequentially then the ancillary statistics based on the data from the preceding observations can be used to determine the design assignment for the current observation. The main results of this work describe the benefits of incorporating ancillary statistics, specifically, the ancillary statistic that constitutes a relevant subset, into an adaptive design.  
\end{abstract}



\end{frontmatter}

\section{Introduction} \label{sec:intro}

A random variable, $\boldsymbol{\mathcal{A}}$, is \textit{ancillary} if its distribution is independent of the model parameters; additionally, it is an \emph{ancillary complement} if $(\boldsymbol{\hat{\theta}},\boldsymbol{\mathcal{A}})$ is a minimal sufficient sufficient, where $\boldsymbol{\hat{\theta}}$ is the maximum likelihood estimate (MLE) of the $p$-dimensional parameter $\boldsymbol{\theta}$. In parametric settings it is common reduce the sample, $\boldsymbol{\mathcal{Y}}  = (\pmb{\mathcal{Y}}_{1}^{T},\ldots,\pmb{\mathcal{Y}}_{d}^{T})^{T}$, to the point estimate, $\boldsymbol{\hat{\theta}} = \boldsymbol{\hat{\theta}}(\boldsymbol{\mathcal{Y}})$, in order to make inferences about $\boldsymbol{\theta}$, where $\boldsymbol{\mathcal{Y}}$ is a random sample of independent responses from an experiment with design $\xi = \{(x_{i},w_{i})\}_{i=1}^{d}$ and known probability density function (pdf) $f_{\boldsymbol{\theta}}(\boldsymbol{\mathcal{Y}}|\xi)$;  $\boldsymbol{x} = (x_{1},\ldots,x_{d})^{T}$ are the support points of the design; $\boldsymbol{\mathcal{Y}}_{i} = (\mathcal{Y}_{i1},\ldots,\mathcal{Y}_{in_{i}})^{T}$, $n_{i}$ and $w_{i} = n_{i}/n$ are the responses, allocation weight, and sample size corresponding to the support point $x_{i}$; and $n = \sum_{i} n_{i}$ is the total sample size. \citet{Fish:TwoN:1934} noted that the use of $\boldsymbol{\hat{\theta}}$, in this setting, results in a loss of information and argued that this information can be recovered by conditioning on the observed value of the ancillary complement. To illustrate Fisher's argument, let $\boldsymbol{I}_{\boldsymbol{\mathcal{Y}}}^{\boldsymbol{x}}(\boldsymbol{\theta}) = -(\partial^{2}/\partial \boldsymbol{\theta}^{2}) \log f_{\boldsymbol{\theta}}(\boldsymbol{\mathcal{Y}}|\xi)$. The Fisher information in $\boldsymbol{\mathcal{Y}}$, denoted $\boldsymbol{F}^{\xi}(\boldsymbol{\theta})= {\rm{E}}[\boldsymbol{I}_{\boldsymbol{\mathcal{Y}}}^{\boldsymbol{x}}(\boldsymbol{\theta})]$,
measures the information about $\boldsymbol{\theta}$ contained in the \emph{entire} sample. This differs from the Fisher information in $\boldsymbol{\hat{\theta}}$ defined as 
$\boldsymbol{G}^{\xi}(\boldsymbol{\theta}) = -{\rm{E}}[(\partial^{2}/\partial \boldsymbol{\theta}^{2}) \log g_{\boldsymbol{\theta}}(\boldsymbol{\hat{\theta}}|\xi)]$, where $g_{\boldsymbol{\theta}}(\boldsymbol{\hat{\theta}}|\xi)$ is the marginal pdf of $\boldsymbol{\hat{\theta}}$. Using Jensen's inequality it can be shown that $\boldsymbol{F}^{\xi}\ge\boldsymbol{G}^{\xi}$ with equality if and only if $\boldsymbol{\hat{\theta}}$ is sufficient, where $\ge$ indicates with respect to Loewner ordering. This result confirms Fisher's claim that $\boldsymbol{\hat{\theta}}$ conveys a loss of information and suggests that if $\boldsymbol{\hat{\theta}}$ is used as a point estimate then it is misleading to report $\boldsymbol{F}^{\xi}$ as the corresponding information since it over-states $\boldsymbol{G}^{\xi}$. This discussion reveals that $\boldsymbol{F}^{\xi}$ and $\boldsymbol{G}^{\xi}$ are unsatisfactory measures of information.

The second part of Fisher's argument, described in \cite{Cox:Some:1958}, \cite{Basu:Reco:1969} and \cite{Ghos:Reid:Fras:Anci:2010}, is that if $(\boldsymbol{\hat{\theta}},\boldsymbol{\mathcal{A}})$ is a minimal sufficient sufficient statistic then $\boldsymbol{\hat{\theta}}$ is a sufficient statistic on the subset of the sample space where $\boldsymbol{\mathcal{A}}=\boldsymbol{A}$. As a consequence, the Fisher information in $\boldsymbol{\hat{\theta}}|(\boldsymbol{\mathcal{A}} = \boldsymbol{A})$ is equivalent to the Fisher information in $\boldsymbol{\mathcal{Y}}|(\boldsymbol{\mathcal{A}} = \boldsymbol{A})$ and is defined as
\begin{align}  \label{eq:JA}
   \boldsymbol{H}_{\boldsymbol{A}}^{\boldsymbol{x}}(\boldsymbol{\theta}) &= {\rm{E}}[\boldsymbol{I}_{\boldsymbol{\mathcal{Y}}}^{\boldsymbol{x}}(\boldsymbol{\theta})|\boldsymbol{\mathcal{A}} = \boldsymbol{A}].
\end{align}
This measure correctly reflects the information about $\boldsymbol{\theta}$ in the point estimate, $\boldsymbol{\hat{\theta}}$, for any given $\boldsymbol{A}$. Additionally, noting that $\boldsymbol{F}^{\xi}= {\rm{E}}[\boldsymbol{H}_{\boldsymbol{\mathcal{A}}}^{\boldsymbol{x}}]$ it can be seen that $\boldsymbol{H}_{\boldsymbol{A}}^{\boldsymbol{x}}$ does not represent a loss of information when compared to $\boldsymbol{F}^{\xi}$. These revelations led \citet{Fish:TwoN:1934} and others to conclude that $\boldsymbol{H}_{\boldsymbol{A}}^{\boldsymbol{x}}$ is a more appropriate measure of information than either $\boldsymbol{F}^{\xi}$ or $\boldsymbol{G}^{\xi}$.  

The observed value of the ancillary complement has been referred to as a ``relevant subset'' or ``reference set'', e.g. \cite{Ghos:Reid:Fras:Anci:2010} use the former and \cite{Basu:Reco:1969} uses the latter. Here the phrase \emph{relevant subset} is adopted to refer to the subset of the sample space such that $\boldsymbol{\mathcal{A}} = \boldsymbol{A}$ and the phrase ``conditional on the relevant subset'' is equivalent to ``conditional on $\boldsymbol{\mathcal{A}} = \boldsymbol{A}$''. Further, analogous to $\boldsymbol{F}^{\xi}$ representing the \emph{Fisher information in the sample}, $\boldsymbol{H}_{\boldsymbol{A}}^{\boldsymbol{x}}$ represents the \emph{Fisher information in the relevant subset}. 

Conditioning on the relevant subset in post-data parametric inference has a long standing history with significant interest and support [\cite{Fish:TwoN:1934,Cox:Some:1958,Basu:Reco:1969,McCu:Cond:1992,Efro:Hink:Asse:1978,Ghos:Reid:Fras:Anci:2010}].  The central objective of this work is to determine the role of ancillary statistics, specifically, the relevant subset defined by the ancillary complement, in the design of experiments.

\subsection{Optimal Design}

The field of optimal design is primarily focused on optimizing the Fisher information in the sample, $\boldsymbol{F}^{\xi}$, with respect to a convex optimality criterion, denoted $\Psi$. Let $\mathcal{S}_{+}^{p}= \{\boldsymbol{F}^{\xi}(\boldsymbol{\theta}):\xi\in\Xi\}$, where $\Xi$ is the set of all possible designs. The criterion $\Psi$ is a mapping from $\mathcal{S}_{+}^{p}$ to $\overline{\mathbb{R}}_{+}:=(0,\infty]$. The Fisher information in the sample is said to be \emph{optimized} if $\Psi(\boldsymbol{F}^{\xi})$ is the minimum for all $\boldsymbol{F}^{\xi}\in\mathcal{S}_{+}^{p}$; the design corresponding to this minimum is referred to as the $\Psi$-\emph{optimal design}, denoted $\xi^{*}(\boldsymbol{\theta})$.  

The optimality criterion relates the experimental objective to the design, e.g. the $D$-optimal design, characterized by the criterion $\Psi(\boldsymbol{F}^{\xi}) = |\boldsymbol{F}^{\xi}|^{-1/2}$, minimizes the volume of the confidence ellipsoid for $\boldsymbol{\theta}$ defined by the interior of $(\boldsymbol{\hat{\theta}} - \boldsymbol{\theta})^{T}\boldsymbol{F}^{\boldsymbol{\xi}}(\boldsymbol{\hat{\theta}})(\boldsymbol{\hat{\theta}} - \boldsymbol{\theta}) = \chi_{p}^{2(1-\alpha)}$, where $\chi_{p}^{2(1-\alpha)}$ represents the $(1-\alpha)$ quantile of a $\chi^{2}$ distribution with $p$ degrees of freedom.. However, if Fisher's conditioning argument is accepted then it is implied that confidence ellipsoids for $\boldsymbol{\theta}$ should defined as the interior of $(\boldsymbol{\hat{\theta}} - \boldsymbol{\theta})^{T}\boldsymbol{H}_{\boldsymbol{A}}^{\boldsymbol{x}}(\boldsymbol{\hat{\theta}})(\boldsymbol{\hat{\theta}} - \boldsymbol{\theta}) = \chi_{p}^{2(1-\alpha)}$ which has volume proportional to $|\boldsymbol{H}_{\boldsymbol{\mathcal{A}}}^{\boldsymbol{x}}|^{-1/2}$. The design that minimizes ${\rm{E}}[|\boldsymbol{H}_{\boldsymbol{\mathcal{A}}}^{\boldsymbol{x}}|^{-1/2}] \ne |\boldsymbol{F}^{\xi}|^{-1/2}$ will minimize the expected volume of this confidence ellipsoid. Smaller confidence ellipsoids represent an improvement in inference. 

For a general $\Psi$ a design optimizes inference if it minimizes ${\rm{E}}[\Psi(\boldsymbol{H}_{\boldsymbol{\mathcal{A}}}^{\boldsymbol{x}})]$, where ${\rm{E}}[\Psi(\boldsymbol{H}_{\boldsymbol{\mathcal{A}}}^{\boldsymbol{x}})]\ge \Psi(\boldsymbol{F}^{\xi})$ by Jensen's inequality. The objective of minimizing ${\rm{E}}[\Psi(\boldsymbol{H}_{\boldsymbol{\mathcal{A}}}^{\boldsymbol{x}})]$ will be referred to as the \emph{inference} objective. It is not desirable for a design to result in a loss of Fisher information in the sample; a second design objective is to ensure that the design remains optimal with respect to $\boldsymbol{F}^{\xi}$. This is referred to as the \emph{information} objective. The primary goal of this work is to develop a design that is optimal with respect to both objectives.  

\subsection{Relevant Subset Designs}

Recall, the observed value of the ancillary complement, $\boldsymbol{\mathcal{A}} = \boldsymbol{A}$, defines a relevant subset of the sample space on which inference is restricted and define a relevant subset design (RSD) as any procedure that incorporates the relevant subset into the design of the experiment. There is a conceptual difference between RSD and optimal design. The Fisher information in the sample is known \emph{a priori}; which allows the optimal design to be determined in advance. The relevant subset is a random variable and it is unknown prior to data collection. However, if observations occur sequentially, in a series of runs, the relevant subset from the preceding runs is known and can be used in  the design assignment of the current run. To realize their full potential RSDs require adaptation. Relevant subset designs are introduced in depth in Section \ref{sec:RSD}.

Adaptive designs have been considered in the context of models where $\boldsymbol{F}^{\xi}$ has an explicit dependence on $\boldsymbol{\theta}$. This results in designs that are \textit{locally optimal} in a neighborhood of the true parameters. Fixed (not adaptive) locally optimal designs (FLOD) have been thoroughly investigated. \cite{Cher:Loca:1953} and \cite{Mela:Opti:1978} are early references where the FLODs in nonlinear models are considered; see \cite{Dett:Bied:Robu:2003}, \cite{Han:Chal:Dandc:2003} and \cite{Dett:Lope:Rodr:Maxi:2006} for more recent advances. The local dependence can be addressed by using an \emph{a priori} guess of the underlying parameters; such designs can be inefficient in practice if this guess is far from the reality. The FLOD evaluated at the true parameters can be viewed as a benchmark for \textit{adaptive optimal designs} (AODs). Use of an AOD eliminates the local dependence by estimating the parameters after each sequential run and computing the FLOD for the current run based on this estimate.  \cite*{Box:Hunt:Sequ:1965} present the earliest example of an AOD which has recently received significant attention particularly in the context of dose finding clinical trials [\cite*{Drag:Fedo:Adap:2005}, \cite*{Drag:Fedo:Wu:Adap:2008}, \cite*{Lane:Yao:Flou:Info:2014} and others].

\subsection{Summary of Contributions}

The current work considers the design of experiments for additive error nonlinear regression models. Nonlinear regression is widely used in practice to model a response as a function the explanatory variables [\cite{Sebe:Wild:Nonl:1989}, \cite{Ratk:Nonl:1983}, etc.]. In this work it is assumed that, given $x_{i}$, responses are observed from the nonlinear model 
\begin{align} \label{eq:model}
    y_{ij} = \eta_{x_{i}}(\boldsymbol{\theta}) + \varepsilon_{ij}, \quad j=1,\ldots,n_{i}, \quad i = 1,\ldots,d
\end{align}
where $\boldsymbol{\varepsilon} = (\varepsilon_{11},\ldots,\varepsilon_{dn_{d}})^{T}$ is a sequence of independent and identically distributed random variables with finite variance; $\boldsymbol{\theta}$ is a $p\times 1$ vector within the parameter space $\Theta$; $\Theta$ is an open subset of $\mathbb{R}^{p}$; $x$ is an $s\times 1$ vector within the design region, $\mathcal{X}$; and $\mathcal{X}$ is a compact subset of $\mathbb{R}^{s}$. The information under model \eqref{eq:model} depends on the parameters and can vary substantially according to the design.

\citet*{Lane:Adap:2019,lane2019optimality} developed designs that implicitly incorporate the relevant subset into the design of experiments in linear models. The aforementioned works optimize the observed information, defined as $\boldsymbol{I}_{\boldsymbol{y}}^{\boldsymbol{x}}$, where $\boldsymbol{y}$ is the observed value of $\boldsymbol{\mathcal{Y}}$. In the location family the observed information is a function of the relevant subset  [\cite{Efro:Hink:Asse:1978}]. There are two issues with extending these methods to nonlinear regression. First, as \citet{Lane:Adap:2019} remarks, a general equivalence theorem for observed information does not appear to hold for nonlinear models. The general equivalence theorem is the basis for many algorithms to construct optimal designs.  A second problem is that the logic for using the observed information does not extend. Specifically, it is not clear if $\boldsymbol{I}_{\boldsymbol{y}}^{\boldsymbol{x}}$ is an accurate approximation of $\boldsymbol{H}_{\boldsymbol{\mathcal{A}}}^{\boldsymbol{x}}$. This raises an additional challenge; $\boldsymbol{H}_{\boldsymbol{\mathcal{A}}}^{\boldsymbol{x}}$ does not, in general, have an analytic solution which makes its use in the design and analysis of experiments impractical. These problems are resolved in the current work.

In Section \ref{sec:CondVar} a measure of information is proposed that approximates $\boldsymbol{H}_{\boldsymbol{\mathcal{A}}}^{\boldsymbol{x}}$ to a higher degree than either the $\boldsymbol{I}_{\boldsymbol{y}}^{\boldsymbol{x}}$ or $\boldsymbol{F}^{\xi}$. The proposed measure has a convenient analytic expression and a general equivalence theorem for this measure is stated in Section \ref{sec:RSD}. This theorem is applied in Section \ref{sec:OSA} to develop a one-step ahead optimal RSD. In Section \ref{sec:sim} a simulation study is presented which includes different nonlinear mean functions, error distributions and optimality criteria. Each simulation is motivated by a real world data set. Recall, in nonlinear models that the FLOD depends on the model parameters. In the simulation study the FLOD is evaluated at the true value of the underlying parameters and should not be considered a competing design but instead an upper bound on the efficiency of fixed designs. Despite this fact the proposed optimal RSD, which has no local dependence, was still more efficient that the FLOD in nearly every case considered. The AOD was also examined; the optimal RSD was uniformly more efficient than the AOD. This simulation study demonstrates that it is possible to use optimal RSD to improve the efficiency of experiments compared to existing alternatives.

\section{Approximating the Fisher information in the Relevant Subset} \label{sec:CondVar}

To begin the information about the one-dimensional parameter $\eta_{i} = \eta_{x_{i}}(\boldsymbol{\theta})$ is considered. Denote the likelihood and its first two derivatives, with respect to $\eta_{i}$, as $l_{\boldsymbol{y}_{i}}(\eta_{i}) = \log f_{\eta_{i}}(\boldsymbol{y}_{i}|x_{i})$, $\dot{l}_{\boldsymbol{y}_{i}}(\eta_{i}) = (\partial/\partial \eta_{i}) l_{\boldsymbol{y}_{i}}(\eta_{i})$ and $\ddot{l}_{\boldsymbol{y}_{i}}(\eta_{i}) = -(\partial^{2}/\partial \eta_{i}^{2}) l_{\boldsymbol{y}_{i}}(\eta_{i})$, where $f_{\eta_{i}}(\boldsymbol{y}_{i}|x_{i})$ is the pdf of $\boldsymbol{y}_{i}$ with support point $x_{i}$. In this section it is assumed that there exists an ancillary $\boldsymbol{a}_{i}$ such that $(\hat{\eta}_{i},\boldsymbol{a}_{i})$ is a minimal sufficient statistic; it is not required that a form for $\boldsymbol{a}_{i}$ is known only that it exists. A notable exception is when responses are normally distributed since $\hat{\eta}_{i}$ is minimally sufficient. The observed information about $\eta_{i}$ is $i_{\boldsymbol{y}_{i}}(\eta_{i}) = -\ddot{l}_{\boldsymbol{y}_{i}}(\eta_{i})$. The Fisher information in the relevant subset about $\eta_{i}$ is $h_{\boldsymbol{a_{i}}}(\eta_{i}) = {\rm{E}}[i_{\boldsymbol{\mathcal{Y}}_{i}}(\eta_{i})|\boldsymbol{\mathcal{A}}_{i} = \boldsymbol{a_{i}}]$, where $\boldsymbol{\mathcal{Y}}_{i}$ and $\boldsymbol{\mathcal{A}}_{i}$ are the random variables associated with the observed realizations $\boldsymbol{y}_{i}$ and $\boldsymbol{a}_{i}$, respectively. Assuming that derivatives and integrals are exchangeable, the Fisher information in the sample about $\eta_{i}$ is $\mathscr{F}_{i} = {{\rm{E}}}[i_{\boldsymbol{\mathcal{Y}}_{i}}(\eta_{i})]$. In this work it is assumed that $\mathscr{F}_{i}$ does not depend on $\eta_{i}$, i.e. $\mu = {\rm{E}}[i_{\mathcal{Y}}(\eta_{i})] =  n_{i}^{-1}\mathscr{F}_{i}$, where $i_{y}(\eta_{i}) = -[\partial^{2}/(\partial\eta_{i}^{2})]f_{\eta_{i}}(y|x_{i})$ is the observed information from a single observation with support $x_{i}$. 

For the below proposition $f_{\eta_{i}}(y|x_{i})$ is restricted to the location family; which is characterized by the relation $f_{\eta_{i}}(y|x_{i}) = f_{0}(y - \eta_{i}|x_{i})$. \citet{Efro:Hink:Asse:1978} show that in the location family the observed information evaluated at the MLE of $\eta_{i}$, denoted $\hat{\eta}_{i}$, can be expressed as a function of $\boldsymbol{a}_{i}$ alone. The notation $i_{\boldsymbol{a}_{i}} = i_{\boldsymbol{y}_{i}}(\hat{\eta}_{i})$ is used to make this dependence explicit. The following proposition shows that $h_{\boldsymbol{a_{i}}}(\eta_{i})$ is more accurately approximated by $i_{\boldsymbol{a_{i}}}$ than $\mathscr{F}_{i}$. 
\begin{prop} \label{prop:info}
Under the conditions stated in Section \ref{sec:Tech}
\begin{align} 
    n^{-1}[i_{\boldsymbol{a_{i}}} - h_{\boldsymbol{a_{i}}}(\eta_{i})] &= O_{p}(n^{-1}) \label{eq:ObsElemDiff} \quad \mbox{and} \quad
    n^{-1}[\mathscr{F}_{i} - h_{\boldsymbol{a_{i}}}(\eta_{i})] = O_{p}(n^{-1/2}). 
\end{align}
\end{prop}
The location family condition is required order to obtain the order of approximation in the proposition, primarily, to ensure $i_{\boldsymbol{a_{i}}}$ is ancillary.  \citet{Efro:Hink:Asse:1978} argue that $i_{\boldsymbol{a_{i}}}$ may be approximately ancillary more generally; to the author's knowledge this remains an open problem. If this were true then the result might hold more generally. Further, it is often argued that the observed information is the appropriate measure of information in the one-dimensional parameter case more generally [\citet{Efro:Hink:Asse:1978}, \citet{McCu:Loca:1984}, \citet{Barn:Sore:ARev:1994}, \citet{Ghos:Reid:Fras:Anci:2010}]. In the remainder of this section Proposition \ref{prop:info} is used to develop an approximation to the Fisher information in the relevant subset for model \eqref{eq:model}.  

\subsection{Information in Nonlinear Models}
In the context of model \eqref{eq:model} the interest is in information about $\boldsymbol{\theta}$, a $p$-dimensional vector. To connect the preceding section to the current setting, recall that the response function of the observations with support $x_{i}$ is $\eta_{i} = \eta_{x_{i}}(\boldsymbol{\theta})$; however, at times $\eta_{i}$ will still be used for shorthand when the meaning is clear. Let $\boldsymbol{y}=(\boldsymbol{y_{1}}^{T},\ldots,\boldsymbol{y_{d}}^{T})^{T}$ and $\boldsymbol{A} = (\boldsymbol{a}_{1},\ldots,\boldsymbol{a}_{d})$ denote the observed response vector and the matrix of ancillary configuration statistics, respectively. 

The information measures introduced in Section \ref{sec:intro} are now explicitly derived for model \eqref{eq:model}. The observed information can be written, after some basic algebra, as
\begin{align} \label{eq:J}
   \boldsymbol{I}_{\boldsymbol{y}}^{\boldsymbol{x}}(\boldsymbol{\theta}) &=  \sum_{i=1}^{d} i_{\boldsymbol{y}_{i}}(\eta_{i}) \dot{\eta}_{i}\dot{\eta}_{i}^{T} - \sum_{i=1}^{d} \dot{l}_{\boldsymbol{y}_{i}}(\eta_{i}) \ddot{\eta}_{i},
\end{align}
where 
$\dot{\eta}_{i} = \dot{\eta}_{x}(\boldsymbol{\theta}) = \left[\partial \eta_{i}/(\partial\theta_{1}),\ldots,\partial \eta_{i}/(\partial\theta_{d})\right]^{T}$ and $\ddot{\eta}_{i} = \ddot{\eta}_{x}(\boldsymbol{\theta}) = \left[\partial \dot{\eta}_{i}/(\partial\theta_{1}),\ldots,\partial \dot{\eta}_{i}/(\partial\theta_{d})\right]^{T}$. 

The Fisher information in the relevant subset is as defined in \eqref{eq:JA} and no further simplification is readily available. Given standard regularity conditions the Fisher information in the sample under model \eqref{eq:model} is
\begin{align} \label{eq:F}
   \boldsymbol{F}^{\xi}(\boldsymbol{\theta}) = {\rm{E}}[ \boldsymbol{H}_{\boldsymbol{\mathcal{A}}}^{\boldsymbol{x}}(\boldsymbol{\theta}) ] = n\mu\sum_{i=1}^{d} w_{i} \dot{\eta}_{i}\dot{\eta}_{i}^{T}.
\end{align}

One advantage of $\boldsymbol{F}^{\xi}$ is that under minimal assumptions it has the form given in \eqref{eq:F}. This has likely contributed to its ubiquitous use in both the design and analysis of experiments. Conversely, it is not a simple task to find a convenient expression for $\boldsymbol{H}_{\boldsymbol{A}}^{\boldsymbol{x}}$. The remainder of this section addresses this challenge by defining a relatively simple measure that accurately approximates $\boldsymbol{H}_{\boldsymbol{A}}^{\boldsymbol{x}}$. 

\begin{table}
\centering
\scriptsize
\begin{tabular}{ ccc }
\noalign{\smallskip}\hline\noalign{\smallskip}
Notation & Brief Description & Equation Number \\
\noalign{\smallskip}\hline\noalign{\smallskip}
$\boldsymbol{H}_{\boldsymbol{A}}^{\boldsymbol{x}}(\boldsymbol{\theta})$ & Fisher information in the relevant subset, $\boldsymbol{\hat{\theta}}|(\boldsymbol{\mathcal{A}} = \boldsymbol{A})$ & \eqref{eq:JA} \\
$\boldsymbol{I}_{\boldsymbol{y}}^{\boldsymbol{x}}(\boldsymbol{\theta})$ & Observed information & \eqref{eq:J} \\
$\boldsymbol{F}^{\xi}(\boldsymbol{\theta})$ & Fisher information in the sample, $\boldsymbol{\mathcal{Y}}$ & \eqref{eq:F} \\
$\boldsymbol{G}^{\xi}(\boldsymbol{\theta})$ & Fisher information in the MLE, $\boldsymbol{\hat{\theta}}$ & \\
$\boldsymbol{J}_{\boldsymbol{y}}^{\boldsymbol{x}}(\boldsymbol{\theta})$ & Hybrid information & \eqref{eq:KAlt} \\
$\boldsymbol{K}_{\boldsymbol{y}}^{\boldsymbol{x}}(\boldsymbol{\theta})$ & First term in the observed information & \eqref{eq:JAlt} \\
$\boldsymbol{M}^{\xi}(\boldsymbol{\theta})$ & Normalized Fisher information in the sample & \eqref{eq:M} \\
\noalign{\smallskip}\hline\noalign{\smallskip}
\end{tabular}
\caption{Notation, brief description and equation number (where available) of different information measures.} 
\end{table}


A \emph{hybrid} measure of information is now introduced where $i_{\boldsymbol{y}_{i}}(\eta_{i})$ and $\dot{l}_{\boldsymbol{y}_{i}}(\eta_{i})$ in \eqref{eq:J}, in \eqref{eq:J}, are evaluated at $\hat{\eta}_{i}$; whereas, $\dot{\eta}_{i}$ and $\ddot{\eta}_{i}$ are left as functions of $\boldsymbol{\theta}$. Specifically, the hybrid measure proposed is
\begin{align} \label{eq:KAlt}
\boldsymbol{J}_{\boldsymbol{A}}^{\boldsymbol{x}}(\boldsymbol{\theta}) =   \sum_{i=1}^{d} i_{\boldsymbol{a_{i}}} \dot{\eta}_{i}\dot{\eta}_{i}^{T}.
\end{align}
This measure accomplishes two things. First, the second term in \eqref{eq:J} equals zero since $\dot{l}_{\boldsymbol{y}_{i}}(\hat{\eta}_{i}) = 0$; this significantly reduces the complexity since the hybrid measure no longer depends on $\ddot{\eta}_{i}$. Second, the first term is a function of $i_{\boldsymbol{a_{i}}}$ which from Proposition \ref{prop:info} is known to best approximate the Fisher information in the relevant subset at each point in the design. The main result regarding the analysis of experiments in nonlinear models establishes the usefulness of this measure.

\begin{theorem} \label{thm:CondInfo}
Under the conditions stated in Section \ref{sec:Tech}
$n^{-1}[\boldsymbol{J}_{\boldsymbol{A}}^{\boldsymbol{x}}(\boldsymbol{\theta}) -\boldsymbol{H}_{\boldsymbol{A}}^{\boldsymbol{x}}(\boldsymbol{\theta})] = O_{p}(n^{-1})$ and $n^{-1}[\boldsymbol{L}(\boldsymbol{\theta}) - \boldsymbol{H}_{\boldsymbol{A}}^{\boldsymbol{x}}(\boldsymbol{\theta})] = O_{p}(n^{-1/2})$ for all $\boldsymbol{\theta}\in\Theta$, where $\boldsymbol{L}(\boldsymbol{\theta})$ can be either $\boldsymbol{F}^{\xi}(\boldsymbol{\theta})$ or $\boldsymbol{I}_{\boldsymbol{y}}^{\boldsymbol{x}}(\boldsymbol{\theta})$.
\end{theorem}
For the analysis of experiments Theorem \ref{thm:CondInfo} has the clear implication that the, relatively, simple measure $\boldsymbol{J}_{\boldsymbol{A}}^{\boldsymbol{x}}$ is a second order approximation to the Fisher information in the relevant subset, whereas $\boldsymbol{F}^{\xi}$ and $\boldsymbol{I}_{\boldsymbol{y}}^{\boldsymbol{x}}$ are accurate only to the first order. First and second order approximations are defined as having error $O_{p}(n^{-1/2})$ and $O_{p}(n^{-1})$, respectively. Accepting that $\boldsymbol{H}_{\boldsymbol{A}}^{\boldsymbol{x}}$ is preferred measure of information implies that $\boldsymbol{J}_{\boldsymbol{A}}^{\boldsymbol{x}}$ is preferred to $\boldsymbol{F}^{\xi}$ and $\boldsymbol{I}_{\boldsymbol{y}}^{\boldsymbol{x}}$. To obtain the second order approximation Theorem \ref{thm:CondInfo} requires location family errors. However, if it is granted that the observed information is generally the best measure of information in the single parameter case, then $\boldsymbol{J}_{\boldsymbol{A}}^{\boldsymbol{x}}$ is applicable more generally. 

The implication for design is more nuanced but just as significant. Theorem \ref{thm:CondInfo} provides a potential scheme where both the \emph{information} and \emph{inference} objectives, described in Section \ref{sec:intro}, can be accomplished. Theorem \ref{thm:CondInfo} indicates that an adaptive procedure that optimizes $\boldsymbol{J}_{\boldsymbol{A}}^{\boldsymbol{x}}$ will optimize $\boldsymbol{H}_{\boldsymbol{A}}^{\boldsymbol{x}}$ and by extension optimize inference. Intuitively, an adaptive design procedure that optimizes the information for every possible relevant subset might also optimize the Fisher information over all relevant subsets, i.e. it will simultaneously optimize the Fisher information in the sample, since $\boldsymbol{F}^{\xi} = {\rm{E}}[\boldsymbol{H}_{\boldsymbol{\mathcal{A}}}^{\boldsymbol{x}}]$. The remainder of this work develops and examines an adaptive design that formalizes this approach. 

\begin{remark}
It is common practice to evaluate the information at the MLE, e.g. $\boldsymbol{J}_{\boldsymbol{A}}^{\boldsymbol{x}}(\boldsymbol{\hat{\theta}})$. Since Theorem \ref{thm:CondInfo} holds for all $\boldsymbol{\theta}\in\Theta$ it also holds for $\boldsymbol{\hat{\theta}}$, i.e. $n^{-1}[\boldsymbol{J}_{\boldsymbol{A}}^{\boldsymbol{x}}(\boldsymbol{\hat{\theta}}) -\boldsymbol{H}_{\boldsymbol{A}}^{\boldsymbol{x}}(\boldsymbol{\hat{\theta}})] = O_{p}(n^{-1})$.
\end{remark}

\begin{remark}
The assumption of repeats is central to Theorem \ref{thm:CondInfo}. In the absence of repeats approximating the Fisher information in the relevant subset to the second order remains an open problem. Without sufficient repeats the first term in the observed information on $\boldsymbol{\theta}$
\begin{align} \label{eq:JAlt}
   \boldsymbol{K}_{\boldsymbol{y}}^{\boldsymbol{x}}(\boldsymbol{\theta}) &= \sum_{i=1}^{d} i_{\boldsymbol{y}_{i}}(\eta_{i}) \dot{\eta}_{i}\dot{\eta}_{i}^{T},
\end{align}
has an intuitive appeal due to its connection to the hybrid measure, $\boldsymbol{J}_{\boldsymbol{A}}^{\boldsymbol{x}}$. It is not shown, or expected, that Theorem \ref{thm:CondInfo} extends to this measure. 
\end{remark}

\section{Relevant Subset Design} \label{sec:RSD}

In the first part of this section optimal design is, briefly, reviewed. The second part formulates relevant subset design (RSD) as an optimal design problem and states an equivalence theorem for optimal RSD. This theorem is applied in Section \ref{sec:OSA} to develop a one-step ahead optimal RSD.

\subsection{Optimal Design}

The most common objective in optimal design is to optimize the Fisher information in the sample, measured by $\boldsymbol{F}^{\xi}$. The constant $n\mu$ is a scalar in $\boldsymbol{F}^{\xi}$ and does not influence the optimization problem. As a result the optimal design is often defined with respect to the normalized Fisher information in the sample, denoted
\begin{align} \label{eq:M}
    \boldsymbol{M}^{\xi}(\boldsymbol{\theta}) = (n\mu)^{-1} \boldsymbol{F}^{\xi}(\boldsymbol{\theta}) = \sum_{i=1}^{d} w_{i} \dot{\eta}_{i}\dot{\eta}_{i}^{T}.  
\end{align}
In most settings no single design is optimal with respect to all objectives; instead, a design is optimal with respect to a specific convex criterion, denoted $\Psi$. Formally, a design, $\xi^{*}(\boldsymbol{\theta})$, is $\Psi$-optimal if
\begin{align} \label{eq:Opt}
\xi^{*}(\boldsymbol{\theta}) = \arg \min_{\xi \in \Xi} \Psi\{\boldsymbol{M}^{\xi}(\boldsymbol{\theta})\}.
\end{align}
In nonlinear models $\xi^{*}(\boldsymbol{\theta})$ depends on the parameters and is optimal only in the neighborhood of $\boldsymbol{\theta}$. Such designs are referred to as fixed locally optimal designs (FLODs); where fixed indicates that it is determined before the experiment (not adaptive) and local refers to its dependence on $\boldsymbol{\theta}$. 

The optimality criterion relates the primary objective of the experiment to the optimal design. For example, the $D$-optimal design minimizes the volume of the confidence ellipsoid around the MLE. In nonlinear models certain functions of the parameters have relevant interpretations. The $c_{\boldsymbol{\theta}}$-optimal design minimizes the MSE of a one-dimensional function of the parameter estimates. The $D$ and  $c_{\boldsymbol{\theta}}$ criteria are defined as $\Psi(\boldsymbol{M}^{\xi})=|\boldsymbol{M}^{\xi}|^{-1/p}$ and $\Psi(\boldsymbol{M}^{\xi}) = c_{\boldsymbol{\theta}}^{T}[\boldsymbol{M}^{\xi}]^{-1}c_{\boldsymbol{\theta}}$, respectively. The $D$ and  $c_{\boldsymbol{\theta}}$ criteria are used as illustrative examples; more generally $\Psi$ is a positive homogeneous convex function.

There are two definitions of optimal design, exact and continuous, defined by the set over which the minimization in \eqref{eq:Opt} takes place. An exact design restricts the scaled allocation weights, $nw_{i}$, to be integers. A solution to \eqref{eq:Opt} is an exact optimal design if the minimization is with respect to all possible exact designs. In this work the primary interest is in \textit{continuous optimal design}. A continuous design relaxes the integer restriction to $0 \le w_{i} \le 1$ and $\sum_{i} w_{i} = 1$. A continuous optimal design is a solution to \eqref{eq:Opt} with respect to the set of all possible continuous designs, denoted $\Xi_{\Delta}$. In nonlinear models, to find continuous optimal designs often requires a numeric search algorithm. \cite*{Wynn:TheS:1970} and \cite*{Fedo:Theo:1972} are early examples of continuous optimal design algorithms. The continuous optimal design problem remains an active area of research and there exist many modern solutions [\cite*{Yu:Mono:2011}, \cite*{Yang:Bied:Tang:2013}, \cite*{Harm:Filo:Rich:ARan:2019} and others].


In Section \ref{sec:intro} it was argued that optimizing the Fisher information in the sample should not be the sole design objective. In the remainder of this section a framework for optimal relevant subset design is developed.

\subsection{Optimal Relevant Subset Design}
Adaptive designs require sequentially observed responses. In the sequential setting the sample is an ordered set of experimental runs. Specifically, some of the data from the preceding runs is available prior to the design assignment of the current run. 

Consider a sequential experiment with $J$ runs, where each run consists of a known number, $m(j)\ge 1$, of independent observations, for $j=1,\ldots,J$. Denote the total sample size up to and including the $j$th run as $n(j) = \sum_{k=1}^{j} m(k)$, where $n = n(J)$ is fixed. Note, $(j)$ is used to indicate a quantity associated with run $j$, e.g. $\boldsymbol{y}(j) = [\boldsymbol{y}_{1}(j),\ldots,\boldsymbol{y}_{d}(j)]^{T}$, $\boldsymbol{A}(j) = [\boldsymbol{a}_{1}(j),\ldots,\boldsymbol{a}_{d}(j)]$ and $\boldsymbol{\hat{\eta}}(j) = [\hat{\eta}_{1}(j),\dots,\hat{\eta}_{d}(j)]^{T}$ represent the data, the relative subset and the MLE of $\boldsymbol{\eta}$ from the first $j$ runs, respectively. After run $j$ these quantities are known and can be used to determine the design of the current run. Let $x(k)$ be the design point for the $k$th observation in the experiment then $\boldsymbol{x}(j) = \{x(1),\ldots,x[n(j)]\}^{T}$ denotes the collection of design points up to and including run $j$ and $\xi(j) = \sum_{k=1}^{n(j)} \delta_{x(k)}$ is the design for the first $j$ runs, where $\delta_{x}$ is a design with support, $x$, and unit allocation. Note $\boldsymbol{x}(j)$ represents the sequential design assignments and are not necessarily unique points. Let $x_{1}(j),\ldots,x_{d}(j)$ represent the unique points in the design space with positive weight after run $j$. Per the described notation $\boldsymbol{y}(J)$, $\boldsymbol{A}(J)$, $\boldsymbol{x}(J)$ and $\boldsymbol{\hat{\eta}}(J)$ correspond to variables from the entire experiment; however, the dependence of $(J)$ is dropped when the meaning is clear.

The objective of this work is to develop an adaptive framework that optimizes the Fisher information in the relevant subset, $\boldsymbol{H}_{\boldsymbol{A}}^{\boldsymbol{x}}$, with respect to a convex criterion, $\Psi$. Unfortunately, this cannot be done directly since $\boldsymbol{A}$ remains unknown until the data from all observations has been collected. Instead, $\boldsymbol{H}_{\boldsymbol{A}}^{\boldsymbol{x}}$ is increased by sequentially optimizing the Fisher information in the relevant subset generated by the first $j$ runs, which, in the location family, can be shown to be  
\begin{align} \label{eq:Tj}
  \boldsymbol{T}_{\boldsymbol{A}(j)}^{\xi}(\boldsymbol{\theta}) = E\left[\boldsymbol{I}_{\boldsymbol{\mathcal{Y}}(j+1)}^{\boldsymbol{x}(j+1)}(\boldsymbol{\theta})|\boldsymbol{\mathcal{A}}(j) =  \boldsymbol{A}(j)\right] = \mu m(j+1)\boldsymbol{M}^{\xi}(\boldsymbol{\theta}) + \boldsymbol{H}_{\boldsymbol{A}(j)}^{\boldsymbol{x}(j)}(\boldsymbol{\theta}),
\end{align}  
where, here, $\xi$ is the design for run $j+1$. In principle, for a sequential experiment the design that optimizes $\boldsymbol{T}_{\boldsymbol{A}(j)}^{\xi}$ can be found. Practically, as previously remarked, there does not exist a closed form expression for $\boldsymbol{H}_{\boldsymbol{A}(j)}^{\boldsymbol{x}(j)}$ which makes optimizing $\boldsymbol{T}_{\boldsymbol{A}(j)}^{\xi}$ directly impractical. For the purpose of optimization Theorem \ref{thm:CondInfo} can be used to obtain the approximate form of \eqref{eq:Tj}
\begin{align} \label{eq:Thatj}
    \boldsymbol{T}_{\boldsymbol{A}(j)}^{\xi}(\boldsymbol{\theta}) \approx \boldsymbol{\hat{T}}_{\boldsymbol{A}(j)}^{\xi}(\boldsymbol{\theta}) = \mu m(j+1)\boldsymbol{M}^{\xi}(\boldsymbol{\theta}) + \boldsymbol{J}_{\boldsymbol{A}(j)}^{\boldsymbol{x}(j)}(\boldsymbol{\theta}).
\end{align}
The above has a closed form expression which allows the optimization problem considered in this work to be clearly stated as to find the design that optimizes $\boldsymbol{\hat{T}}_{\boldsymbol{A}(j)}^{\xi}$. This motivates the following definition of the $\Psi_{\boldsymbol{A}(j)}$-optimal design.
\begin{defn}
The design $\xi_{\boldsymbol{A}(j)}^{*}(\boldsymbol{\theta})$ is $\Psi_{\boldsymbol{A}(j)}$-optimal if 
\begin{align}
    \xi_{\boldsymbol{A}(j)}^{*}(\boldsymbol{\theta}) = \arg\min_{\xi \in \Xi_{\Delta}} \Psi\{\boldsymbol{\hat{T}}_{\boldsymbol{A}(j)}^{\xi}(\boldsymbol{\theta})\}.
\end{align} 
\end{defn}
In principle, the design $\xi_{\boldsymbol{A}(j)}^{*}(\boldsymbol{\theta})$ can be computed following run $j-1$ and iteratively used as the design for $j=2,\ldots,J$. The next section addresses how to find $\xi_{\boldsymbol{A}(j)}^{*}(\boldsymbol{\theta})$ in practice. 

\subsection{General Equivalence Theorem}

One of the most important tools in the construction of continuous optimal designs is the general equivalence theorem [\cite{Kief:Wolf:TheE:1960, Whit:AnEx:1973,Keif:Opti:1975}]. Denote the derivative of $\Psi$ in the direction of $\delta_{x}$ as
\begin{align} \label{eq:phi}
\phi^{\xi}(x,\boldsymbol{\theta}) = \nabla_{\delta_{x}}\Psi\{\boldsymbol{M}^{\xi}(\boldsymbol{\theta})\}.
\end{align}
The \textit{general equivalence theorem} states that the following are equivalent (1) $\xi^{*}(\boldsymbol{\theta})$ is the continuous optimal design (2) $\xi^{*}(\boldsymbol{\theta}) =\arg\max_{\xi \in \Xi_{\Delta}}\min_{x\in\mathscr{X}}\phi^{\xi}(x,\boldsymbol{\theta})$ and (3) the $\min_{x\in\mathscr{X}}\phi^{\xi^{*}(\boldsymbol{\theta})}(x,\boldsymbol{\theta}) = 0$ with the minimum occurring at the support points of $\xi^{*}(\boldsymbol{\theta})$.

The importance of $\boldsymbol{J}_{\boldsymbol{A}(j)}^{\boldsymbol{x}(j)}$ was discussed in Section \ref{sec:CondVar}. An additional property is that by defining $\boldsymbol{\hat{T}}_{\boldsymbol{A}(j)}^{\xi}$ as a function of $\boldsymbol{J}_{\boldsymbol{A}(j)}^{\boldsymbol{x}(j)}$ the following general equivalence theorem holds.

\begin{theorem} \label{thm:Gen_Eq} 
Under the conditions in the Section \ref{sec:Tech} (1) $\xi_{\boldsymbol{A}(j)}^{*}(\boldsymbol{\theta})$ is a $\Psi_{\boldsymbol{A}(j)}$-optimal design; (2) $\xi_{\boldsymbol{A}(j)}^{*}(\boldsymbol{\theta})=\arg\max_{\xi \in \Xi_{\Delta}}\min_{x\in\mathscr{X}}\nu_{\boldsymbol{A}(j)}^{\xi}(x,\theta)$ and (3) $\min_{x\in\mathscr{X}}\nu_{\boldsymbol{A}(j)}^{\xi_{\boldsymbol{A}(j)}^{*}(\boldsymbol{\theta})}(x,\theta) = 0$ for all $x\in\mathcal{X}$, with the minimum occurring at the support points of $\xi_{\boldsymbol{A}(j)}^{*}(\boldsymbol{\theta})$ are equivalent, where
\begin{align}
\nu_{\boldsymbol{A}(j)}^{\xi}(x,\theta) &= \dot{\eta}_{x}(\boldsymbol{\theta})^{T} \{R_{\boldsymbol{A}(j)}^{\xi}(\boldsymbol{\theta})\}^{-1}\dot{\eta}_{x}(\boldsymbol{\theta}) - tr[\boldsymbol{M}^{\xi}(\boldsymbol{\theta})\{R_{\boldsymbol{A}(j)}^{\xi}(\boldsymbol{\theta})\}^{-1}],
\end{align}
\begin{align}
R_{\boldsymbol{A}(j)}^{\xi}(\boldsymbol{\theta}) = \mu\beta(j+1) \boldsymbol{M}^{\xi}(\boldsymbol{\theta}) + [1-\mu\beta(j+1)]\boldsymbol{M}^{\tau_{\boldsymbol{A}(j)}}(\boldsymbol{\theta}),
\end{align}
$\beta(j+1) =  m(j+1)/[\mu m(j+1) + Q_{\boldsymbol{A}(j)}]$,
$\tau_{\boldsymbol{A}(j)} = \{(x_{i}(j),\omega_{i}[\boldsymbol{A}(j)])\}_{i=1}^{d}$,
$Q_{\boldsymbol{A}(j)} = \sum_{i} i_{\boldsymbol{a}_{i}(j)}$ and $\omega_{i}[\boldsymbol{A}(j)] = i_{\boldsymbol{a}_{i}(j)}/Q_{\boldsymbol{A}(j)}$. 
\end{theorem}

The importance of this theorem is that it establishes that many existing optimal design algorithms to find FLODs can be adapted to find the optimal RSDs. For illustrative purposes in the next section a one-step ahead approach is developed.

\begin{remark}
Theorem \ref{thm:Gen_Eq} holds for all $\boldsymbol{\theta}\in\Theta$. One could use a fixed initial guess at the value of $\boldsymbol{\theta}$ to find the optimal design after every run. This approach would be analogous to the FLOD which also uses an initial guess to compute the design completely in advance of data collection. In this work the primary interest is determining the optimal design for $\boldsymbol{\theta}$ evaluated at the MLE from the preceding $j$ runs, denoted $\boldsymbol{\hat{\theta}}(j)$. Such a design is developed in Section \ref{sec:OSA}.
\end{remark}

\begin{remark} \label{rem:nonloc}
Equation \eqref{eq:Thatj} requires location family errors. For general error distributions the justification of $\boldsymbol{\hat{T}}_{\boldsymbol{A}(j)}^{\xi}$ is slightly different. The expected value of $\boldsymbol{I}_{\boldsymbol{y}}^{\boldsymbol{x}}$ conditional on the responses from the first $j$ runs, evaluated at $\boldsymbol{\hat{\eta}}(j)$, is
\begin{align}
\boldsymbol{\hat{T}}_{\boldsymbol{A}(j)}^{\xi}(\boldsymbol{\theta}) &= E\left[\boldsymbol{I}_{\boldsymbol{y}}^{\boldsymbol{x}}(\boldsymbol{\theta})|\boldsymbol{\mathcal{Y}}(j) = \boldsymbol{y}(j)\right]_{\boldsymbol{\eta} = \boldsymbol{\hat{\eta}}(j)} = \mu m(j+1)\boldsymbol{M}^{\xi}(\boldsymbol{\theta}) + \boldsymbol{J}_{\boldsymbol{A}(j)}^{\boldsymbol{x}(j)}(\boldsymbol{\theta}) \label{eq:Tj2}.
\end{align}
The form of the above is the same as \eqref{eq:Thatj}; as a result Theorem \ref{thm:Gen_Eq} holds without the location condition; however, its justification is not as strong as in the location error setting. Equation \eqref{eq:Tj2} is obtained by conditioning on the entire data from the preceding runs rather than only the relevant subset. As a result the design may contain information about the parameters. However, as discussed in Section \ref{sec:CondVar} the use of $\boldsymbol{J}_{\boldsymbol{A}(j)}^{\boldsymbol{x}(j)}(\boldsymbol{\theta})$ is valid beyond location errors.    
\end{remark}

\section{One-Step Ahead Optimal Relevant Subset Design} \label{sec:OSA}

In this section Theorem \ref{thm:Gen_Eq} is applied to develop a one-step ahead optimal RSD for nonlinear models. To begin, the one-step ahead approach to adaptive optimal design (AOD) is reviewed. Let $\xi(j)$ be any arbitrary design from the first $j$ runs. The basis of the one-step ahead approach is to update $\xi(j)$ with the design of run $j+1$, i.e. to determine $\xi(j+1) = [1-\alpha(j)]\xi(j) + \alpha(j)\xi$, where $\alpha(j)$ is the step size for run $j+1$. For a given $\alpha(j)>0$ the optimal design for run $j+1$ is 
\begin{align} \label{eq:FirstOrderOpt}
    \xi^{*} = \arg\min_{\xi\in\Xi_{\Delta}} \Psi\{\boldsymbol{M}^{\alpha(j)\xi + [1 - \alpha(j)]\xi(j)}(\boldsymbol{\theta})\}.
\end{align}
Equation \eqref{eq:FirstOrderOpt} does not simplify the design problem stated in \eqref{eq:Opt}. However, for a small enough $\alpha(j)$ the first order approximation $\xi^{*} \approx \delta_{x(j+1)}$, where
\begin{align}
x(j+1) = \arg\min_{x\in\mathcal{X}} \phi^{\xi(j)}(x,\boldsymbol{\theta})
\end{align}
is valid. The first order approach to FLOD is obtained by setting $\xi(j+1) =\alpha(j)\delta_{x(j+1)} + [1-\alpha(j)]\xi(j)$ and iterating for $j=m(1)+1,\ldots,J$. The general equivalence theorem guarantees the existence of an $x(j+1)$ such that $\Psi\{\boldsymbol{M}^{\xi(j+1)}(\boldsymbol{\theta})\} < \Psi\{\boldsymbol{M}^{\xi(j)}(\boldsymbol{\theta})\}$ [\cite{Fedo:Opti:2010}]. Convergence of the first order approach is discussed in \citet{Wu:Wynn:1978}. The optimal step size, $\alpha(j)$, has been studied [\cite{Fedo:Theo:1972}, \cite{Cook:Nach:Comp:1989}, and others]. 

Selecting $\alpha(j) = (j+1)^{-1}$ mimics a fully sequential experiment; where each run consists of a single observation, i.e., $m(j) = 1$ for all $j=m(1) + 1,\ldots,J$ and $J = n-m(1)$. The one-step ahead AOD exploits this approach by allocating the first $m(1)$ observations according to a fixed design and then allocating observation $j+1$ to
\begin{align} \label{eq:xj}
\hat{x}(j+1) = \arg\min_{x\in\mathcal{X}} \phi^{\xi(j)}[x,\boldsymbol{\hat{\theta}}(j)],
\end{align}
for $j=m(1)+1,\ldots,J$. Denote the design following a one-step ahead AOD as $\hat{\xi}_{AOD}$.

Finally, the one-step ahead AOD is extended to obtain a one-step ahead optimal RSD. Substituting $\delta_{x}$ for $\xi$ in the right hand side of \eqref{eq:Thatj} yields, see the proof of Theorem \ref{thm:Gen_Eq},
\begin{align} \label{eq:TequalM}
\boldsymbol{\hat{T}}_{\boldsymbol{A}(j)}^{\delta_{x}}(\boldsymbol{\theta}) &= \beta(j+1)^{-1} \boldsymbol{M}^{\tau(j+1)}(\boldsymbol{\theta}),
\end{align}
where $\tau(j+1) = \beta(j+1) \delta_{x} + [1 - \beta(j+1)]\tau_{\boldsymbol{A}(j)}$ and note that if $m(j+1)=1$ then $\beta(j+1) = [\mu + Q_{\boldsymbol{A}(j)}]^{-1}$, where $Q_{\boldsymbol{A}(j)}$ is defined in Theorem \ref{thm:Gen_Eq}. After the initial run $\beta(j+1)$ is a known constant and the design that minimizes $\Psi[\boldsymbol{\hat{T}}_{\boldsymbol{A}(j)}^{\delta_{x}}(\boldsymbol{\theta})]$ minimizes $\Psi[\boldsymbol{M}^{\tau(j+1)}(\boldsymbol{\theta})]$. Further, since $i_{\boldsymbol{a}_{i}(j)}\ge 0$ for all $i=1,\ldots,d$ $\tau_{\boldsymbol{A}(j)}\in \Xi_{\Delta}$ for all $j=m(1)+1,\ldots,J$. The consequence of this discussion is that the one-step ahead optimal RSD is obtained by substituting $\tau_{\boldsymbol{A}(j)}$ for $\xi$ in \eqref{eq:xj}. Every other step in the one-step ahead AOD remains the same. The following algorithm, referred to as the optimal RSD in nonlinear models, is derived from the preceding discussion.
\begin{algorithm} \label{alg:RSD} One-step ahead optimal relevant subset design
\begin{enumerate}[nolistsep]
\item{Place the first $m(1)$ observations according to a pre-determined fixed design.}
\item{For $j = m(1)$ calculate $\tau_{\boldsymbol{A}(j)}$ and $\boldsymbol{\hat{\theta}}(j)$ based on the data available from the first $j$ observations.}
\item{Allocate observation $j+1$ to the point
$\hat{x}(j+1) = \arg\min_{x\in\mathcal{X}} \phi^{\tau_{\boldsymbol{A}(j)}}[x,\boldsymbol{\hat{\theta}}(j)]$.
}
\item{Repeat step 2 and 3 for $j=m(1)+1,\ldots,J$.}
\end{enumerate}
\end{algorithm}
The optimal RSD obtained from this algorithm is denoted $\hat{\xi}_{RSD}$. This design is independent of the underlying parameters.

\begin{remark}
In nonlinear models the existence of the MLE is not guaranteed. If after $m(1)$ runs the MLE does not exist additional points can be added to the initial design.
\end{remark}

\begin{remark}
The extension to $m(j)>1$ is straightforward. Equation \eqref{eq:xj} in the optimal RSD algorithm can be replaced with $\xi_{\boldsymbol{A}(j)}^{*}[\boldsymbol{\hat{\theta}}(j)]$. Modern algorithms that rely on the general equivalence theorem can be adapted to find $\xi_{\boldsymbol{A}(j)}^{*}[\boldsymbol{\hat{\theta}}(j)]$ for any $m(j)$.
\end{remark}

\begin{remark} \label{rem:repeat}
As previously stated, the justification of $\boldsymbol{J}_{\boldsymbol{A}(j)}^{\boldsymbol{x}(j)}$, and by extension the justification of $\boldsymbol{\hat{T}}_{\boldsymbol{A}(j)}^{\xi}$, assumes sufficient repeats. In cases where sufficient repeats do not exists replace $\boldsymbol{\hat{T}}_{\boldsymbol{A}(j)}^{\xi}$ with
\begin{align}
\boldsymbol{S}_{\boldsymbol{y}(j)}^{\xi}(\boldsymbol{\theta}) = \mu m(j+1)\boldsymbol{M}^{\xi}(\boldsymbol{\theta}) + \boldsymbol{K}_{\boldsymbol{y}(j)}^{\boldsymbol{x}(j)}(\boldsymbol{\theta})
\end{align}
to develop Algorithm \ref{alg:RSD}. Specifically, to accommodate a design with limited repeats replace $\hat{x}_{j+1}$ with $\hat{x}'(j+1) = \min_{x\in\mathcal{X}} \phi^{\tau_{\boldsymbol{y}(j)}[\boldsymbol{\hat{\theta}}(j)]}[x,\boldsymbol{\hat{\theta}}(j)]$
in Step 3 in Algorithm \ref{alg:RSD}, where $\tau_{\boldsymbol{y}(j)}(\boldsymbol{\theta}) = \{(x_{i}(j),\omega_{i}[\boldsymbol{y}(j)])\}_{i=1}^{d}$,
$\omega_{i}[\boldsymbol{y}(j)] = i_{\boldsymbol{y}_{i}(j)}/Q_{\boldsymbol{y}(j)}$ and $Q_{\boldsymbol{y}(j)}(\boldsymbol{\theta}) = \sum_{i} i_{\boldsymbol{y}_{i}}(\eta_{i})$. A general equivalence theorem for the design that minimizes $\Psi\{\boldsymbol{S}_{\boldsymbol{y}(j)}^{\xi}(\boldsymbol{\theta})\}$ can also be developed if $\boldsymbol{K}_{\boldsymbol{y}(j)}^{\boldsymbol{x}(j)}(\boldsymbol{\theta})$ is non-negative definite. However, this is not guaranteed in general. One solution for this is to regularize $\boldsymbol{K}_{\boldsymbol{y}(j)}^{\boldsymbol{x}(j)}(\boldsymbol{\theta})$ by adding $c\boldsymbol{I}_{p}$, where $c$ is a small positive constant and $\boldsymbol{I}_{p}$ is the identity matrix to ensure non-negative definiteness.
\end{remark}


\section{Simulation Study} \label{sec:sim}
In this section a simulation study is conducted to compare the efficiency of the optimal RSD and the AOD relative to the FLOD. For each example the FLOD is computed based on the true value of the parameters. As a result, it is not possible to use this design in practice and it should be viewed as a benchmark for fixed designs. This study includes different nonlinear mean functions, error distributions and optimality criteria.

\subsection{Mean Functions and Fixed Locally Optimal Designs}
In this simulation study three different mean functions are considered. Each of them are well known examples of nonlinear models that have been extensively considered both in analysis and design. For each of the three mean functions the $D$- and $c_{\boldsymbol{\theta}}$- optimal design are presented. Each design is based on a real world data set.

\emph{Example 1:} The first mean function is the Michaeles-Menten enzyme kinetic function with 
$\eta_{x}(\boldsymbol{\theta}) = \theta_{1}x/(\theta_{2} + x)$,
where $x\in\mathcal{X}=[0,x_{\max}]$, the response $y$ is the reaction velocity, $\theta_{1}>0$ is the maximum velocity and $\theta_{2}>0$ is the Michaeles-Menten constant and corresponds to the value of $x$ such that $y$ is half of the maximum velocity. This model has been extensively studied in the design literature [\cite{Lope:Wong:Desi:2002}, \cite{Dett:Bied:Robu:2003} and others]. \cite{Cres:Keig:1979} used the Michaeles-Menten mean function  to model estrogen bound to receptors, $y$, to the amount of hormone not bound to receptors, $x$, in a study of human breast cancer. The design space for their experiment was $\mathcal{X}=[0,2000]$ and MLEs $\hat{\theta}_{1} = 43.95$ and $\hat{\theta}_{2} = 236.53$ were obtained. The variance of $\varepsilon$, denoted $\sigma^{2}$, does not influence the design; however, it does impact the variability of the responses. In the \cite{Cres:Keig:1979} data set $\hat{\sigma}=1.39$. This example was also considered in \cite{Dett:Bied:Robu:2003}. 

\cite{Lope:Wong:Desi:2002} find that the local $D$-optimal design places equal weight on the two points $x_{\max}$ and $x_{\max}/[(x_{\max}/\theta_{2}) + 2]$. The Michaeles-Menten constant is often of particular interest. This corresponds to a $c_{\boldsymbol{\theta}}$-optimal design with $c_{\boldsymbol{\theta}} = (0,1)$.  \cite{Lope:Wong:Desi:2002} show that the local $c_{\boldsymbol{\theta}}$-optimal design for the Michaeles-Menten constant has support $x_{\max}$ and $\theta_{2}b(\sqrt{2} - 1)/[1 + b\sqrt{2} (\sqrt{2} - 1)]$, where $b=x_{\max}/\theta_{2}$, with corresponding weights $1/\sqrt{2}$ and $1-1/\sqrt{2}$. 

Specifically, using the value of the parameter estimates from \cite{Cres:Keig:1979} as the true values the local $D$- and $c_{\boldsymbol{\theta}}$-optimal designs are
$\xi_{D} = \{(191.285.1/2), (2000,1/2)\}$ and $\xi_{c_{\boldsymbol{\theta}}} = \{(139.157,1/\sqrt{2} ), (2000, 1 - 1/\sqrt{2}) \}$, respectively.

\emph{Example 2:} The second mean function, $\eta_{x}(\boldsymbol{\theta}) = \theta_{1}(1-e^{-\theta_{2}x})$, is from the class of exponential decay models, where $\theta_{1}>0$ represents the initial amount of the substance and $\theta_{2}>0$ is the decay rate. Optimal designs for exponential decay models have been studied in \cite{Han:Chal:Dandc:2003}, \cite{Dett:Lope:Rodr:Maxi:2006}, among others. \cite{Dett:Lope:Rodr:Maxi:2006} present the data from an experiment measuring metabolism of glucose in sheep conducted by Allan Danfaer, Department of Animal Nutrition and Physiology, Danish Institute of Agricultural Sciences.  The design space for this experiment was $\mathcal{X}=[0,500]$ and MLEs $\hat{\theta}_{1} = 1.215$, $\hat{\theta}_{2} = 0.01539$ and $\hat{\sigma}=0.054$ were obtained.

\cite{Dett:Lope:Rodr:Maxi:2006} state that the $D$-optimal design places equal weight on the support points $x_{\max}$ and $1/\theta_{2} - x_{\max}e^{-\theta_{2}}/(1+e^{-\theta_{2}})$. In exponential decay models the decay parameter is often of particular interest.  \cite{Han:Chal:Dandc:2003} found the local $c_{\boldsymbol{\theta}}$-optimal designs for the decay parameter, $\theta_{2}$, for a variety of exponential decay models. However, this specific model was not included. The local $c_{\boldsymbol{\theta}}$-optimal design for $\theta_{2}$ for this model was found numerically.

The FLODs considered for this mean function are based on the data for the experiment presented in \cite{Dett:Lope:Rodr:Maxi:2006} as the true values, which yields local $D$- and $c_{\boldsymbol{\theta}}$-optimal designs of
$\xi_{D} = \{(70.972,1/2),(500,1/2) \}$ and $\xi_{c_{\boldsymbol{\theta}}} = \{(54.551,0.652),(500, 0.348)\}$, 
respectively. 

\emph{Example 3:} The final mean function considered is the three parameter compartmental model with mean function 
$\eta_{x}(\boldsymbol{\theta}) = \theta_{1}(e^{-\theta_{2}x} - e^{-\theta_{3}x})$, where $\theta_{3}$ is the initial concentration $\theta_{2}(>\theta_{3})$ governs the elimination rate and $\theta_{1}>0$ determines the height of the curve. \citet{Fres:Aspe:1984} used this mean function to model the concentration of theophylline in the blood of a horse over time.  In the \citet{Fres:Aspe:1984} experiment the design space was $\mathcal{X}=[0,48]$ and the parameters were estimated as $\hat{\theta_{1}} = 21.8$, $\hat{\theta_{2}} = 0.059$ and $\hat{\theta}_{3} = 4.29$ with $\hat{\sigma}=1.13$. \cite{Atki:Chal:Herz:Opti:1993} and \cite{Pazm:Pron:2014} both consider this example in the context of design.

Analytic forms for designs are not currently available. \cite{Atki:Chal:Herz:Opti:1993} gives numeric solutions for the local $D$-optimal design and the local $c_{\boldsymbol{\theta}}$-optimal design for the time to maximum concentration based on the estimates from the \citet{Fres:Aspe:1984} data. The time to maximum concentration is of practical interest and it is given by $\log(\theta_{1}/\theta_{2})/(\theta_{1} - \theta_{2})$. The local $c_{\boldsymbol{\theta}}$-optimal design for time to maximum concentration is a two-point design and as a result it is not possible to estimate all three of the model parameters using this design. Rather than use the two point $c_{\boldsymbol{\theta}}$-optimal design, in the simulation study a Bayesian design, proposed in \cite{Atki:Chal:Herz:Opti:1993}, is used. This design uses a prior distribution centered around the estimated parameter values in Fresen (1984). Specifically, the local $D$-optimal design and Bayesian $c_{\boldsymbol{\theta}}$-optimal design considered are
$\xi_{D} = \{(0.229,1/3),(1.389,1/3), (18.417,1/3)\}$ and $\xi_{c_{\boldsymbol{\theta}}} = \{(0.183,0.6023),(2.464,0.298), (8.854, 0.010)\}$, respectively. The Bayesian $c_{\boldsymbol{\theta}}$-optimal design has an efficiency of approximately 99\% with respect to the local $c_{\boldsymbol{\theta}}$-criterion. 

Each of the designs given are continuous optimal designs. In real experiments, and simulated experiments,  continuous designs must be rounded in order to ensure the total sample size equals $n$. For the simulation study the Adams appropriation method was used; see \cite{Puke:Ried:Effi:1992}.

The FLOD is computed at the parameter values that are used to generate the simulated data. In practice these values are unknown and cannot be used to initialize the adaptive designs. In nonlinear models it is important to initialize the adaptive designs to ensure the existence of the MLE with high probability. This is even more important in a simulation since it is required that the MLE exists for every iteration. The selection of the initial designs for the adaptive designs, the optimal RSD and AOD, used in the simulation study is discussed in the supplemental materials. 

\subsection{Error Distributions}
In addition to various mean functions, different error distributions are considered. In the simulation study each of the three examples will be examined for three different error distributions. 

The first error distribution considered is the standard Cauchy distribution.  \cite{McCu:Cond:1992} extensively studied conditional inference in the Cauchy model. Linear regression with Cauchy errors is a popular method to analyze data with heavy tails and is considered in \cite{Esti:Kadi:Murt:1977}, \cite{He:Simp:Wang:2000}, \cite{Mize:Mull:Brea:2002}, among others. The probability density function (p.d.f.) of a Cauchy distribution is proportional to 
$f(\varepsilon) \propto [1 + (\varepsilon/\sigma)^{2}]^{-1}$.

The second error distribution is the exponential power distribution. This distribution was proposed by \cite{Subb:OnTh:1923} and is a popular method to deal with non-normal data. The properties of this distribution have been extensively studied [\cite{Box:Anot:1953}, \cite{Turn:OnHe:1960} and others]. In the extreme case it provides an approximation to the uniform distribution. The p.d.f. is proportional to $f(\varepsilon) \propto e^{-|\varepsilon/\sigma|^{\zeta}/\zeta}$.
In the simulation study $\zeta=4$ is used. 

The final error distribution considered is the $q$-Gaussian distribution. This distribution arises in statistical mechanics [\cite{Tsal:Nona:2009}] and has been used to model financial data [\cite{Borl:2002}] among other scientific disciplines. \cite{Borl:2002} found that for $q=3/2$ this model closely fits the empirically observed distribution for many financial time series. For $1<q<3$ the $q$-Gaussian has a p.d.f. proportional to
$f(\varepsilon) \propto \left[1 -  (1 - q) \varepsilon^2/(2\sigma^{2})\right]^{1/(1 - q)}$. A value of $q=3/2$ is used in the simulation studies.

\subsection{Results}

This section summarizes the results from the outlined simulation study for a variety of metrics. Recall from the introduction the primary goal of this work was to develop a design that accomplishes two objectives, to optimize the Fisher information in the sample and to optimize inference. As previously stated, the primary motivation for optimizing the Fisher information in the sample is to minimize the mean square error (MSE) of $\boldsymbol{\hat{\theta}}$. The competing methods will also be assessed by MSE. The results with respect to these objectives will be compared for both the optimal RSD and the AOD relative to the FLOD. Specifically, in this section the adaptive methods are compared relative to the FLOD for the following measures of efficiency
\begin{align}
    {\rm{RM}}_{\xi}^{\Psi} = \frac{\Psi[\boldsymbol{M}^{\xi^{*}}(\boldsymbol{\theta})]}{\Psi[\boldsymbol{M}^{\xi}(\boldsymbol{\theta})]}, {\rm{RJ}}_{\xi}^{\Psi}  = \frac{{\rm{E}}[\Psi[\boldsymbol{J}_{\mathcal{A^{*}}}^{\boldsymbol{x}}(\boldsymbol{\hat{\theta}}_{\xi^{*}})]]}{{\rm{E}}[\Psi[\boldsymbol{J}_{\boldsymbol{\mathcal{A}}_{\xi}}^{\boldsymbol{x}}(\boldsymbol{\hat{\theta}}_{\xi})]]}, \mbox{ and } {\rm{RMSE}}_{\xi}^{\Psi}  = \frac{\Psi[\{\rm{MSE}(\boldsymbol{\hat{\theta}}_{\xi^{*}})\}^{-1}]}{\Psi[\{\rm{MSE}(\boldsymbol{\hat{\theta}}_{\xi})\}^{-1}]}
\end{align}
obtained via simulation, where $\boldsymbol{A}_{\xi}$ and $\boldsymbol{\hat{\theta}}_{\xi}$ are the relevant subset and the MLE corresponding to the design $\xi$. Recall $\xi^{*}$, $\hat{\xi}_{RSD}$ and $\hat{\xi}_{AOD}$ represent the FLOD, the optimal RSD and the AOD, respectively. As stated, each measure above is defined relative to the FLOD. For example, ${\rm{RM}}_{\hat{\xi}_{RSD}}^{\Psi}$ represents the Fisher information in the sample from the optimal RSD relative to the Fisher information in the sample from the optimal design. Values of ${\rm{RM}}_{\hat{\xi}_{RSD}}^{\Psi}>1$ indicate cases where the optimal RSD results in more information than the FLOD. The measure ${\rm{RJ}}_{\hat{\xi}_{RSD}}^{\Psi}$ represents the expected value of $\Psi$ evaluated at the Fisher information in the relevant subset relative to the FLOD. As argued throughout $\boldsymbol{J}_{\boldsymbol{\mathcal{A}}_{\xi}}^{\boldsymbol{x}}$ should be used in inference which implies that values ${\rm{RJ}}_{\hat{\xi}_{RSD}}^{\Psi}>1$ indicate that inference (confidence ellipsoids/intervals, power, etc) is better following the optimal RSD than following the FLOD. The measure ${\rm{RMSE}}_{\hat{\xi}_{RSD}}^{\Psi}$ represents the relative efficiency of the MSE of the estimates from the optimal RSD relative to the FLOD. As with the previous measures ${\rm{RMSE}}_{\hat{\xi}_{RSD}}^{\Psi}>1$ indicate cases were the optimal RSD is more efficient with respect to MSE than the FLOD. The interpretation of each measure is the same for the AOD. 

\subsubsection{Inference Efficiency} 

\begin{sidewaysfigure}
\setlength{\tempwidth}{.31\linewidth}
\settoheight{\tempheight}{\includegraphics[width=\tempwidth]{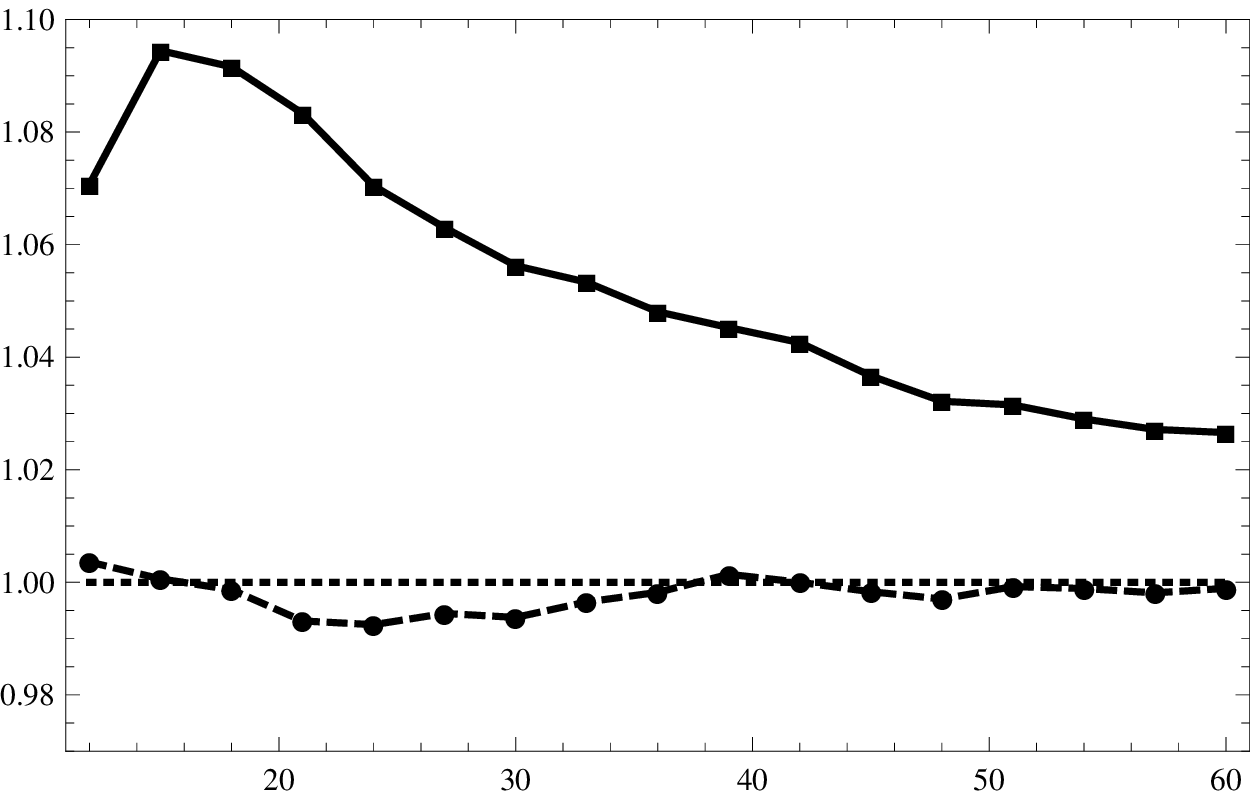}}
\centering
\hspace{\baselineskip} 
\columnname{{\normalfont\scriptsize Cauchy}}\hfil
\columnname{{\normalfont\scriptsize Exponential Power}}\hfil
\columnname{{\normalfont\scriptsize q-Gaussian}}\\
\rowname{{\normalfont\scriptsize Michaeles-Menten}}
\begin{subfigure}[b]{\tempwidth}
       \centering
       \includegraphics[width=\tempwidth]{CauchyMMOID.eps}
\end{subfigure}
\begin{subfigure}[b]{\tempwidth}
       \centering
       \includegraphics[width=\tempwidth]{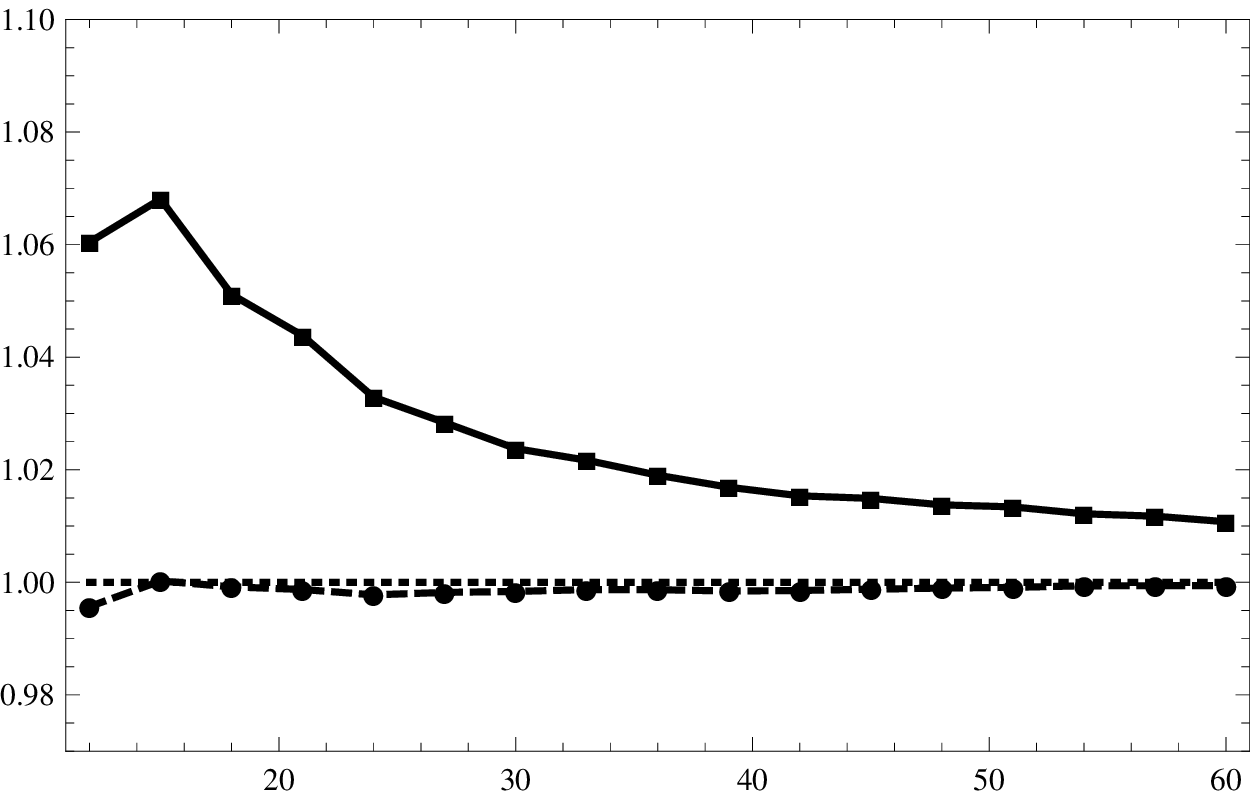}
\end{subfigure}
\begin{subfigure}[b]{\tempwidth}
       \centering
       \includegraphics[width=\tempwidth]{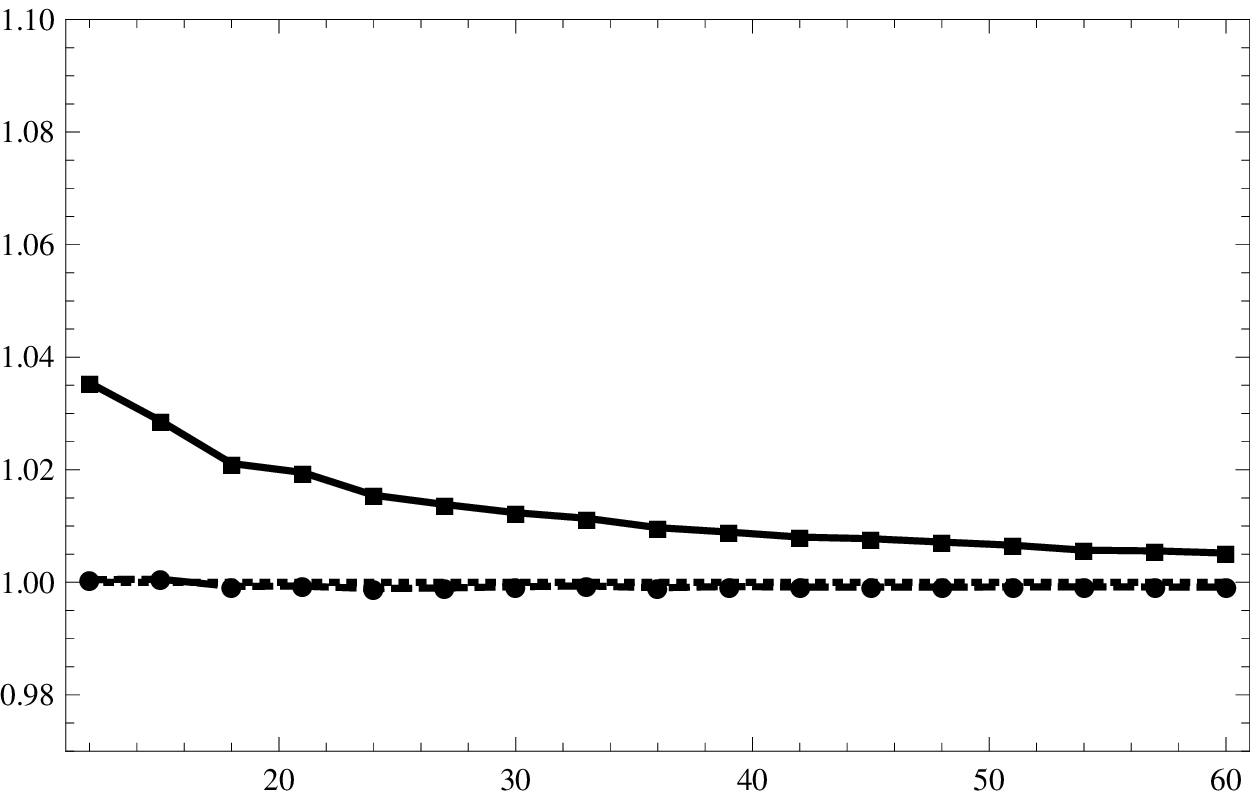}
\end{subfigure} \\
\rowname{{\normalfont\scriptsize Decay}}
\begin{subfigure}[b]{\tempwidth}
       \centering
       \includegraphics[width=\tempwidth]{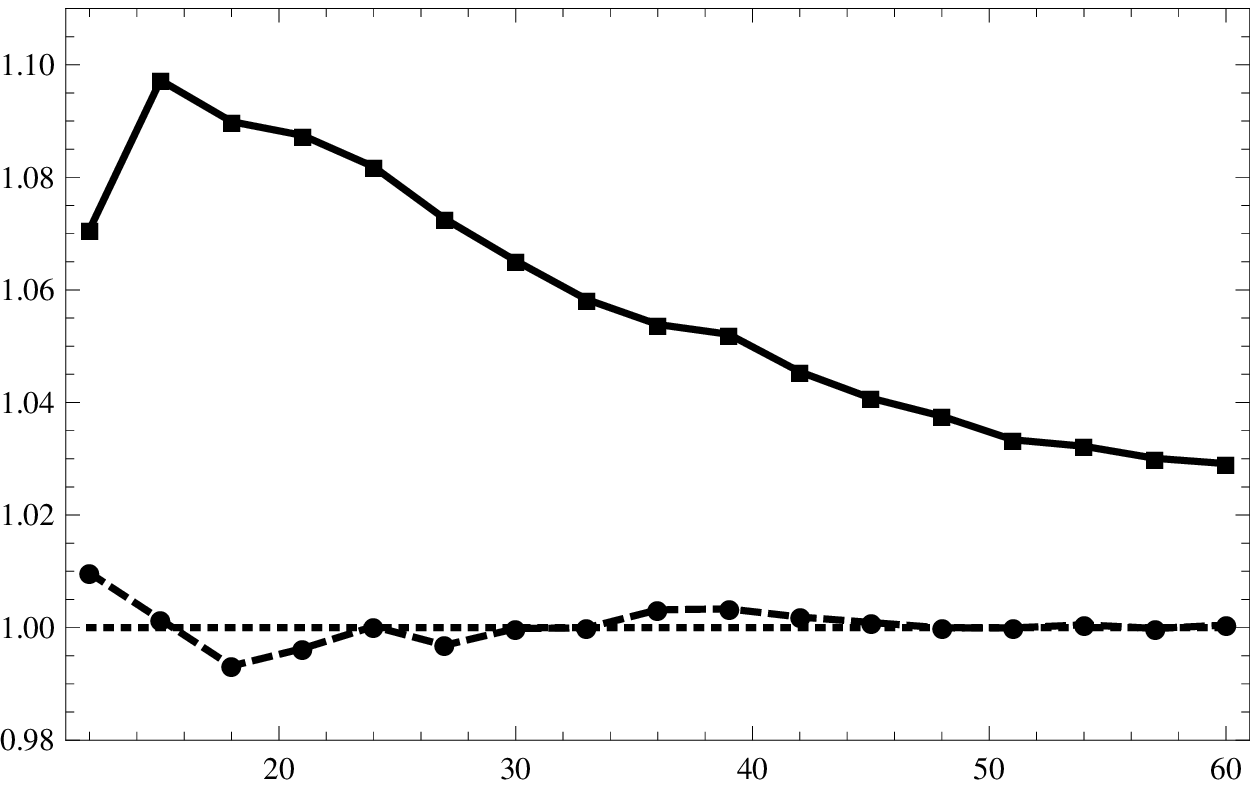}
\end{subfigure}
\begin{subfigure}[b]{\tempwidth}
       \centering
       \includegraphics[width=\tempwidth]{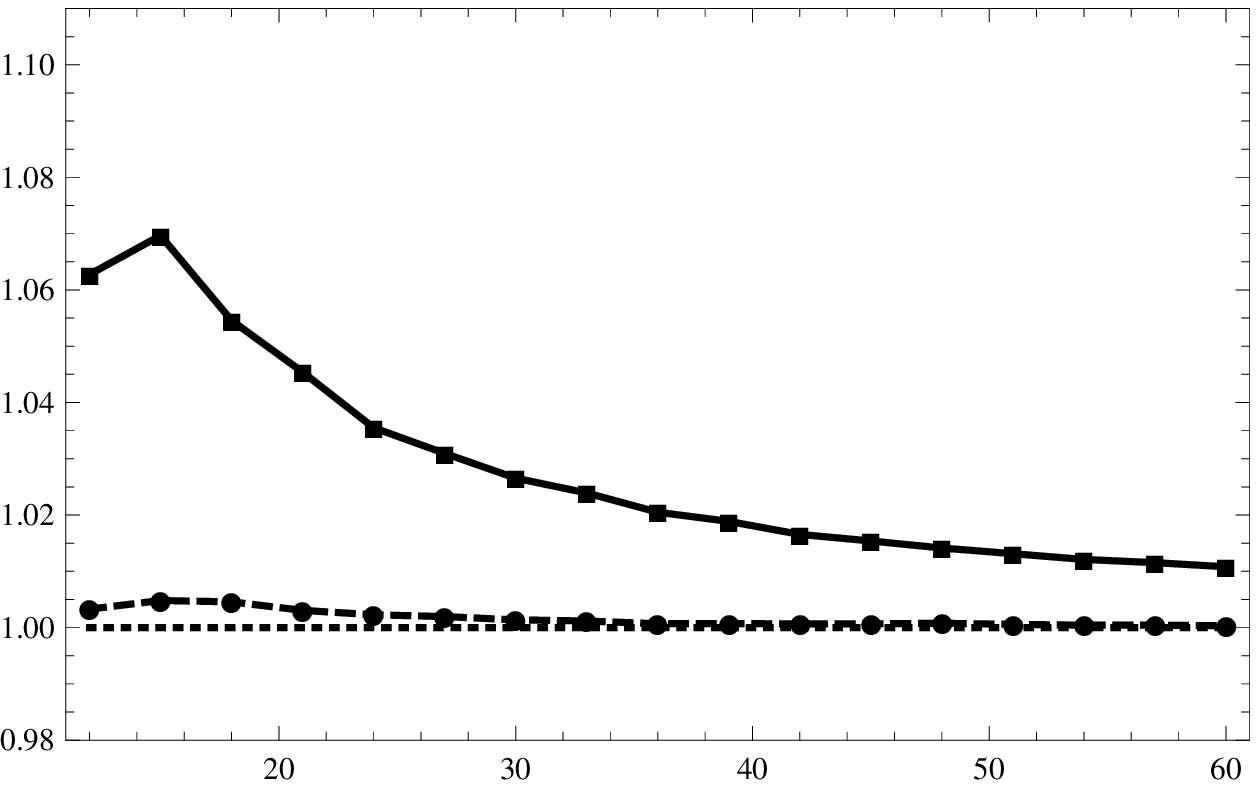}
\end{subfigure}\begin{subfigure}[b]{\tempwidth}
       \centering
       \includegraphics[width=\tempwidth]{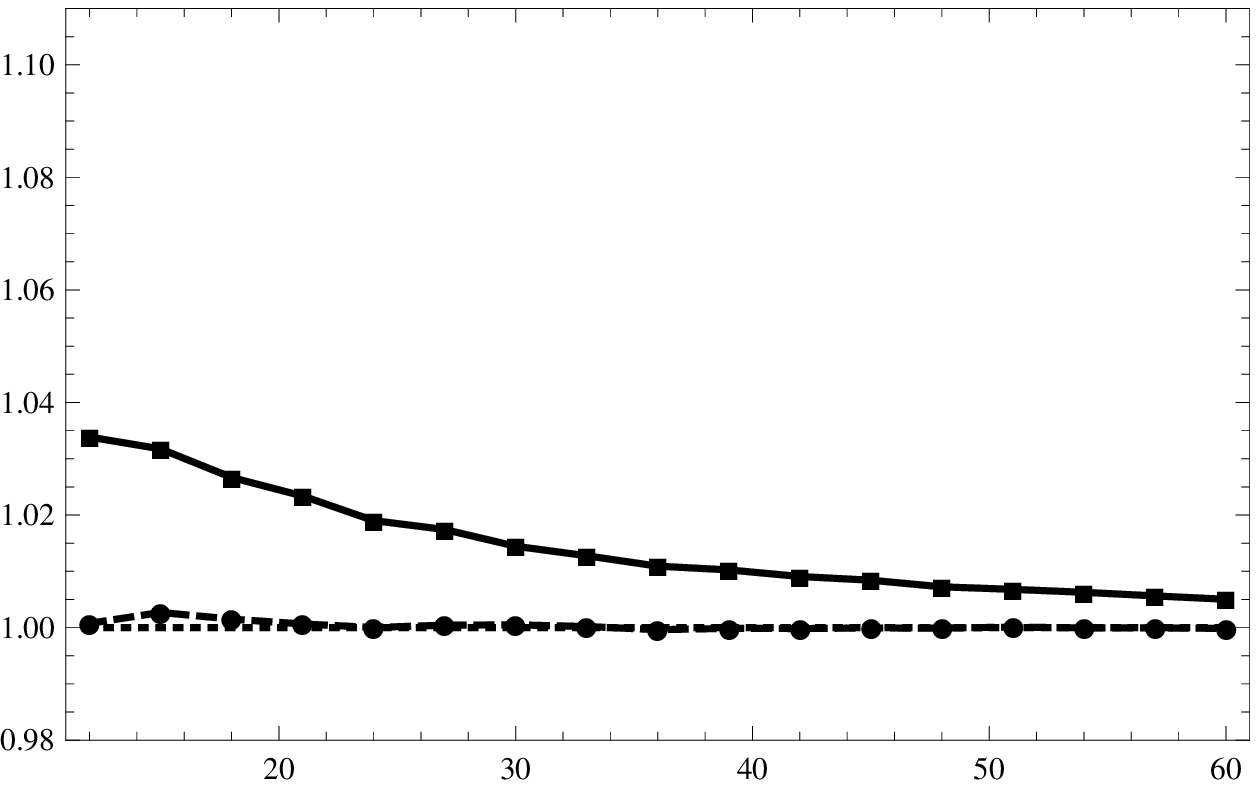}
\end{subfigure}\\
\rowname{{\normalfont\scriptsize Compartmental}}
\begin{subfigure}[b]{\tempwidth}
       \centering
       \includegraphics[width=\tempwidth]{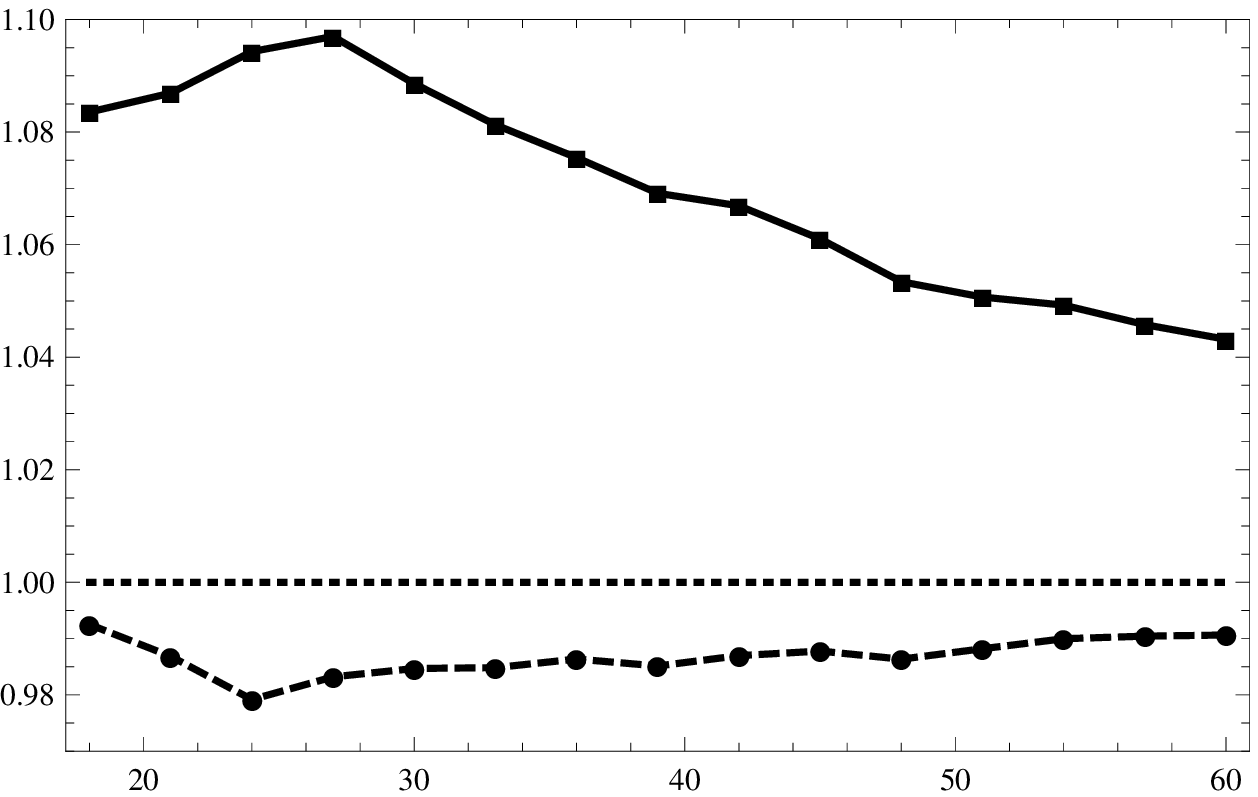}
\end{subfigure} 
\begin{subfigure}[b]{\tempwidth}
       \centering
       \includegraphics[width=\tempwidth]{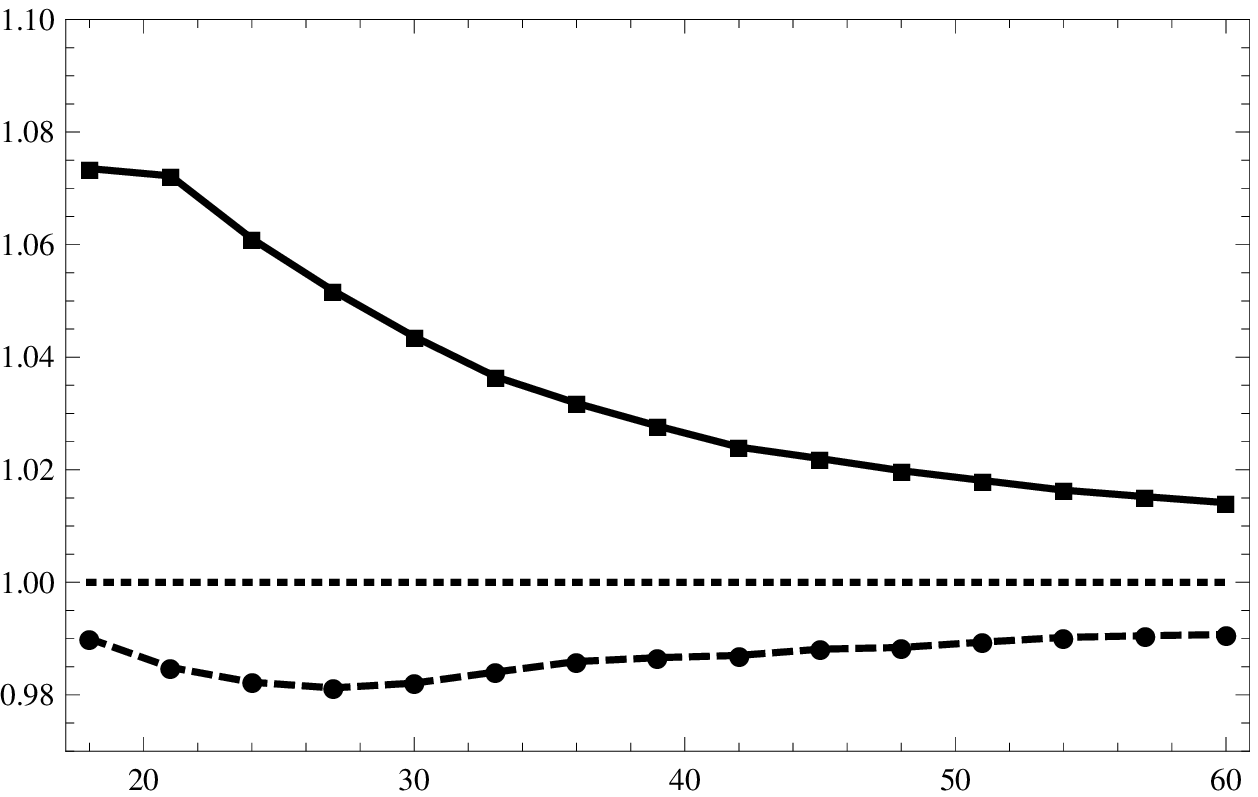}
\end{subfigure} 
\begin{subfigure}[b]{\tempwidth}
       \centering
       \includegraphics[width=\tempwidth]{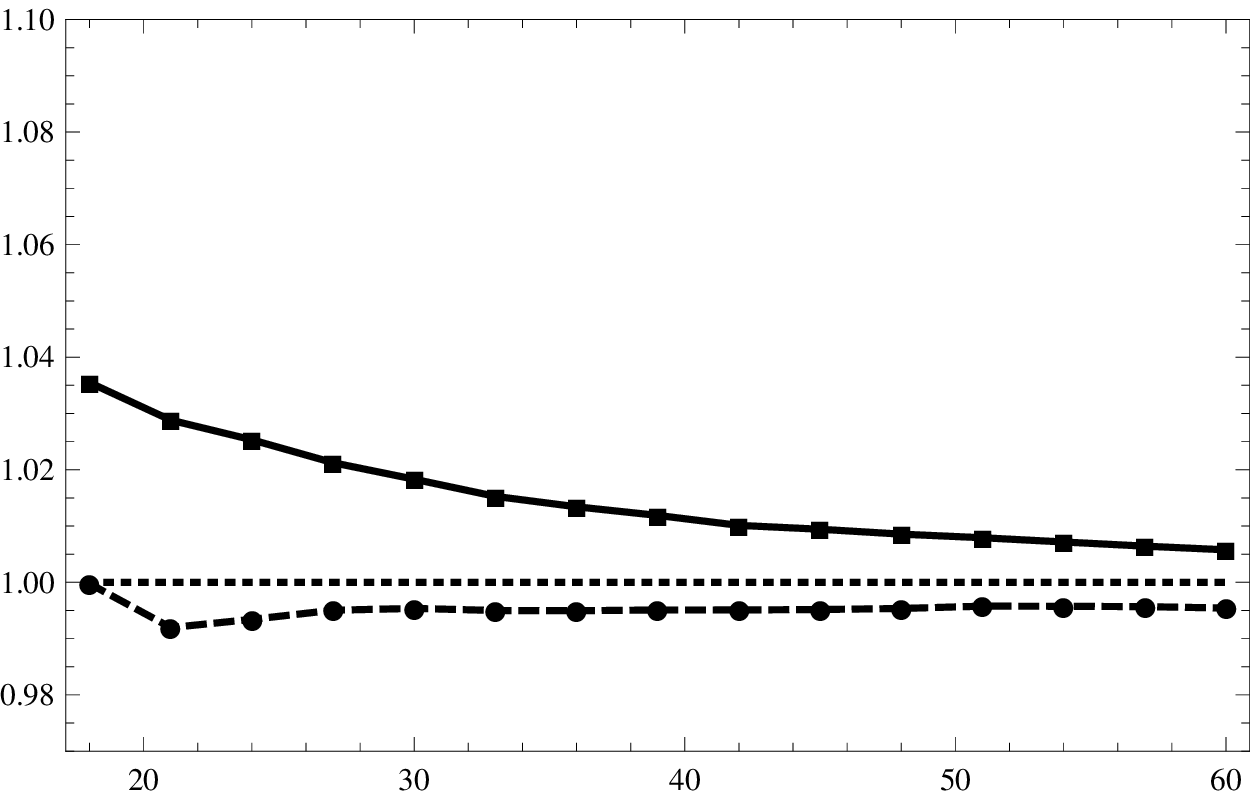}
\end{subfigure} 
\caption{$D$-efficiencies with respect to the Fisher information in the relevant subset of the optimal RSD, ${\rm{RJ}}_{\hat{\xi}_{RSD}}^{D}$ (solid line), and the AOD, ${\rm{RJ}}_{\hat{\xi}_{AOD}}^{D}$ (dashed line), relative to the FLOD, represented by the dotted line at 1, for the Michaeles-Menten, Decay and Compartmental model (top to bottom) and the Cauchy, Exponential Power and $q$-Gaussian error distributions (left to right) as a function of the total sample size $n$.}\label{fig:OID}
\end{sidewaysfigure}

\begin{sidewaysfigure}
\setlength{\tempwidth}{.31\linewidth}
\settoheight{\tempheight}{\includegraphics[width=\tempwidth]{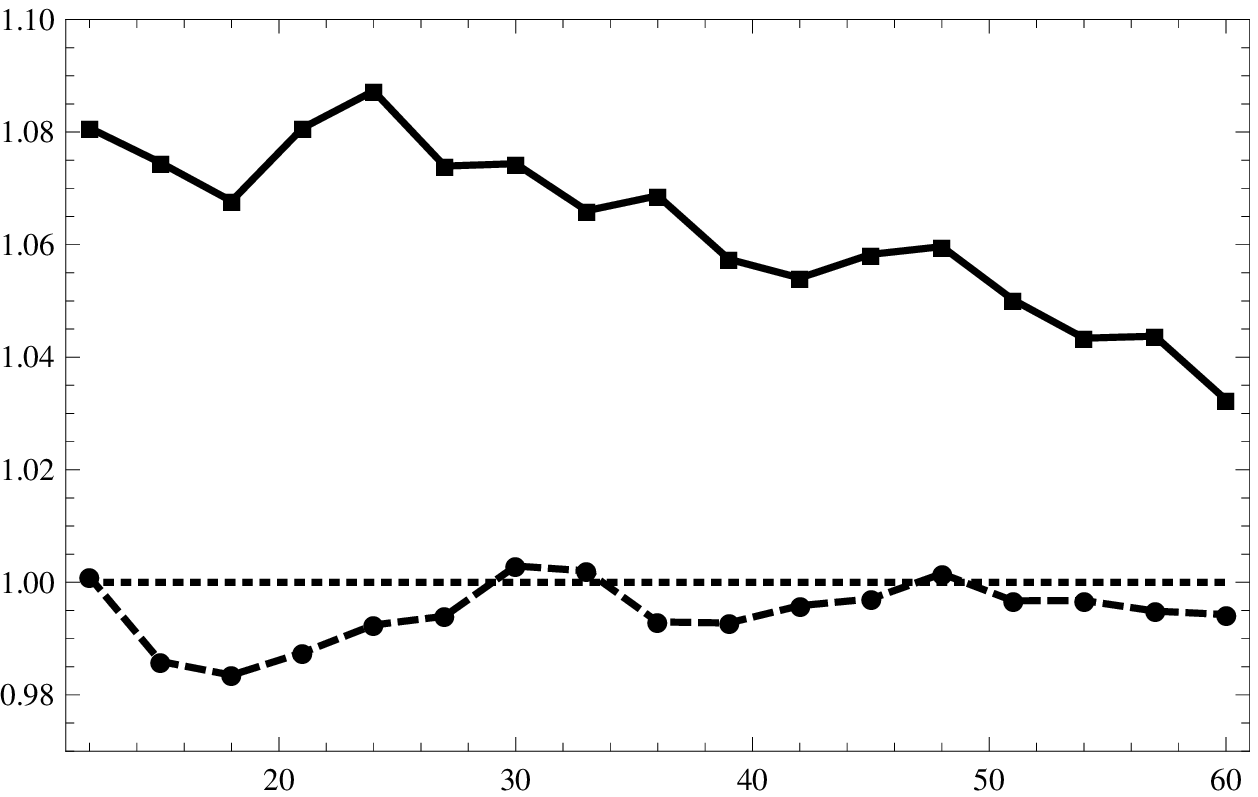}}
\centering
\hspace{\baselineskip}
\columnname{{\normalfont\scriptsize Cauchy}}\hfil
\columnname{{\normalfont\scriptsize Exponential Power}}\hfil
\columnname{{\normalfont\scriptsize q-Gaussian}}\\
\rowname{{\normalfont\scriptsize Michaeles-Menten}}
\begin{subfigure}[b]{\tempwidth}
       \centering
       \includegraphics[width=\tempwidth]{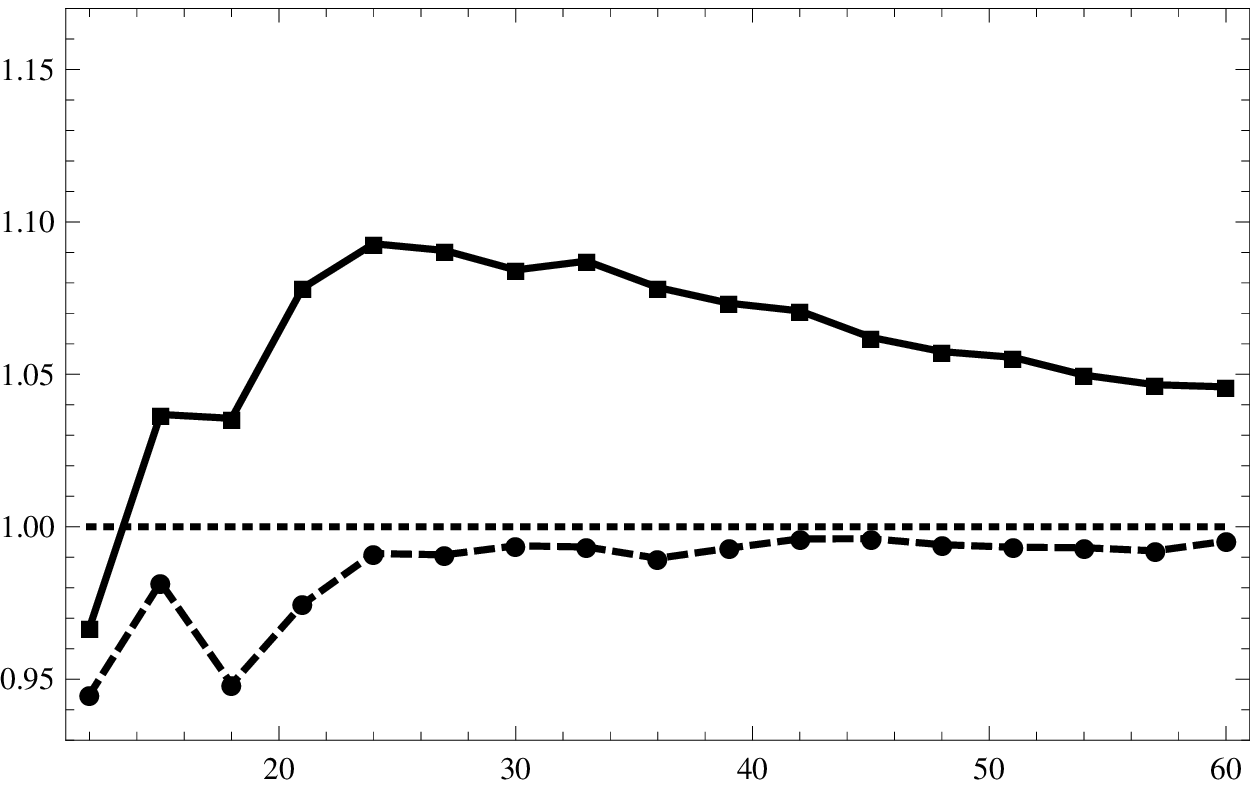}
\end{subfigure}
\begin{subfigure}[b]{\tempwidth}
       \centering
       \includegraphics[width=\tempwidth]{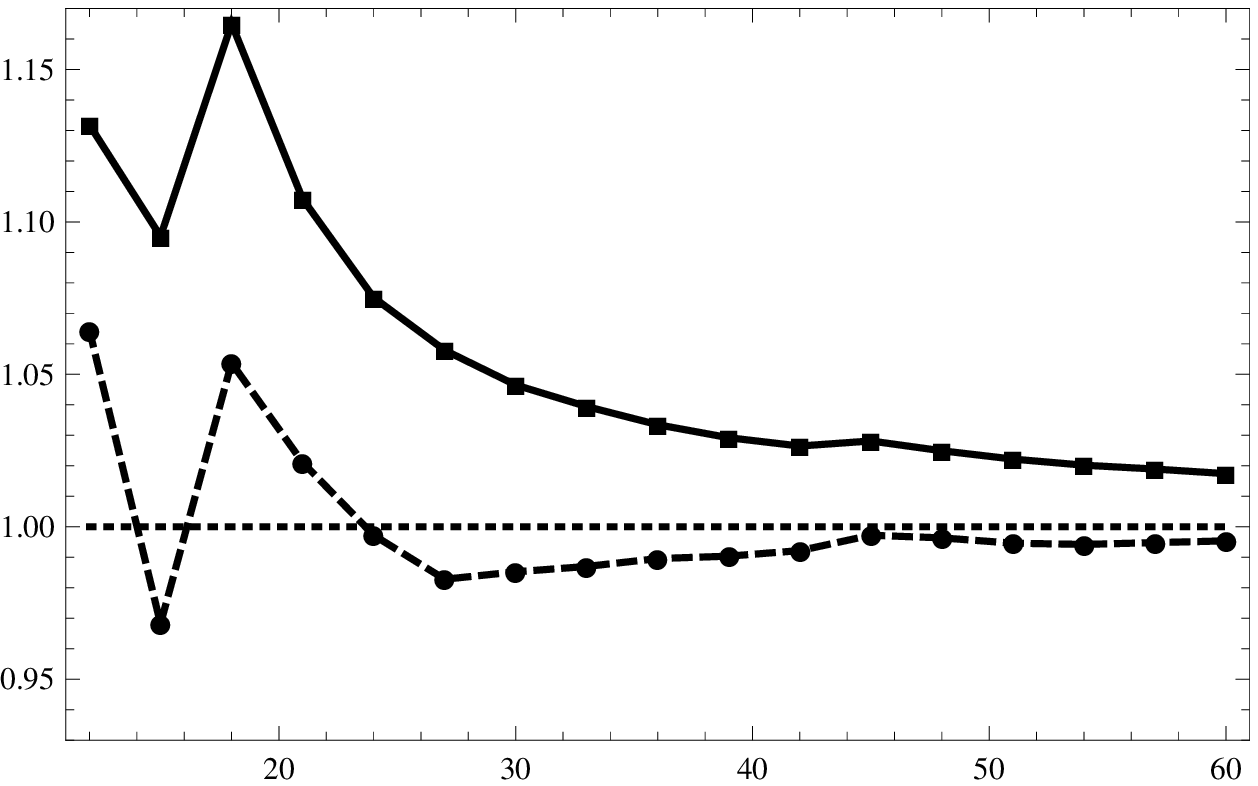}
\end{subfigure}\begin{subfigure}[b]{\tempwidth}
       \centering
       \includegraphics[width=\tempwidth]{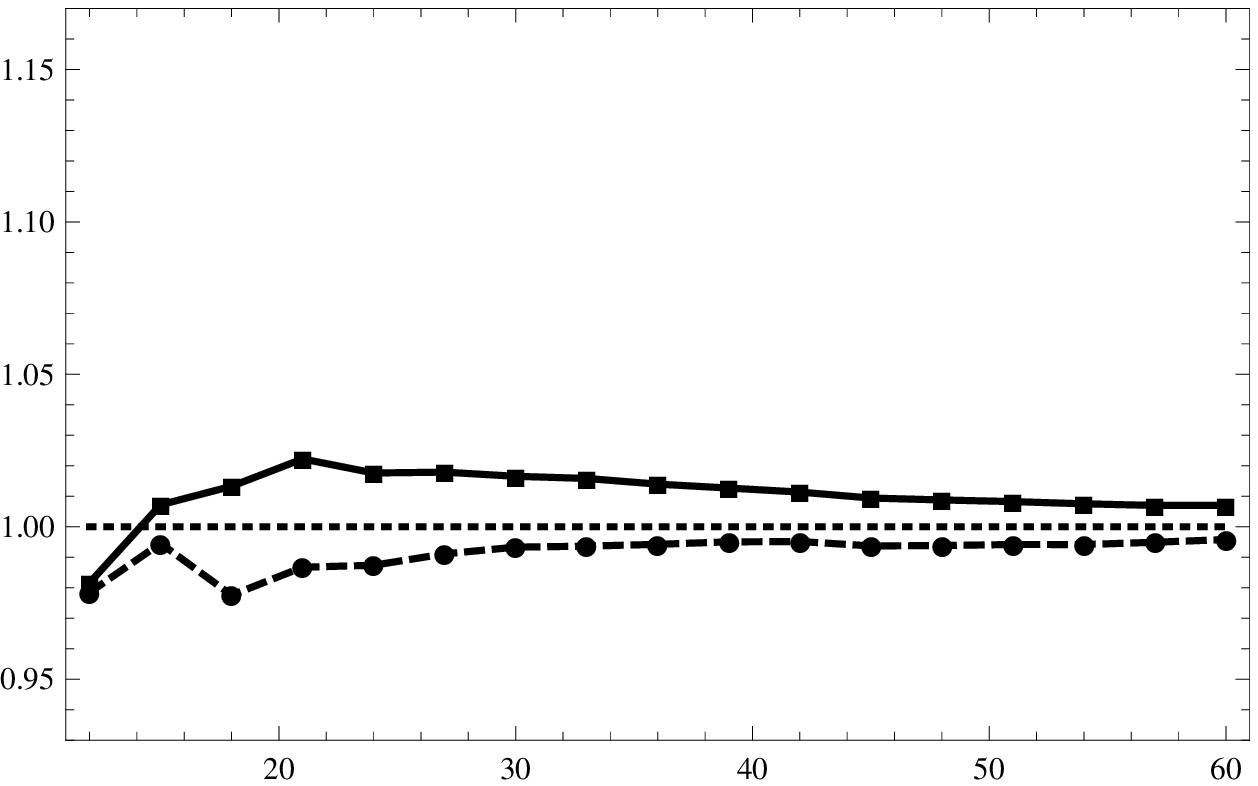}
\end{subfigure}\\
\rowname{{\normalfont\scriptsize Decay}}
\begin{subfigure}[b]{\tempwidth}
       \centering
       \includegraphics[width=\tempwidth]{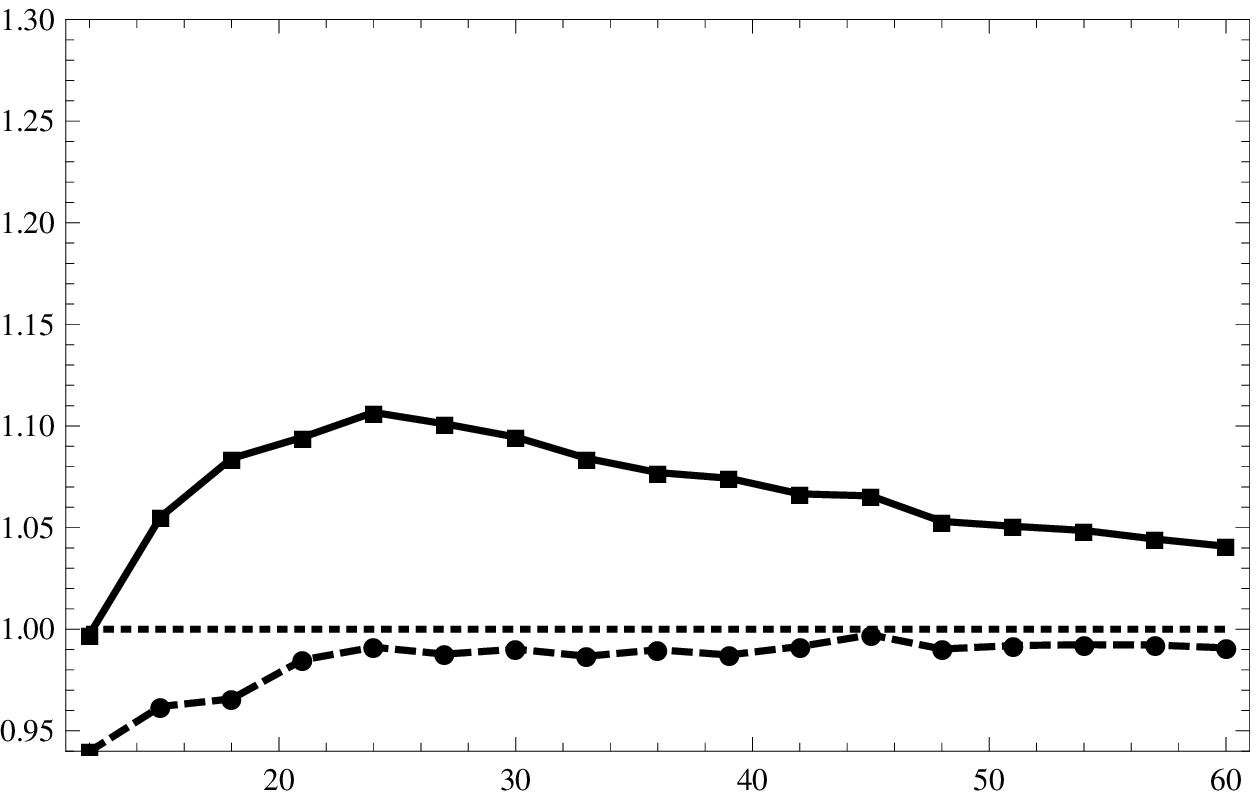}
\end{subfigure}
\begin{subfigure}[b]{\tempwidth}
       \centering
       \includegraphics[width=\tempwidth]{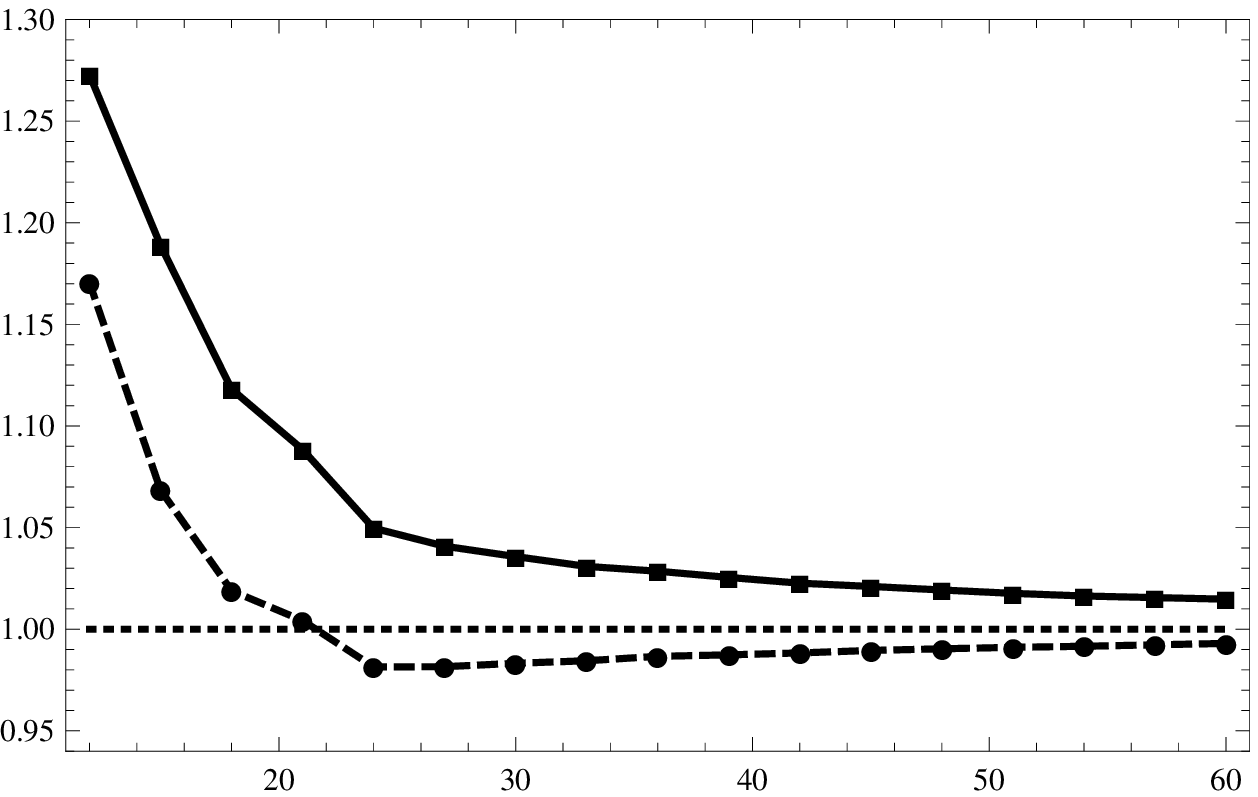}
\end{subfigure}\begin{subfigure}[b]{\tempwidth}
       \centering
       \includegraphics[width=\tempwidth]{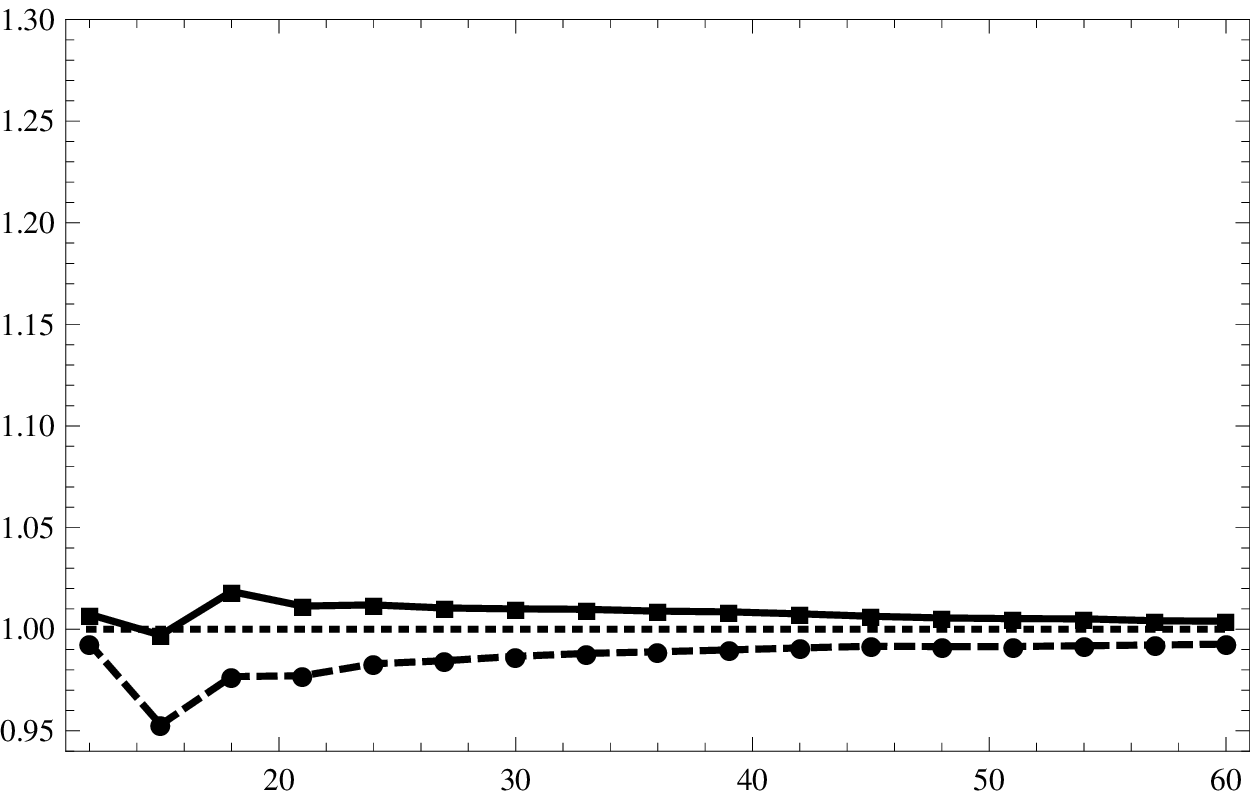}
\end{subfigure}\\
\rowname{{\normalfont\scriptsize Compartmental}}
\begin{subfigure}[b]{\tempwidth}
       \centering
       \includegraphics[width=\tempwidth]{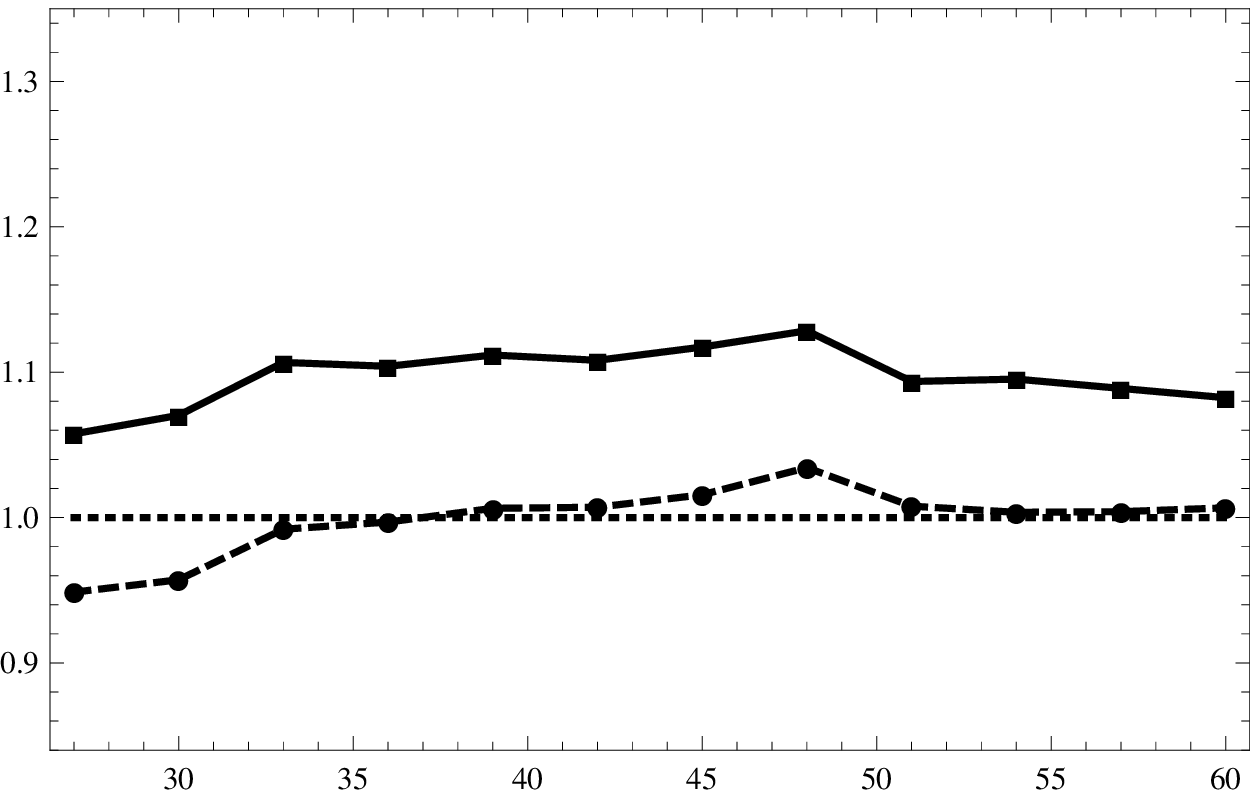}
\end{subfigure}
\begin{subfigure}[b]{\tempwidth}
       \centering
       \includegraphics[width=\tempwidth]{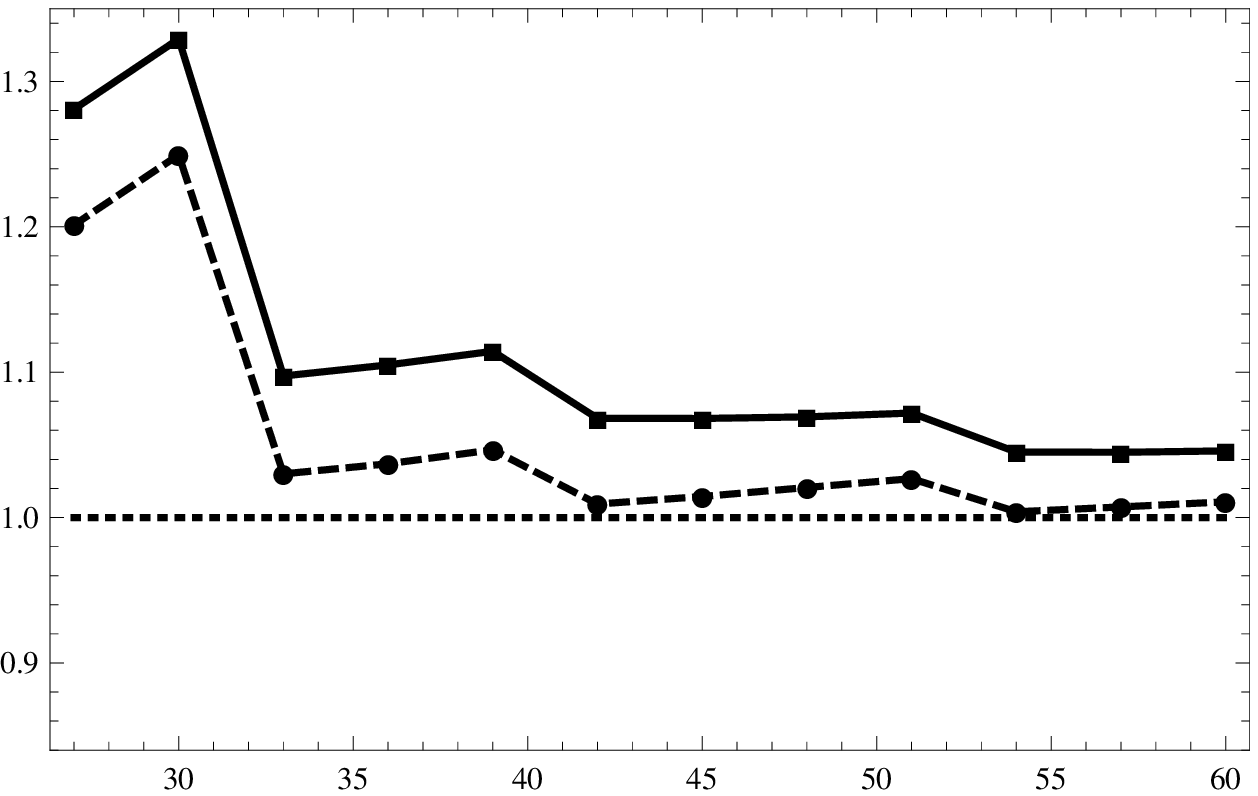}
\end{subfigure}\begin{subfigure}[b]{\tempwidth}
       \centering
       \includegraphics[width=\tempwidth]{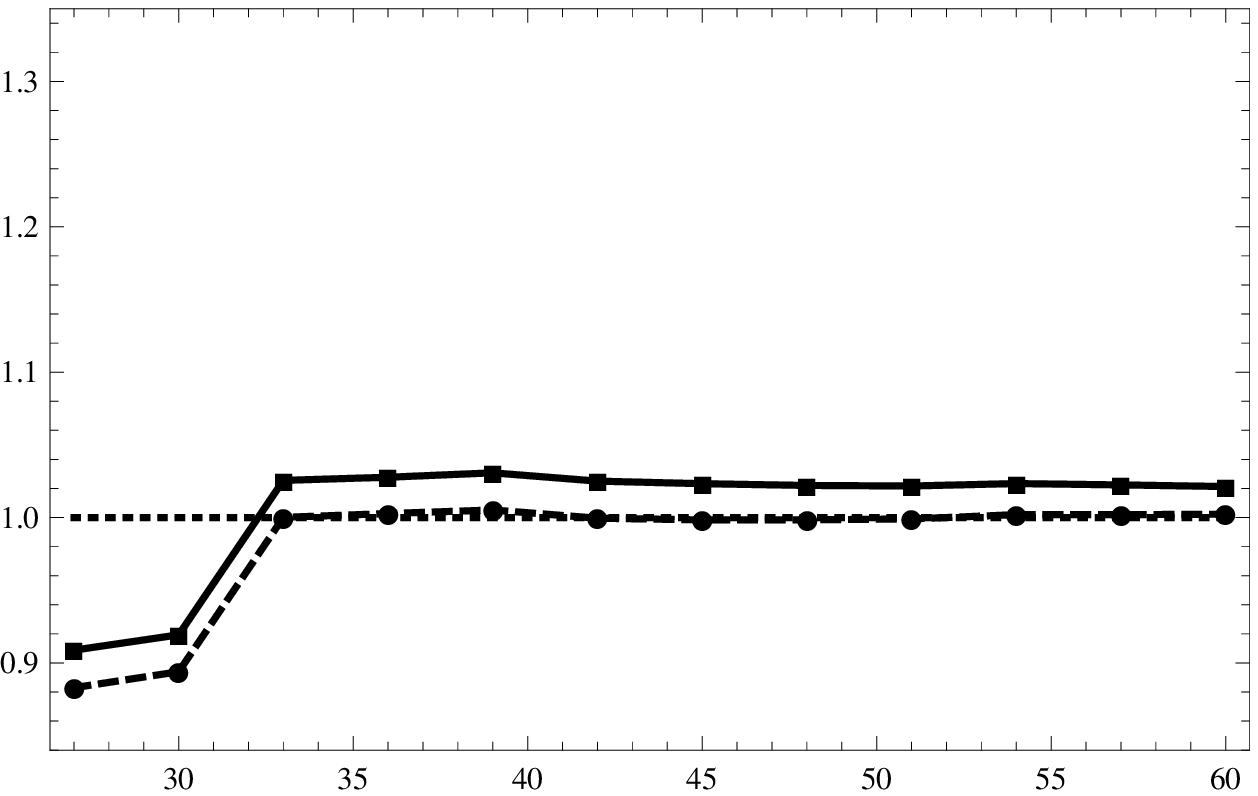}
\end{subfigure}\\
\caption{$c_{\boldsymbol{\theta}}$-efficiencies with respect to the Fisher information in the relevant subset of the optimal RSD, ${\rm{RJ}}_{\hat{\xi}_{RSD}}^{c_{\boldsymbol{\theta}}}$ (solid line), and the AOD, ${\rm{RJ}}_{\hat{\xi}_{AOD}}^{c_{\boldsymbol{\theta}}}$ (dashed line), relative to the FLOD, represented by the dotted line at 1, for the Michaeles-Menten, Decay and Compartmental model (top to bottom) and the Cauchy, Exponential Power and $q$-Gaussian error distributions (left to right) as a function of the total sample size $n$.}\label{fig:OIC}
\end{sidewaysfigure}

The primary objective of the optimal RSD is to improve inference without sacrificing information. Figure \ref{fig:OID} plots the $D$-efficiencies with respect to the Fisher information in the relevant subset of the optimal RSD, ${\rm{RJ}}_{\hat{\xi}_{RSD}}^{D}$ (solid line), and the AOD, ${\rm{RJ}}_{\hat{\xi}_{AOD}}^{D}$ (dashed line), relative to the FLOD, represented by the dotted line at 1, for the Michaeles-Menten, Decay and Compartmental model (top to bottom) and the Cauchy, Exponential Power and $q$-Gaussian error distributions (left to right) as a function of the total sample size $n$. Recall, values of ${\rm{RJ}}_{\hat{\xi}_{RSD}}^{D}$ and ${\rm{RJ}}_{\hat{\xi}_{AOD}}^{D}$ greater than 1 indicate cases where the corresponding design is more efficient than the FLOD with respect to inference.  Figure \ref{fig:OIC} is the same as Figure \ref{fig:OID} except the relative efficiencies are defined in terms of the $c_{\boldsymbol{\theta}}$-criterion. Note the range of the sample sizes presented in \ref{fig:MSED} and \ref{fig:MSEC} is $m(1)+3$ to 60. The exception to this is for Compartmental model for the $c_{\boldsymbol{\theta}}$-optimal criterion. For this mean function the first few sample sizes resulted in  ${\rm{E}}[\Psi[\boldsymbol{J}_{\mathcal{A^{*}}}^{\boldsymbol{x}}(\boldsymbol{\hat{\theta}}_{\xi^{*}})]]$ could not be computed due to singularity issues with $\boldsymbol{J}_{\mathcal{A^{*}}}^{\boldsymbol{x}}$. These cases were excluded for this reason. Interestingly, ${\rm{E}}[\Psi[\boldsymbol{J}_{\boldsymbol{\mathcal{A}}_{\xi}}^{\boldsymbol{x}}(\boldsymbol{\hat{\theta}}_{\xi})]]$, for $\xi$ equal to either $\hat{\xi}_{RSD}$ or $\hat{\xi}_{AOD}$ did suffer from similar computational issues.  

For nearly every model, error distribution, optimality criterion and sample size the relative efficiency of the optimal RSD is greater than the FLOD. The few exceptions all occur due to the sub-optimality of the initial design. Recall, both the optimal RSD and the AOD are initialized based on a incorrect guesses of the parameters and are sub-optimal. Depending on the severity of the sub-optimality of the initial design there exists a "burn in" period where the optimal RSD was less efficient than the FLOD in a few cases; after this period the optimal RSD was uniformly more efficient than the FLOD. An additional possible explanation for the burn in period is that the optimal RSD requires the estimation of $\boldsymbol{\theta}$. A burn in might be required to ensure the precision of the estimates are sufficient to produce reliable updates to the initial design.  The optimal relevant subset design was better with respect to inference than the AOD, i.e., ${\rm{RJ}}_{\hat{\xi}_{RSD}}^{\Psi}>{\rm{RJ}}_{\hat{\xi}_{AOD}}^{\Psi}$, for $\Psi=D$ or $c_{\boldsymbol{\theta}}$, in every case considered.

Comparing the AOD to the FLOD and it can be seen that the FLOD tended to be more efficient with respect to this measure; however, there were a few instances where the AOD was more efficient. 

One pattern that emerges is that the benefit of the optimal RSD is the most pronounced for Cauchy, followed by exponential power and then $q$-Gaussian errors. This can be explained, in part, by statistical curvature. \citet{Efro:Defi:1975} defines statistical curvature as $\gamma = \left(\nu_{02}\nu_{20} - \nu_{11}/\nu_{20}^{3}\right)^{1/2}$, where $\nu_{kl} = {{\rm{E}}}[\dot{l}_{\mathcal{Y}}^{k}(\eta_{i})\{\ddot{l}_{\mathcal{Y}}(\eta_{i}) + {{\rm{E}}}[\dot{l}_{\mathcal{Y}}(\eta_{i})]^{2}\}^{l}]$ and $\dot{l}_{y}$ and $\ddot{l}_{y}$ are the first and second derivatives of the likelihood for a single observation. \cite{lane2019optimality} shows that the larger the statistical curvature the greater the expected benefit of the optimal RSD. Statistical curvature is invariant to linear transformations therefore the curvature in the responses is the same as the curvature in the errors. For the Cauchy, exponential power and $q$-Gamma distributions $\gamma^{2} = 2.50, 1.18$ and 0.63, which matches the observed efficiency of the optimal RSD relative to the FLOD. 


In summary, this simulation study clearly demonstrates that implementing an optimal RSD leads to a significant increase in the efficiency of an experiment with respect to inference.  


\subsubsection{Information and MSE Efficiency}

The second objective of the proposed optimal RSD is to optimize the Fisher information in the sample. As outlined in Section \ref{sec:intro} it was not anticipated the optimal RSD would improve the Fisher information in the sample, only that it would not perform worse than the two competing methods with respect to this measure. The Fisher information in the sample is a surrogate for the MSE. It is more important that the optimal RSD not increase the MSE than it is for it to maintain the Fisher information in the sample. For this reason, in this section the MSE is examined; comparisons of the Fisher information in the sample are presented in the supplemental materials. The primary conclusion from the comparisons of the Fisher information in the sample is that there is very little difference between the three designs with respect to this measure. 

\begin{sidewaysfigure}
\setlength{\tempwidth}{.31\linewidth}
\settoheight{\tempheight}{\includegraphics[width=\tempwidth]{CauchyMMVD.eps}}
\centering
\hspace{\baselineskip}
\columnname{{\normalfont\scriptsize Cauchy}}\hfil
\columnname{{\normalfont\scriptsize Exponential Power}}\hfil
\columnname{{\normalfont\scriptsize q-Gaussian}}\\
\rowname{{\normalfont\scriptsize Michaeles-Menten}}
\begin{subfigure}[b]{\tempwidth}
       \centering
       \includegraphics[width=\tempwidth]{CauchyMMVD.eps}
\end{subfigure}
\begin{subfigure}[b]{\tempwidth}
       \centering
       \includegraphics[width=\tempwidth]{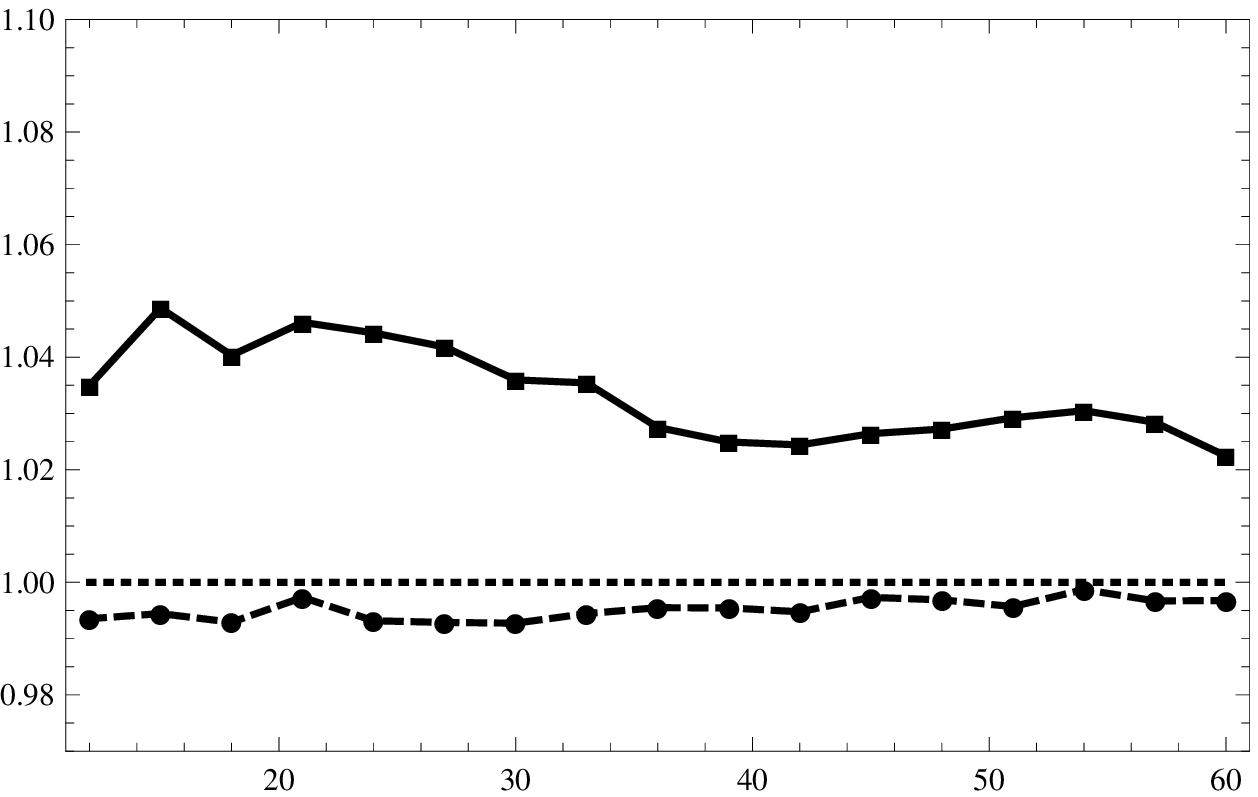}
\end{subfigure}\begin{subfigure}[b]{\tempwidth}
       \centering
       \includegraphics[width=\tempwidth]{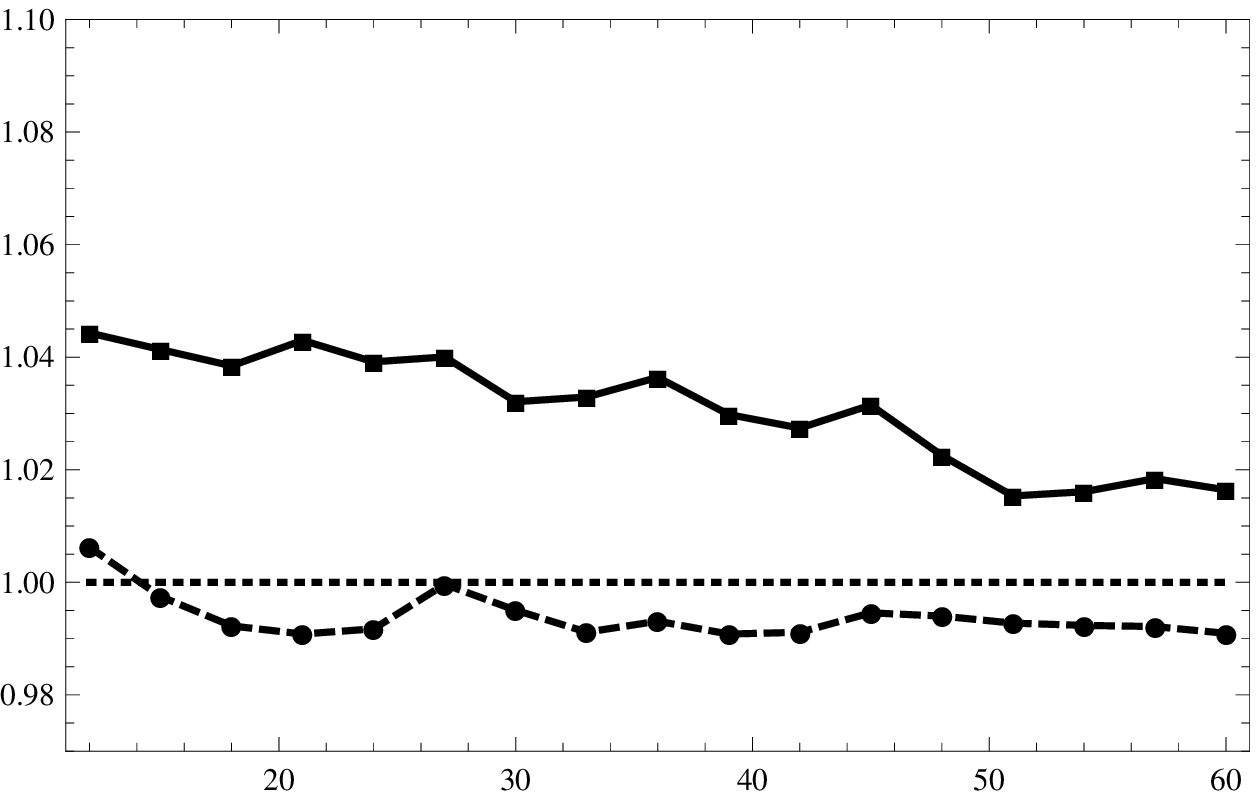}
\end{subfigure}\\
\rowname{{\normalfont\scriptsize Decay}}
\begin{subfigure}[b]{\tempwidth}
       \centering
       \includegraphics[width=\tempwidth]{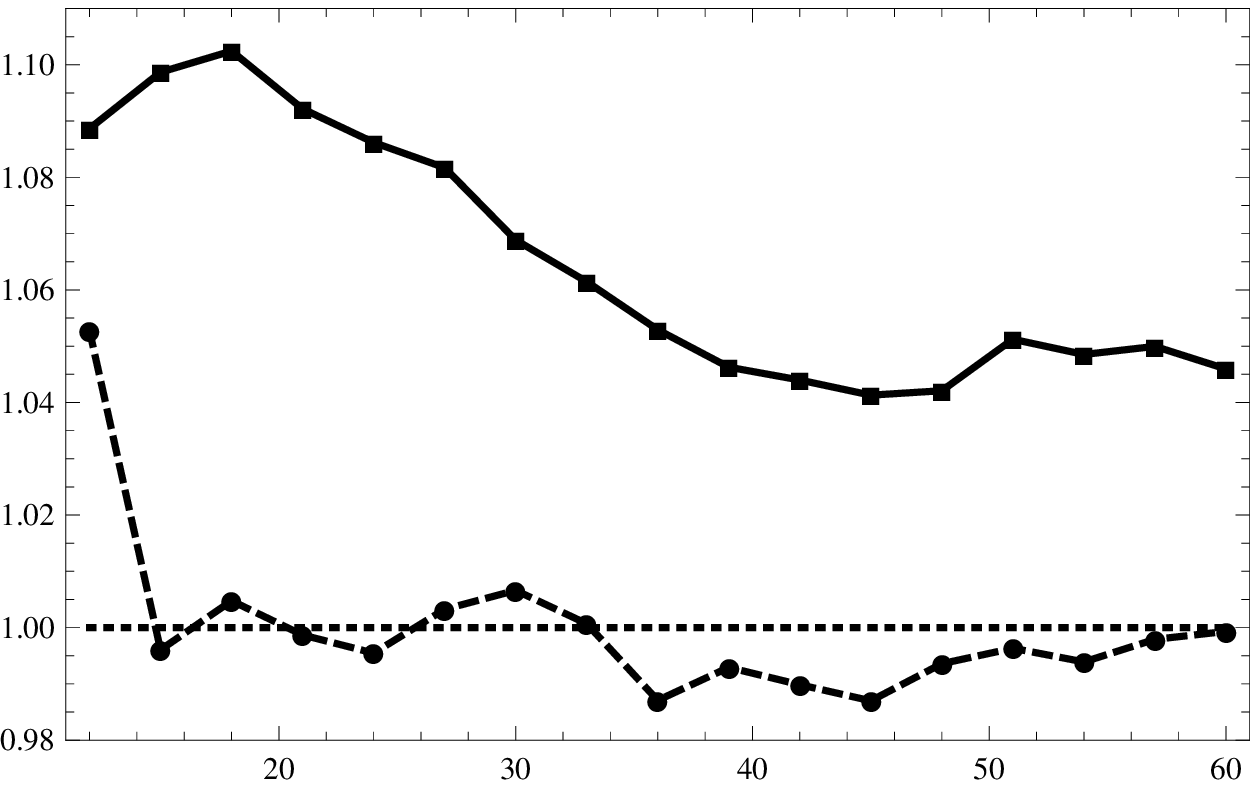}
\end{subfigure}
\begin{subfigure}[b]{\tempwidth}
       \centering
       \includegraphics[width=\tempwidth]{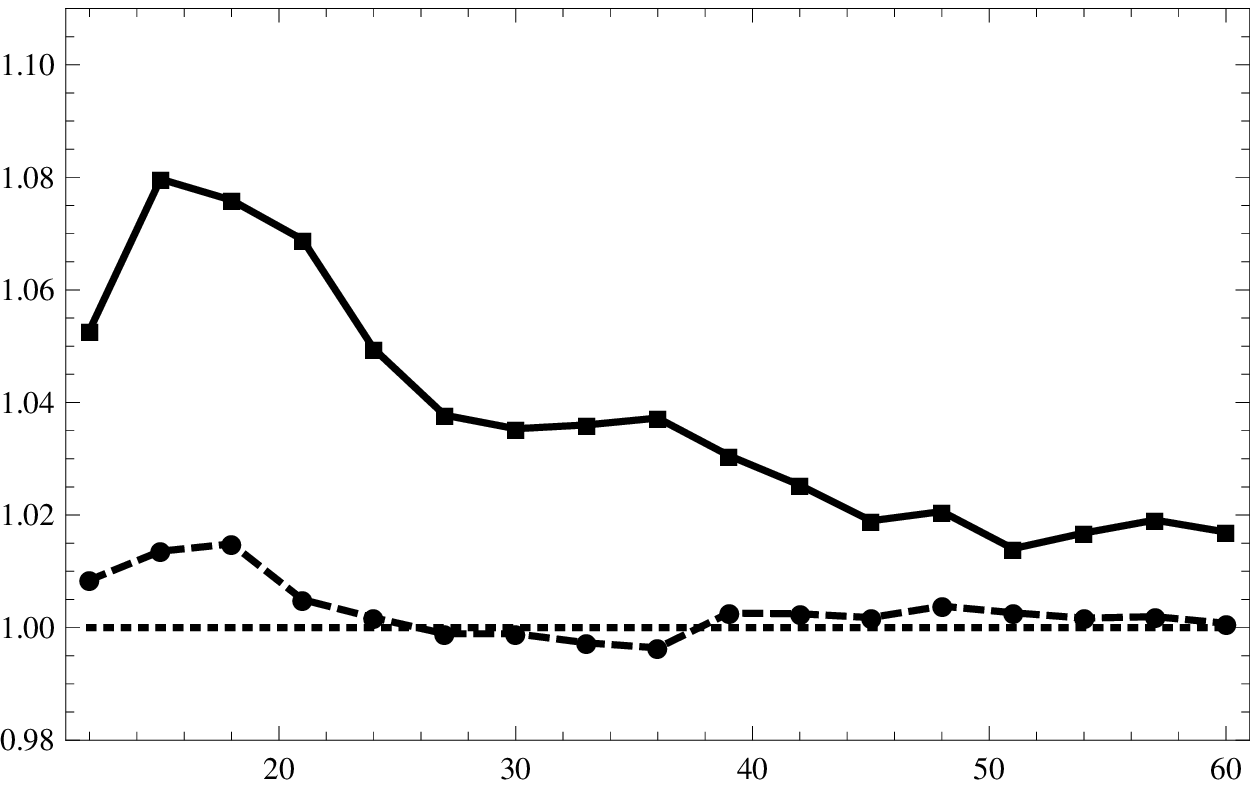}
\end{subfigure}\begin{subfigure}[b]{\tempwidth}
       \centering
       \includegraphics[width=\tempwidth]{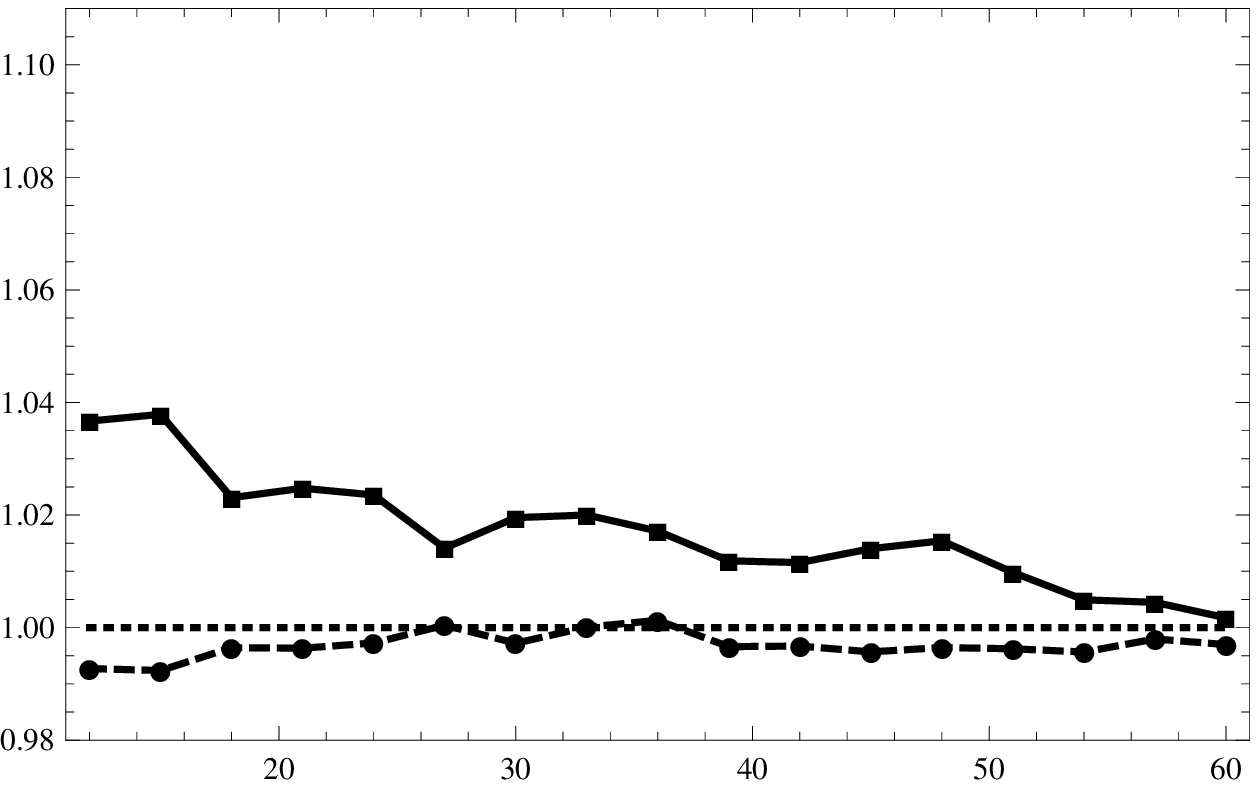}
\end{subfigure}\\
\rowname{{\normalfont\scriptsize Compartmental}}
\begin{subfigure}[b]{\tempwidth}
       \centering
       \includegraphics[width=\tempwidth]{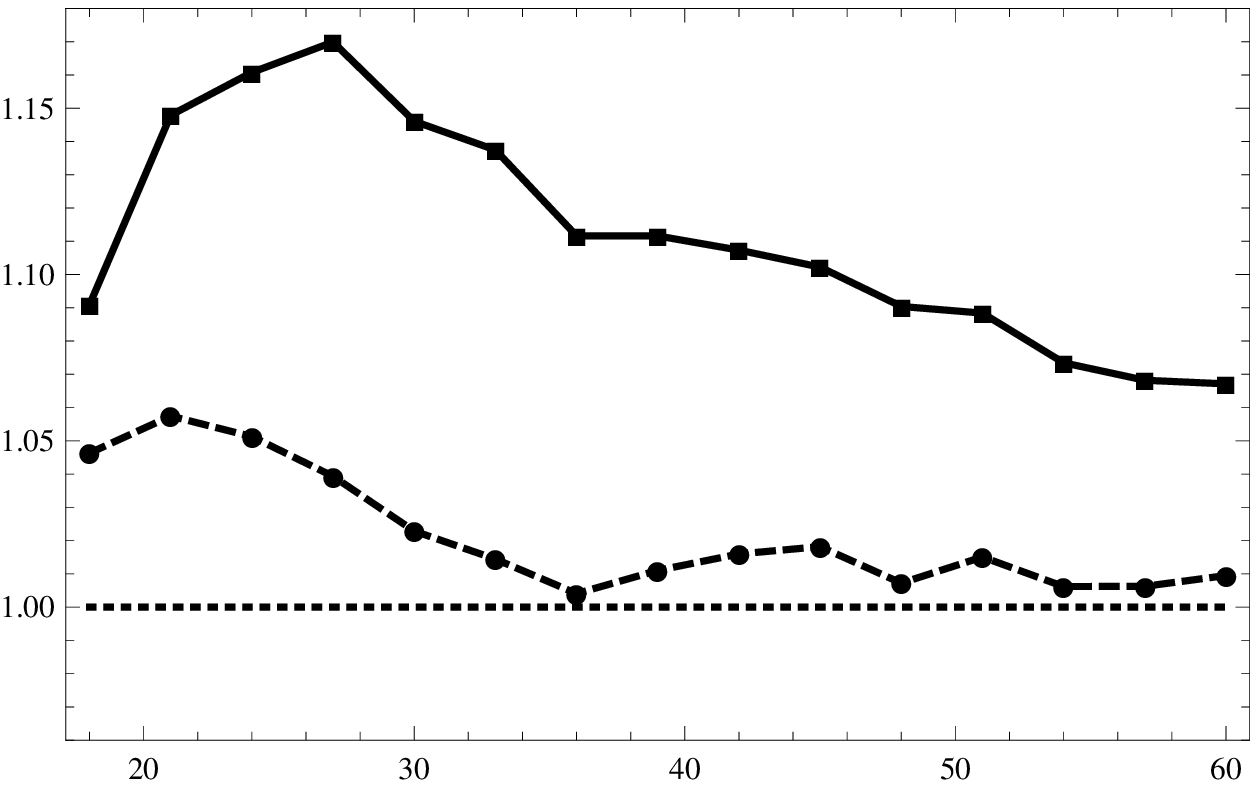}
\end{subfigure}
\begin{subfigure}[b]{\tempwidth}
       \centering
       \includegraphics[width=\tempwidth]{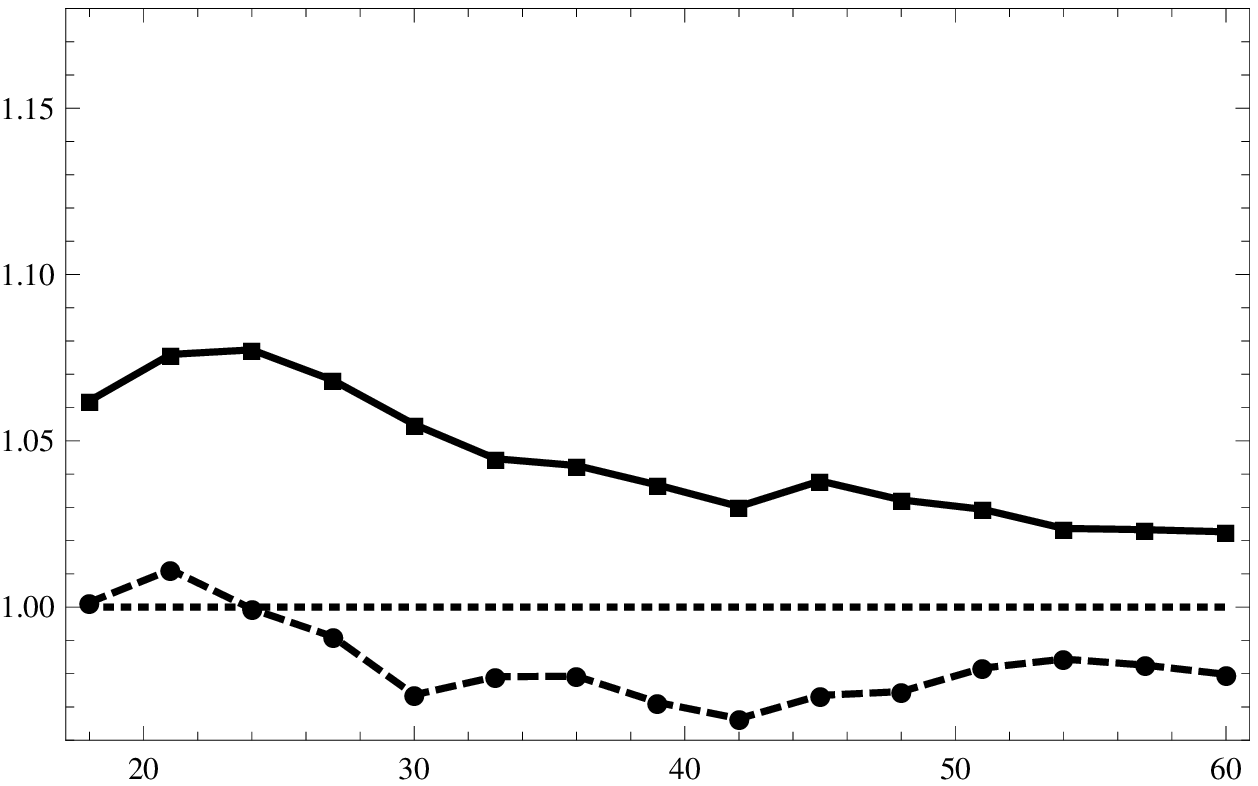}
\end{subfigure}\begin{subfigure}[b]{\tempwidth}
       \centering
       \includegraphics[width=\tempwidth]{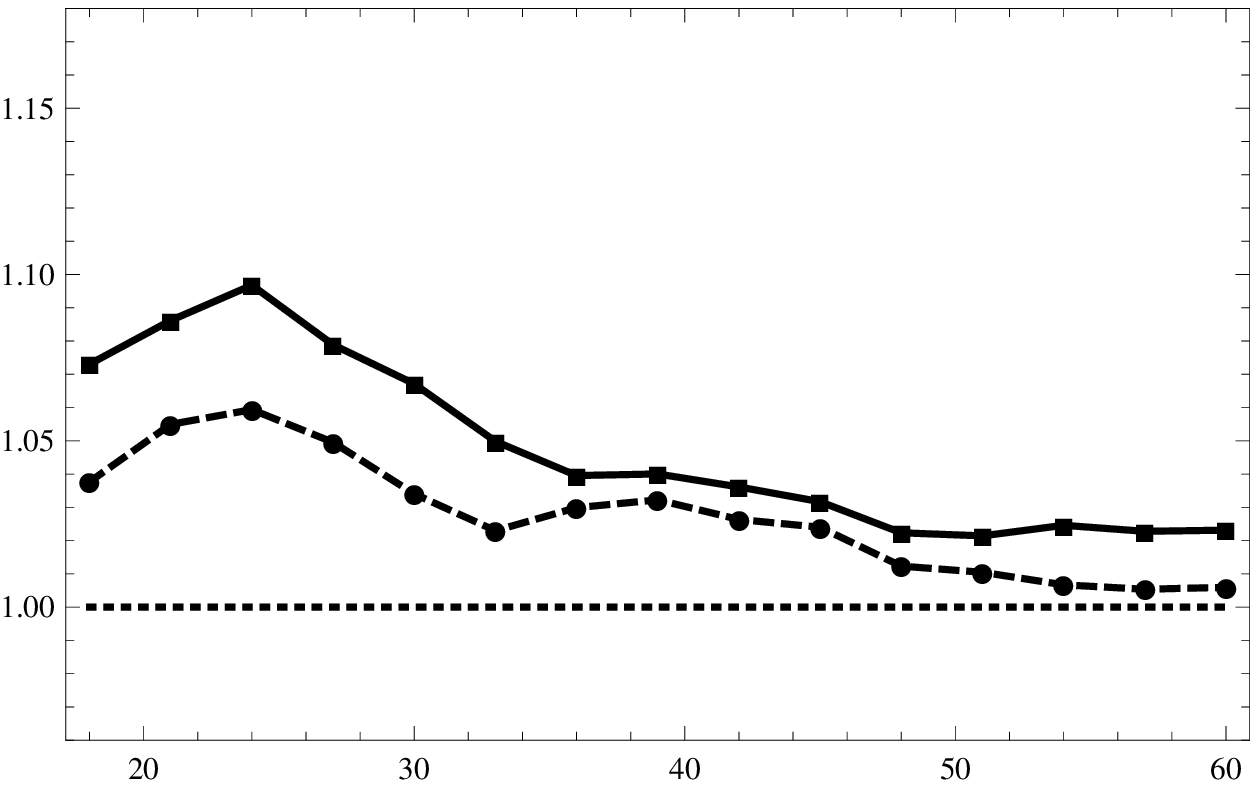}
\end{subfigure}\\
\caption{$D$-efficiencies with respect to MSE of the optimal RSD, ${\rm{RMSE}}_{\hat{\xi}_{RSD}}^{D}$ (solid line), and the AOD, ${\rm{RMSE}}_{\hat{\xi}_{AOD}}^{D}$ (dashed line), relative to the FLOD, represented by the dotted line at 1, for the Michaeles-Menten, Decay and Compartmental model (top to bottom) and the Cauchy, Exponential Power and $q$-Gaussian error distributions (left to right) as a function of the total sample size $n$.}\label{fig:MSED}
\end{sidewaysfigure}

\begin{sidewaysfigure}
\setlength{\tempwidth}{.31\linewidth}
\settoheight{\tempheight}{\includegraphics[width=\tempwidth]{CauchyMMVD.eps}}
\centering
\hspace{\baselineskip}
\columnname{{\normalfont\scriptsize Cauchy}}\hfil
\columnname{{\normalfont\scriptsize Exponential Power}}\hfil
\columnname{{\normalfont\scriptsize q-Gaussian}}\\
\rowname{{\normalfont\scriptsize Michaeles-Menten}}
\begin{subfigure}[b]{\tempwidth}
       \centering
       \includegraphics[width=\tempwidth]{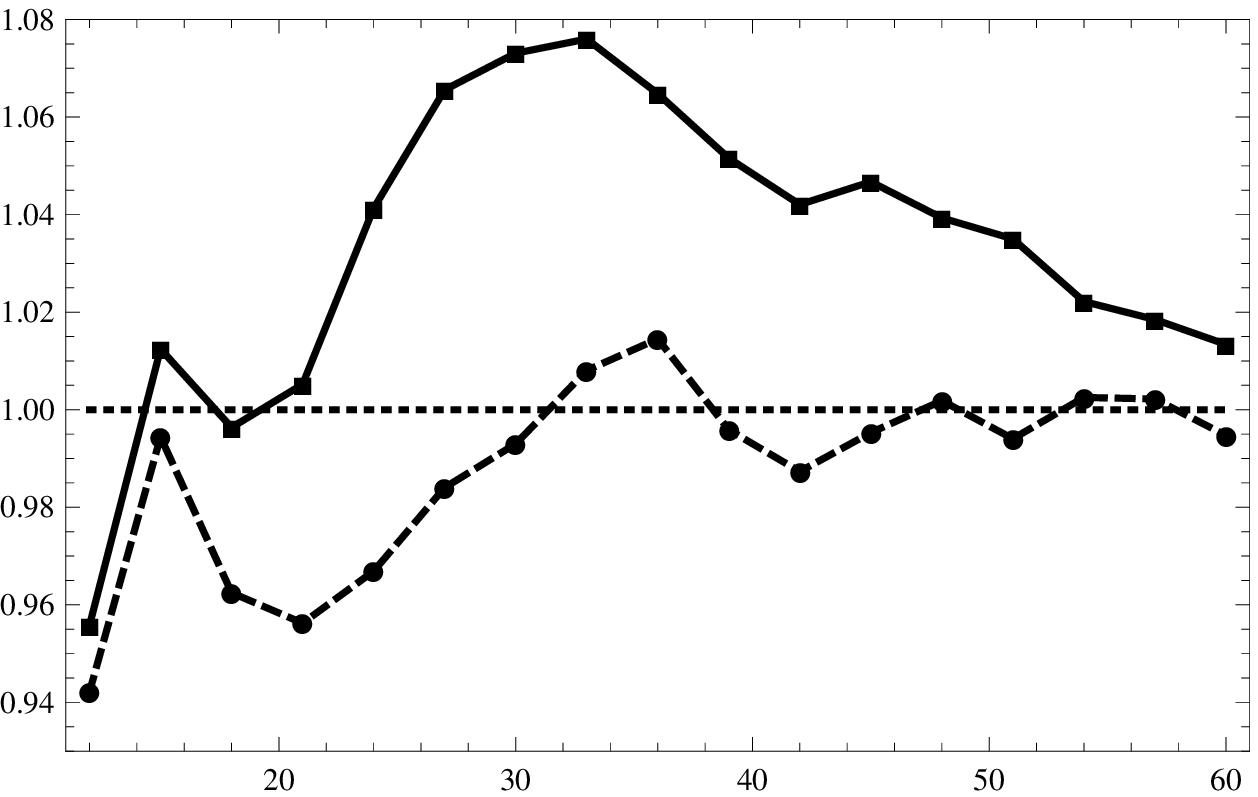}
\end{subfigure}
\begin{subfigure}[b]{\tempwidth}
       \centering
       \includegraphics[width=\tempwidth]{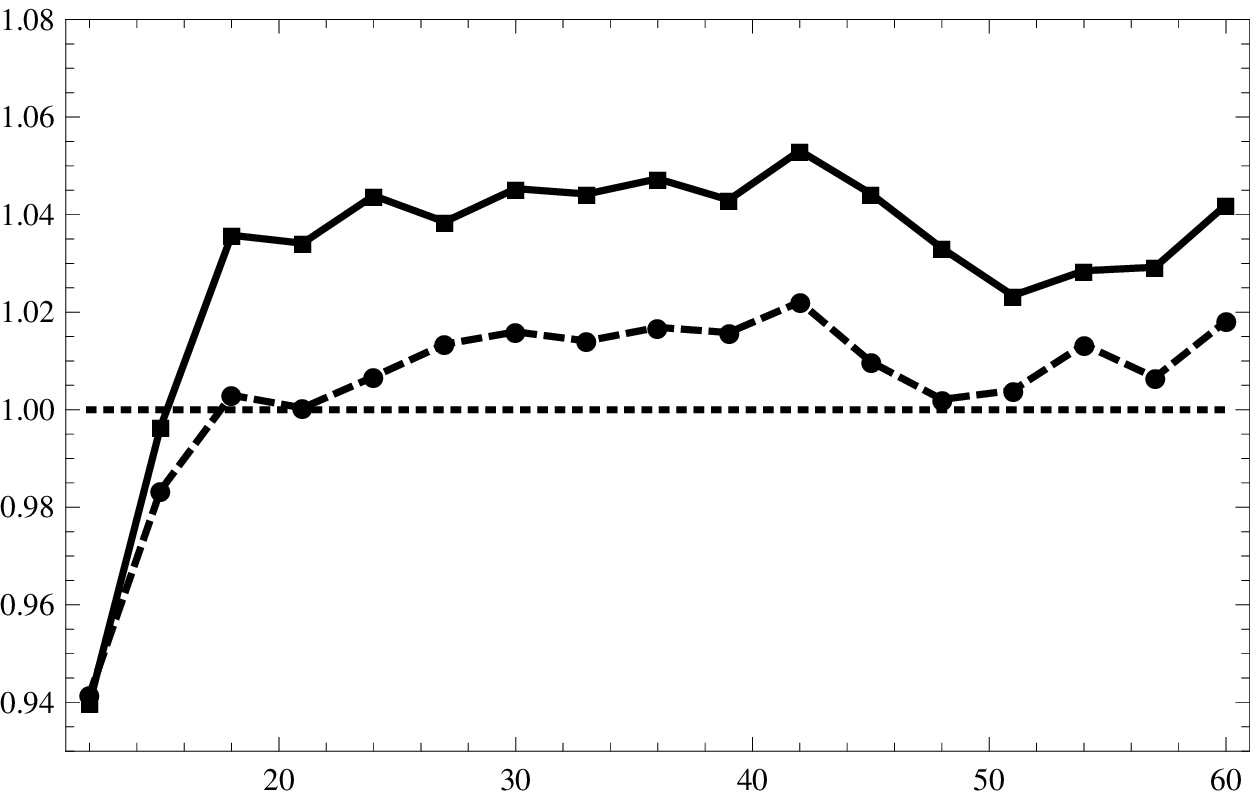}
\end{subfigure}\begin{subfigure}[b]{\tempwidth}
       \centering
       \includegraphics[width=\tempwidth]{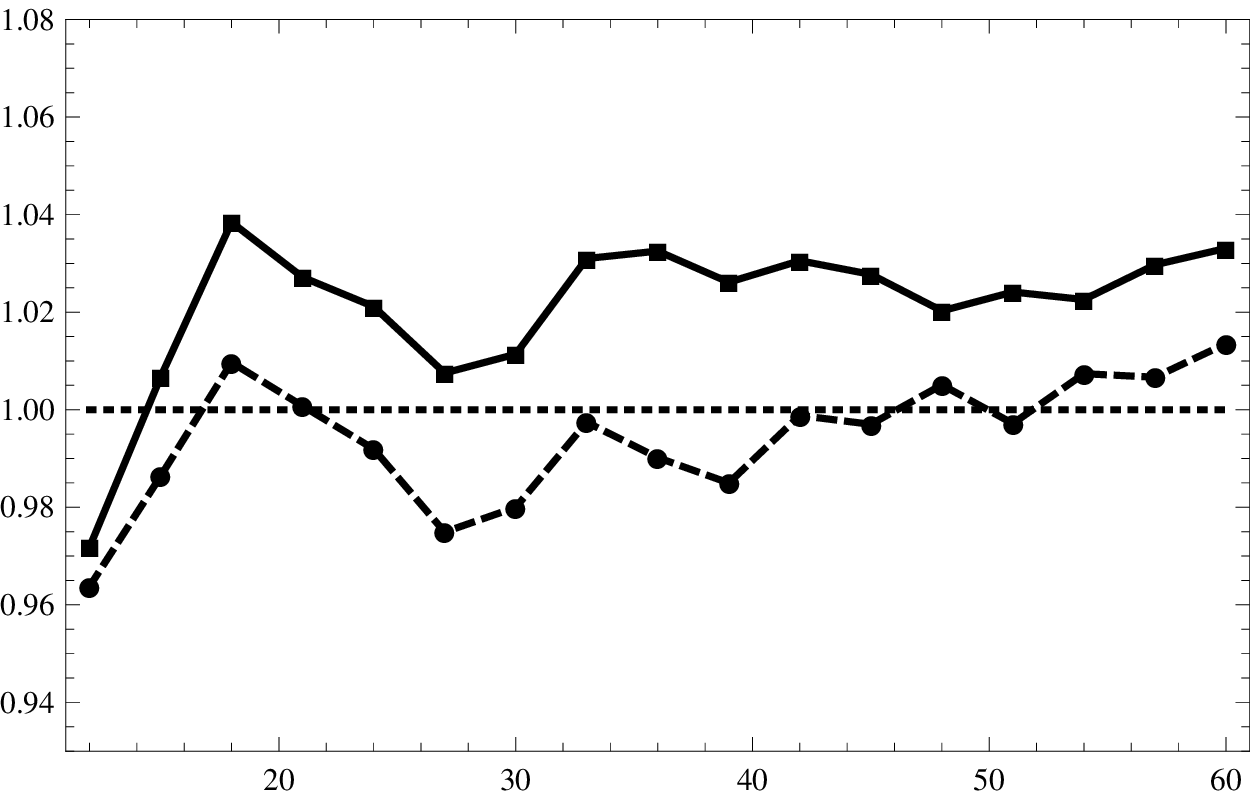}
\end{subfigure}\\
\rowname{{\normalfont\scriptsize Decay}}
\begin{subfigure}[b]{\tempwidth}
       \centering
       \includegraphics[width=\tempwidth]{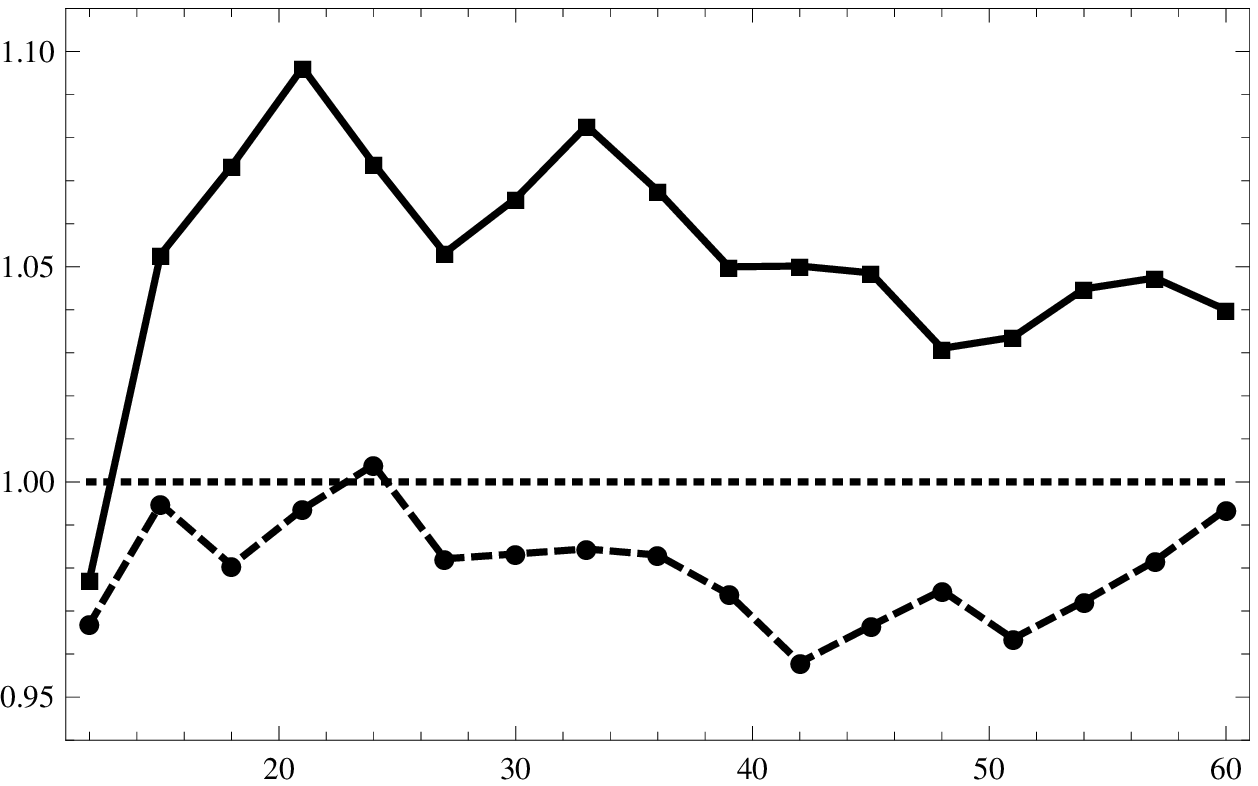}
\end{subfigure}
\begin{subfigure}[b]{\tempwidth}
       \centering
       \includegraphics[width=\tempwidth]{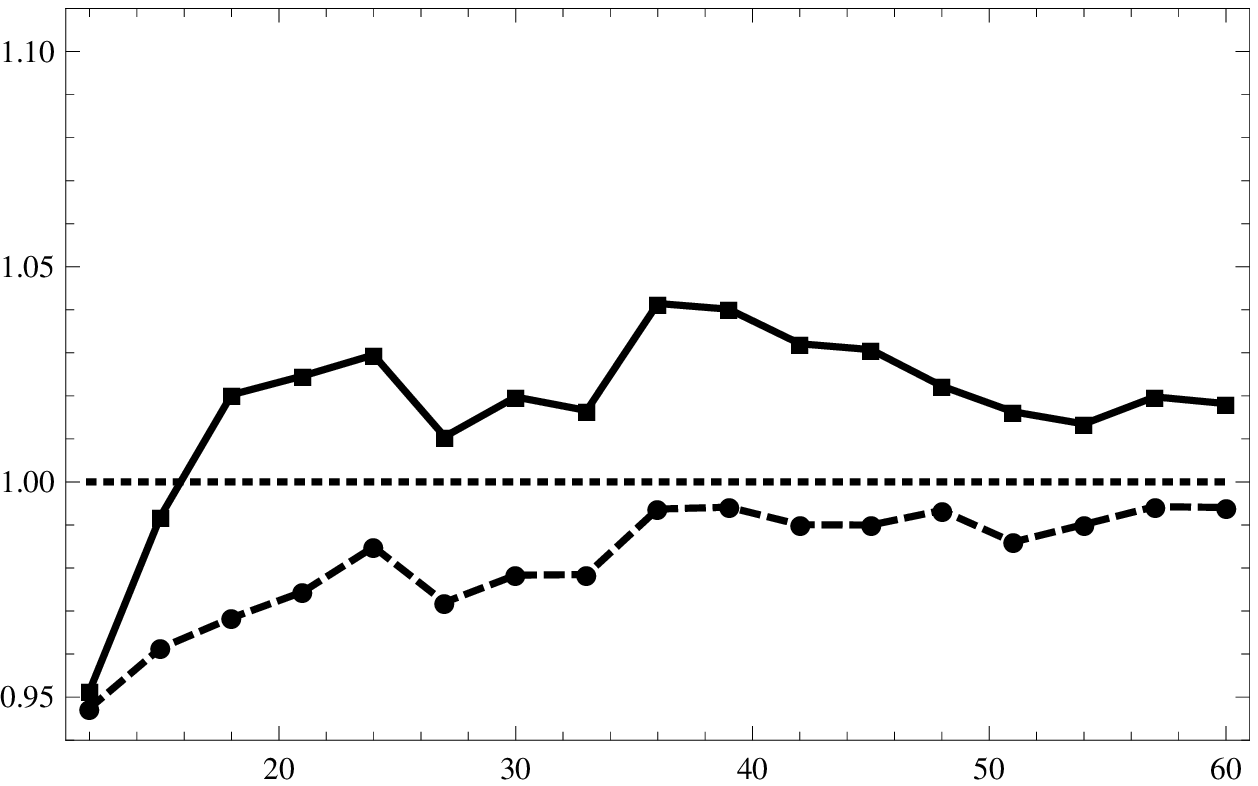}
\end{subfigure}\begin{subfigure}[b]{\tempwidth}
       \centering
       \includegraphics[width=\tempwidth]{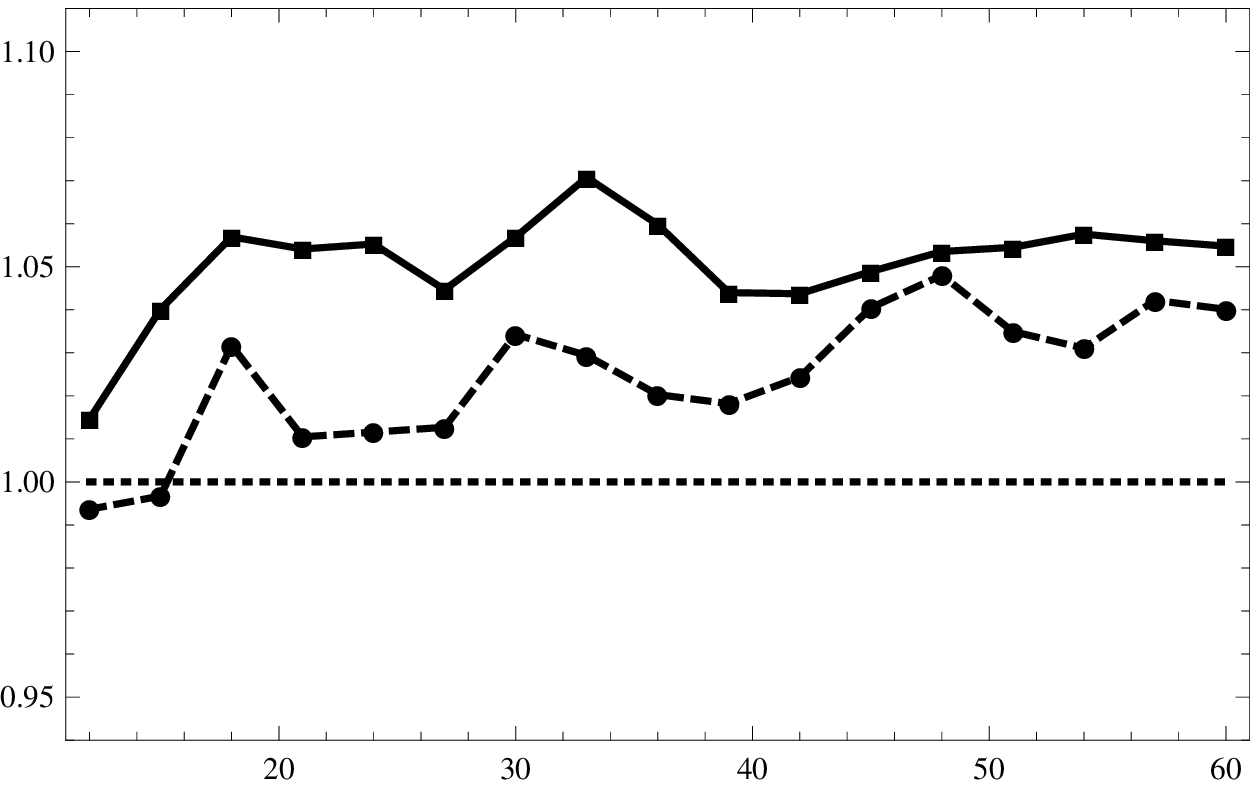}
\end{subfigure}\\
\rowname{{\normalfont\scriptsize Compartmental}}
\begin{subfigure}[b]{\tempwidth}
       \centering
       \includegraphics[width=\tempwidth]{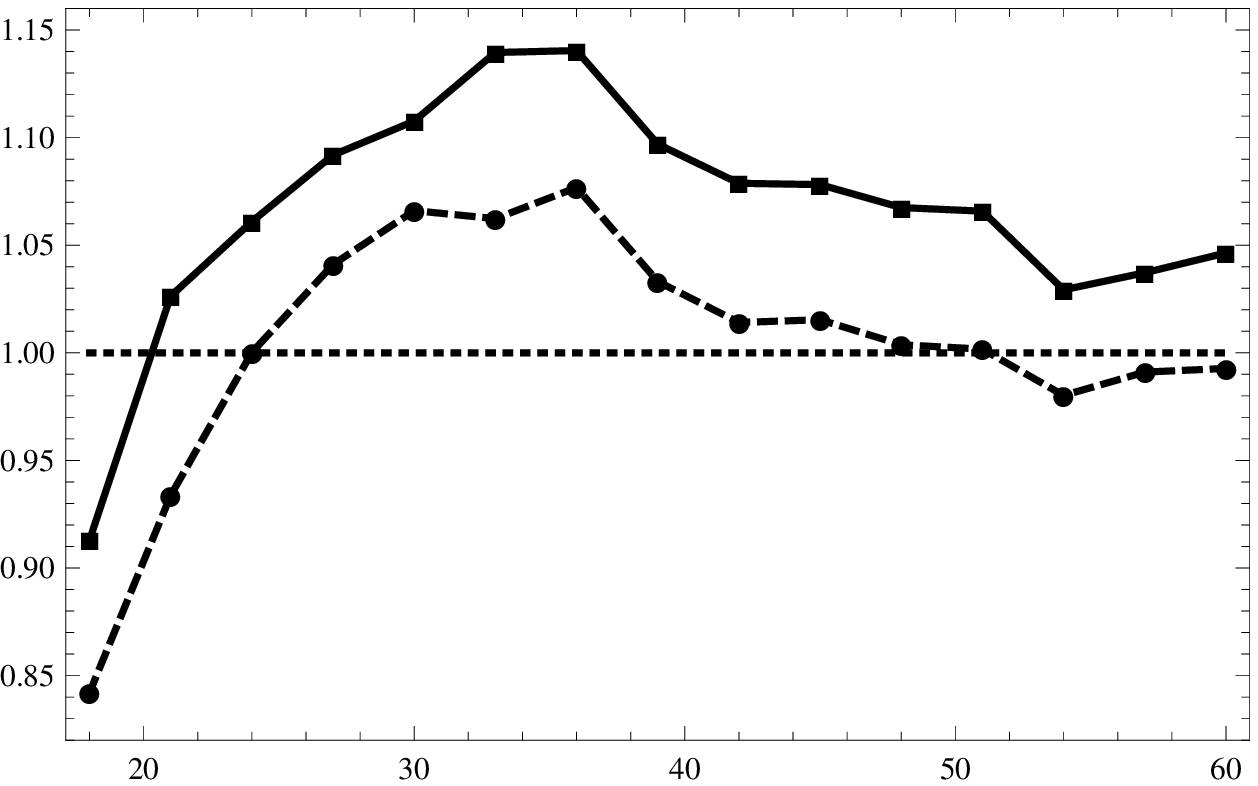}
\end{subfigure}
\begin{subfigure}[b]{\tempwidth}
       \centering
       \includegraphics[width=\tempwidth]{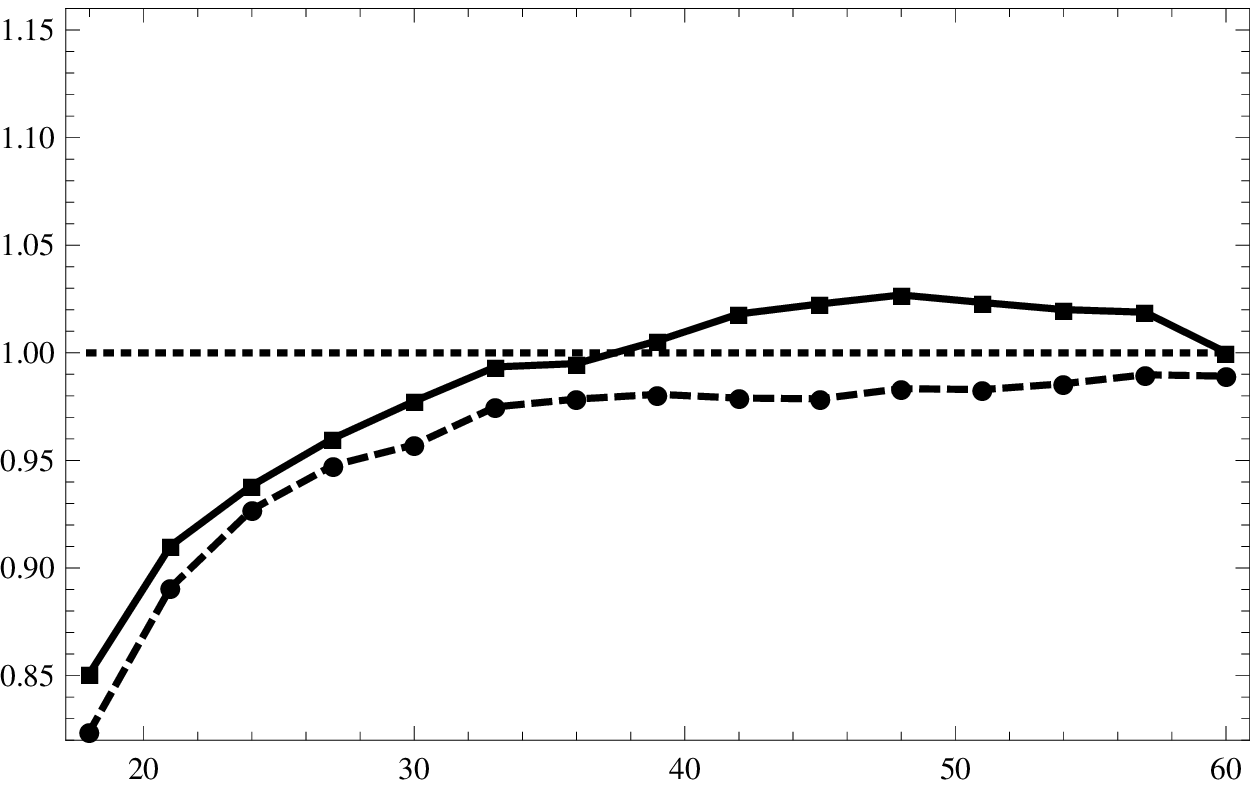}
\end{subfigure}\begin{subfigure}[b]{\tempwidth}
       \centering
       \includegraphics[width=\tempwidth]{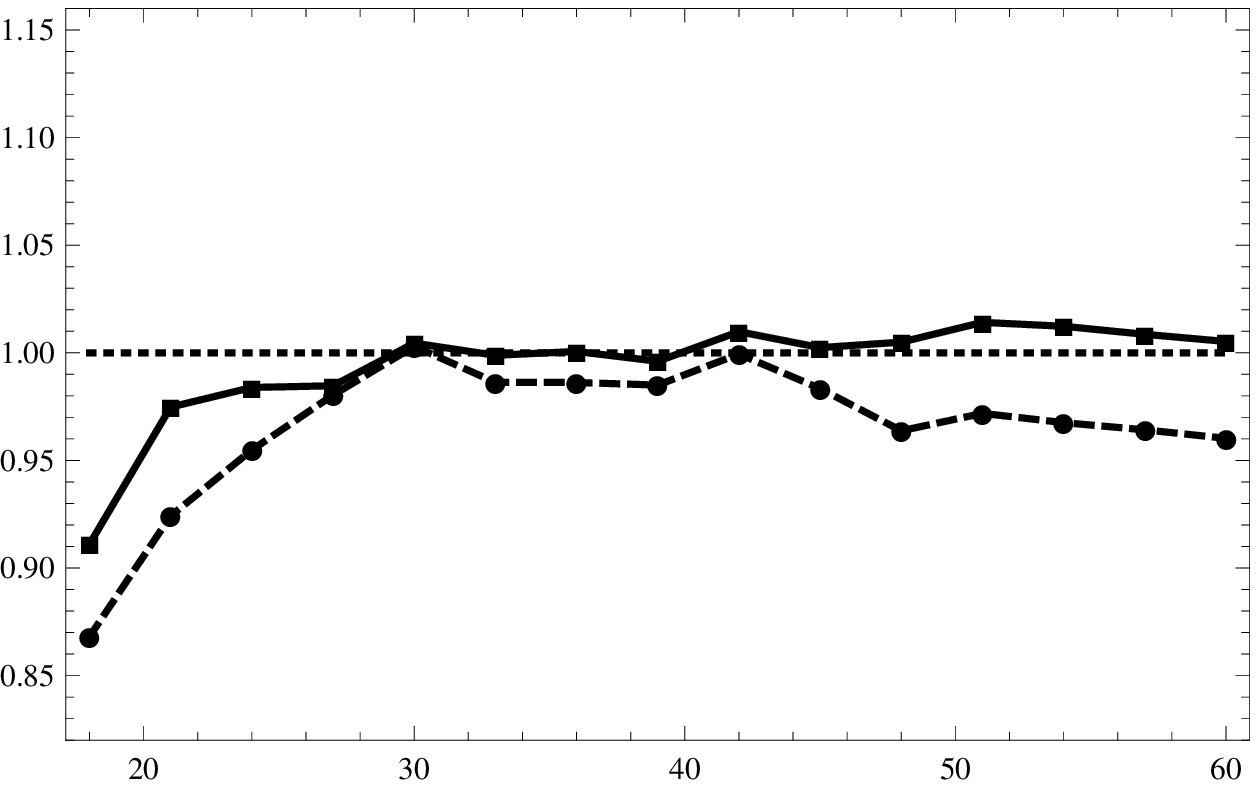}
\end{subfigure} \\
\caption{$c_{\boldsymbol{\theta}}$-efficiencies with respect to the MSE of the optimal RSD, ${\rm{RMSE}}_{\hat{\xi}_{RSD}}^{c_{\boldsymbol{\theta}}}$ (solid line), and the AOD, ${\rm{RMSE}}_{\hat{\xi}_{AOD}}^{c_{\boldsymbol{\theta}}}$ (dashed line), relative to the FLOD, represented by the dotted line at 1, for the Michaeles-Menten, Decay and Compartmental model (top to bottom) and the Cauchy, Exponential Power and $q$-Gaussian error distributions (left to right) as a function of the total sample size $n$.}\label{fig:MSEC}
\end{sidewaysfigure}

Figure \ref{fig:MSED} plots the $D$-efficiencies with respect to MSE of the optimal RSD, ${\rm{RMSE}}_{\hat{\xi}_{RSD}}^{D}$ (solid line), and the AOD, ${\rm{RMSE}}_{\hat{\xi}_{AOD}}^{D}$ (dashed line), relative to the FLOD, represented by the dotted line at 1, for the Michaeles-Menten, Decay and Compartmental model (top to bottom) and the Cauchy, Exponential Power and $q$-Gaussian error distributions (left to right) as a function of the total sample size $n$. Figure \ref{fig:MSEC} is the same as \ref{fig:MSED} except the relative efficiencies are defined in terms of the $c_{\boldsymbol{\theta}}$-optimal criterion.

An interesting feature of Figure \ref{fig:MSED} and \ref{fig:MSEC} is that the optimal RSD is more efficient that the FLOD in the vast majority of cases. In fact the behavior is similar to what was observed in Figures \ref{fig:OID} and \ref{fig:OIC}, there is a "burn in" period where the FLOD is better, in some cases; after the burn in the optimal RSD was uniformly more efficient with respect to MSE than the FLOD. As before the optimal RSD was uniformly more efficient that the AOD. This indicates that not only is the optimal RSD potentially equivalent with respect to MSE, but may in fact be more efficient than the FLOD and AOD. 

The preceding feature can, potentially, be explained by \citet{Efro:Hink:Asse:1978} where it is shown that the difference between the conditional MSE, MSE$[\hat{\eta}_{i}]$ = E$[(\hat{\eta}_{i} - \eta_{i})^{2}|\boldsymbol{\mathcal{A}}_{i} = \boldsymbol{a}_{i}]$, and $i_{\boldsymbol{a}_{i}}^{-1}$ and $\mathscr{F}_{i}$ are $O_{p}(n^{-2})$ and $O_{p}(n^{-3/2})$, respectively. From Proposition \ref{prop:info} this implies that the Fisher information in the relevant subset is a more precise approximation to the conditional MSE than the Fisher information. If this result extends, in some fashion, to the difference between the conditional MSE of $\boldsymbol{\hat{\theta}}$, defined as MSE$[\boldsymbol{\hat{\theta}}|\boldsymbol{\mathcal{A}} = \boldsymbol{A}]$= E$[(\boldsymbol{\hat{\theta}} - \boldsymbol{\theta} )(\boldsymbol{\hat{\theta}} - \boldsymbol{\theta} )^{T}|\boldsymbol{\mathcal{A}} = \boldsymbol{A}]$, and $(\boldsymbol{J}_{\boldsymbol{A}}^{\boldsymbol{x}})^{-1}$ it would indicate that the optimal RSD optimizes the conditional MSE for every relevant subset. Intuitively, this could indicate that the optimal RSD also optimizes the unconditional MSE,  MSE$[\boldsymbol{\hat{\theta}}] =  {\rm{E}}[\rm{MSE}[\boldsymbol{\hat{\theta}}|\boldsymbol{\mathcal{A}} = \boldsymbol{A}]]$. However, a more in depth theoretical investigation is required. 

In summary, this simulation study indicates that the optimal RSD does not lose information and, appears, to reduce the MSE of the parameter estimates.


\section{Discussion}
This work developed a framework for incorporating the relevant subset into the design of experiments. The motivation for such designs is the common argument that the Fisher information in the relevant subset is a more appropriate measure of the variability of the MLE than the Fisher information in the sample. It was found that relevant subset designs (RSD) optimize inference and the Fisher information in the sample subject to convex optimality criteria. 

In order to develop the proposed design the issue that the Fisher information in the relevant subset does not have a known analytical form needed to be resolved. Without an analytic expression it is an impractical optimization target. This was resolved by defining a hybrid measure of information that more accurately approximates the Fisher information in the relevant subset than either the Fisher information in the sample or the observed information. The proposed hybrid measure has a tractable form and can be used in optimization. A version of the general equivalence theorem was given for the hybrid measure which allows many existing optimal design algorithms to be adapted to incorporate the relevant subset. 

As an illustration, in this work a one-step ahead optimal RSD was developed. This design was compared to the fixed locally optimal design (FLOD) and the adaptive optimal design (AOD) in a simulation study. The FLOD depends on the underlying parameters; in the simulation study the FLOD was evaluated at the true value of the parameters and as a result it is not a competing method but instead represents a benchmark for the efficiency of fixed designs. Despite this fact the optimal RSD was still more efficient than the FLOD in the vast majority of the cases. This indicates that the FLOD does not represent a benchmark for RSD as it is for fixed designs. The AOD and optimal RSD do not have a local dependence and are directly comparable. The optimal RSD was uniformly more efficient that the AOD both with respect to the Fisher information in the relevant subset and the MSE. This suggests that the optimal RSD has general applicability in the design of experiments in nonlinear models.

\section{Technical Details} \label{sec:Tech}
Let $l_{\boldsymbol{y_{i}}}^{(\cdot k)}(\eta_{i}) =  (\partial^{k}/\partial\eta_{i}^{k}) \log f(\boldsymbol{y_{i}}|\eta_{i})$
represent the $k$th derivative of the log likelihood for all observations with support $x_{i}$. Further, when evaluated at $\eta_{i} = \hat{\eta}_{i}$ the notation $l_{\boldsymbol{y_{i}}}^{(\cdot k)}(\hat{\eta}_{i}) = l_{\boldsymbol{a_{i}}}^{(\cdot k)}$ is used to highlight the fact that $l_{\boldsymbol{a_{i}}}^{(\cdot k)}$ is  $\boldsymbol{a_{i}}$ measurable.

\subsection{Proof of Proposition \ref{prop:info}}

Required conditions:  (1) $\eta_{i} \in H$, where $H$ is an open subset of $\mathbb{R}$; (2) $\boldsymbol{\varepsilon}_{i}$ is a vector of independent and identically distributed random variables such that $f_{\eta}(y) = f_{0}(\varepsilon)$; (3) $l_{0}^{(\cdot k)}(\varepsilon)$ exists, E$[|l_{0}^{(\cdot k)}(\varepsilon)|]<\infty$ and the derivative $\partial^{k}/\partial\eta_{i}^{k}$ can be exchanged with the expectation for $k=1,\ldots,3$; (4) E$[\ddot{l}_{0}(\varepsilon)]<0$; (5) $l_{0}^{(\cdot 4)}(\varepsilon)$ is bounded in probability; (6) ${\rm{E}}[(\hat{\eta}_{i} - \eta_{i})^{2}] < \infty$; and (7) $n_{i}/n\rightarrow w_{i}>0$.

Lemmas 1 and 2 from \citep{Efro:Hink:Asse:1978} establish ${\rm{E}}[(\hat{\eta}_{i} - \eta_{i})|\boldsymbol{\mathcal{A}}_{i} = \boldsymbol{a_{i}}] = -\frac{1}{2}l_{\boldsymbol{a_{i}}}^{(\cdot 3)} \boldsymbol{i}_{\boldsymbol{a_{i}}}^{-2} + o_{p}(n^{-1})$, ${\rm{E}}[(\hat{\eta}_{i} - \eta_{i})^{2}|\boldsymbol{\mathcal{A}}_{i} = \boldsymbol{a_{i}}] = \boldsymbol{i}_{\boldsymbol{a_{i}}}^{-1} + o_{p}(n^{-1})$,
$n^{-1/2}(\boldsymbol{i}_{\boldsymbol{a_{i}}} - \mathscr{F}) \rightarrow N(0,w_{i}^{2}\gamma^{2})$ and $\boldsymbol{i}_{\boldsymbol{a_{i}}}^{1/2}(\hat{\eta}_{i} - \eta_{i}) \rightarrow N(0,1)$, where convergence is in distribution. The proof given in \citep{Efro:Hink:Asse:1978} requires conditions (1)-(7). From this it can be shown that a Taylor expansion yields
\begin{align} \label{eq:TaylorExp}
    l_{\boldsymbol{y}_{i}}^{(\cdot k)}(\eta_{i}) = l_{\boldsymbol{a_{i}}}^{(\cdot k)} - (\hat{\eta}_{i} - \eta_{i})l_{\boldsymbol{a_{i}}}^{ (\cdot k+1)} + (\hat{\eta}_{i} - \eta_{i})^{2}l_{\boldsymbol{a_{i}}}^{ (\cdot k+2)} + o_{p}(1)
\end{align}
for $k=1,2$.  These results imply
\begin{align} \label{eq:InfoExp}
    h_{\boldsymbol{a}_{i}}(\eta_{i}) &= {\rm{E}}[i_{\boldsymbol{y}_{i}}(\eta_{i})|\boldsymbol{\mathcal{A}}_{i} = \boldsymbol{a_{i}}] = {\rm{E}}[i_{\boldsymbol{a_{i}}} - (\hat{\eta}_{i} - \eta_{i})l_{\boldsymbol{a_{i}}}^{ (\cdot 3)} - (\hat{\eta}_{i} - \eta_{i})^{2}l_{\boldsymbol{a_{i}}}^{ (\cdot 4)}|\boldsymbol{\mathcal{A}}_{i} = \boldsymbol{a_{i}}] + o_{p}(1) \\
    &= i_{\boldsymbol{a_{i}}} + \frac{1}{2}[l_{\boldsymbol{a_{i}}}^{(\cdot 3)}]^{2} i_{\boldsymbol{a_{i}}}^{-2} - l_{\boldsymbol{a_{i}}}^{ (\cdot 4)}\boldsymbol{i}_{\boldsymbol{a_{i}}}^{-1} + o_{p}(1).
\end{align}
The above implies $n^{-1}[i_{\boldsymbol{a_{i}}} - h_{\boldsymbol{a}_{i}}(\eta_{i})] = O_{p}(n^{-1}$ as stated. The preceding statement and the Lemmas of \citep{Efro:Hink:Asse:1978} imply that $n^{-1}(\boldsymbol{i}_{\boldsymbol{a_{i}}} - \mathscr{F}) = O_{p}(n^{-1/2})$ as stated.


\subsection{Proof of Theorem \ref{thm:CondInfo}}
It is required that the conditions of Proposition \ref{prop:info} hold for all $i=1,\ldots,d$; and additionally that (8) $\theta\in\Theta$, where $\Theta$ is an open subset of $\mathbb{R}^{p}$; and (9) $\eta_{\theta'}(x) = \eta_{\theta}(x) \implies \theta' = \theta$ for all $x\in\mathcal{X}$. From \eqref{eq:TaylorExp} and  it can be seen that 
\begin{align}
    n^{-1}[h_{\boldsymbol{a}_{i}}(\eta_{i}) - i_{\boldsymbol{y}_{i}}(\eta_{i})] &= n^{-1}[h_{\boldsymbol{a}_{i}}(\eta_{i}) - \{i_{\boldsymbol{a}_{i}} + (\hat{\eta}_{i} - \eta_{i}) l_{\boldsymbol{a_{i}}}^{ (\cdot 3)} \}] + O_{p}(n^{-1}) = O_{p}(n^{1/2}),
\end{align}
since $(\hat{\eta}_{i} - \eta_{i}) = O_{p}(n^{-1/2})$ per Lemmas 1 and 2 from \citep{Efro:Hink:Asse:1978}. Next, the conditional expectation
\begin{align} \label{eq:ScoreExp}
    {\rm{E}}[\dot{l}_{\boldsymbol{y}_{i}}(\hat{\eta}_{i}) | \boldsymbol{\mathcal{A}}_{i} = \boldsymbol{a_{i}}] &= {\rm{E}}[\dot{l}_{\boldsymbol{a_{i}}} + (\hat{\eta}_{i} - \eta_{i})i_{\boldsymbol{a_{i}}} + (\hat{\eta}_{i} - \eta_{i})^{2}l_{\boldsymbol{a_{i}}}^{(\cdot3)}|\boldsymbol{\mathcal{A}}_{i} = \boldsymbol{a_{i}}] + o_{p}(1) \\
    &= \frac{3}{2}l_{\boldsymbol{a_{i}}}^{(\cdot 3)} \boldsymbol{i}_{\boldsymbol{a_{i}}}^{-1} + o_{p}(1).
\end{align} 
implies that $\dot{l}_{\boldsymbol{y}_{i}}(\eta_{i}) = O_{p}(1)$. These points imply $n^{-1}[H_{\boldsymbol{A}}^{\boldsymbol{x}}(\boldsymbol{\theta}) - \boldsymbol{I}_{\boldsymbol{y}}^{\boldsymbol{x}}(\boldsymbol{\theta})] = O_{p}(n^{-1/2})$. Using \eqref{eq:InfoExp} and \eqref{eq:ScoreExp} we can now write
\begin{align}
   H_{\boldsymbol{A}}^{\boldsymbol{x}}(\boldsymbol{\theta}) &=   {\rm{E}}[I_{\boldsymbol{y}}^{\boldsymbol{x}}(\boldsymbol{\theta})|\boldsymbol{\mathcal{A}}_{i} = \boldsymbol{A}_{i}] = \sum_{i=1}^{d} h_{\boldsymbol{a}_{i}} \dot{\eta}_{i}\dot{\eta}_{i}^{T} + \sum_{i=1}^{d} {\rm{E}}[\dot{l}_{\boldsymbol{y}_{i}}(\eta_{i}) |\boldsymbol{\mathcal{A}}_{i} = \boldsymbol{a}_{i} ] \ddot{\eta}_{i} \\
   &= \sum_{i=1}^{d} \left[i_{\boldsymbol{a}_{i}} + \frac{1}{2}[l_{\boldsymbol{a_{i}}}^{(\cdot 3)}]^{2} \boldsymbol{i}_{\boldsymbol{a_{i}}}^{-2} - l_{\boldsymbol{a_{i}}}^{ (\cdot 4)}\boldsymbol{i}_{\boldsymbol{a_{i}}}^{-1}\right] \dot{\eta}_{i}\dot{\eta}_{i}^{T} + \sum_{i=1}^{d}\frac{3}{2} l_{\boldsymbol{a_{i}}}^{(\cdot 3)} \boldsymbol{i}_{\boldsymbol{a_{i}}}^{-1} \ddot{\eta}_{i} + o_{p}(1)\\
   &= \boldsymbol{J}_{\boldsymbol{A}}^{\boldsymbol{x}}(\boldsymbol{\theta}) + \sum_{i=1}^{d}\left[\{\frac{1}{2}[l_{\boldsymbol{a_{i}}}^{(\cdot 3)}]^{2} \boldsymbol{i}_{\boldsymbol{a_{i}}}^{-2} - l_{\boldsymbol{a_{i}}}^{ (\cdot 4)}\boldsymbol{i}_{\boldsymbol{a_{i}}}^{-1}\} \dot{\eta}_{i}\dot{\eta}_{i}^{T} + \frac{3}{2} l_{\boldsymbol{a_{i}}}^{(\cdot 3)} \boldsymbol{i}_{\boldsymbol{a_{i}}}^{-1}\ddot{\eta}_{i}\right] + o_{p}(1).
\end{align}
Each term in the square brackets is $O_{p}(1)$ which directly implies $n^{-1}[H_{\boldsymbol{A}}^{\boldsymbol{x}}(\boldsymbol{\theta}) - \boldsymbol{J}_{\boldsymbol{A}}^{\boldsymbol{x}}(\boldsymbol{\theta})] = O_{p}(n^{-1})$ as stated. Finally, note that $n^{-1}[\boldsymbol{J}_{\boldsymbol{A}}^{\boldsymbol{x}}(\boldsymbol{\theta}) - \boldsymbol{F}^{\xi}(\boldsymbol{\theta})] =  n^{-1}\sum_{i=1}^{d} [i_{\boldsymbol{a}_{i}} - \mathscr{F}_{i}] \dot{\eta}_{i}\dot{\eta}_{i}^{T} = O_{p}(n^{-1/2})$ from Lemmas 1 and 2 from \citep{Efro:Hink:Asse:1978}. Therefore, $n^{-1}[H_{\boldsymbol{A}}^{\boldsymbol{x}}(\boldsymbol{\theta}) - \boldsymbol{F}(\boldsymbol{\theta})] = O_{p}(n^{-1/2})$ as stated. 

\subsection{Proof of Theorem \ref{thm:Gen_Eq}}
For this theorem it is required that (1) $\Psi(\cdot)$ is a non-negative convex positive-homogeneous function; (2) $\mathcal{X}$ is a compact subspace of $\mathbb{R}^{s}$; and (3) $\boldsymbol{\hat{\eta}}(j)$ exists.  Writing $f(x_{i}) = \eta_{i}$ then $\boldsymbol{M}^{\xi}$ can be written as $\boldsymbol{M}^{\xi}(\boldsymbol{\theta}) = \sum_{i=1}^{d}w_{i}f(x_{i})f(x_{i})^{T}$. The normalized expected Fisher information is written this to demonstrate that it has the same form as that of a linear model.

To define an augmented design let $\xi(1)$ be an initial design with sample size $m(1)$. This initial design will be augmented with a design of size $m(2)$. The normalized expected Fisher information for an augmented design, $\xi$, given the initial design $\xi(1)$, is
$\boldsymbol{M}^{\xi}_{\alpha}(\boldsymbol{\theta}) = \alpha \boldsymbol{M}^{\xi}(\boldsymbol{\theta})  + (1-\alpha)\boldsymbol{M}^{\xi(1)}(\boldsymbol{\theta})$, where $\alpha = m(2)/[m(1) + m(2)]$. The continuous $\Psi-$optimal augmented design is defined as $\xi_{\alpha}^{*}(\boldsymbol{\theta}) = \arg\min_{\xi \in \Xi_{\Delta}} \Psi\{\boldsymbol{M}^{\xi}_{\alpha}(\boldsymbol{\theta})\}$. As pointed out by \cite{Atki:Done:Tobi:opti:2007} ch. 19, Theorem 11.6 and Lemma 6.16 from \cite*{Puke:Opti:1993} imply a general equivalence theorem for augmented designs which states that the following are equivalent (1) $\xi_{\alpha}^{*}(\boldsymbol{\theta})$ is the continuous optimal design (2) $\xi_{\alpha}^{*}(\boldsymbol{\theta})=\arg\max_{\xi \in \Xi_{\Delta}}\min_{x\in\mathscr{X}}\phi_{\alpha}^{\xi}(x,\boldsymbol{\theta})$ and (3) the $\min_{x\in\mathscr{X}}\phi_{\alpha}^{\xi_{\alpha}^{*}(\boldsymbol{\theta})}(x,\boldsymbol{\theta})=0$ and the equality occurs only at the support points of $\xi_{\alpha}^{*}(\boldsymbol{\theta})$, where
$\phi_{\alpha}^{\xi}(x,\boldsymbol{\theta}) = f^{T}(x)[\boldsymbol{M}^{\xi}_{\alpha}(\boldsymbol{\theta})]^{-1}f(x) - tr\{\boldsymbol{M}^{\xi}(\boldsymbol{\theta})[\boldsymbol{M}^{\xi}_{\alpha}(\boldsymbol{\theta})]^{-1}\}$. As defined $\tau_{\boldsymbol{A}(j)}\in\Xi_{\Delta}$, since $i_{\boldsymbol{a}_{i}(j)} \ge 0$. Equation \eqref{eq:Tj} can be written, after some basic algebra, as
\begin{align}
\boldsymbol{\hat{T}}_{\boldsymbol{A}(j)}^{\xi}(\boldsymbol{\theta}) &= [m(j+1)/ \beta(j+1)] \boldsymbol{R}_{\boldsymbol{A}(j)}^{\xi}(\boldsymbol{\theta}).
\end{align}
Since the term $m(j+1) + Q_{\boldsymbol{A}(j)}$ is a known constant after run $j$ the following holds
\begin{align} \label{eq:Aug_Opt_Adapt}
\xi_{\boldsymbol{A}(j)}^{*}(\boldsymbol{\theta}) = \arg\min_{\xi \in \Xi_{\Delta}} \Psi\{\boldsymbol{\hat{T}}_{\boldsymbol{A}(j)}^{\xi}(\boldsymbol{\theta})\} = \arg\min_{\xi \in \Xi_{\Delta}} \Psi\{\boldsymbol{R}_{\boldsymbol{A}(j)}^{\xi}(\boldsymbol{\theta})) \}.
\end{align}
As written $\boldsymbol{R}_{\boldsymbol{A}(j)}^{\xi}(\boldsymbol{\theta})$ is equivalent to an augmented design in the linear model setting and therefore the above referenced general equivalence theorem holds with sensitivity function
\begin{align}
\nu_{\boldsymbol{A}(j)}^{\xi}(x,\boldsymbol{\theta}) &= \dot{\eta}_{\boldsymbol{\theta}}^{T}(x) \{\boldsymbol{R}_{\boldsymbol{A}(j)}^{\xi}(\boldsymbol{\theta})\}^{-1}\dot{\eta}_{\boldsymbol{\theta}}(x) - tr[\boldsymbol{M}^{\xi}(\boldsymbol{\theta})\{\boldsymbol{R}_{\boldsymbol{A}(j)}^{\xi}(\boldsymbol{\theta})\}^{-1}],
\end{align}
which completes the proof.

\begin{supplement}

\begin{center}
\textbf{Supplemental Materials for Optimal Relevant Subset Designs in Nonlinear Models}
\end{center}

\section{Initializing the Adaptive Procedures}
The FLOD is computed at the parameter values that are used to generate the simulated data. In practice these values are unknown and cannot be used to initialize the adaptive designs. In nonlinear models it is important to initialize the adaptive designs to ensure the existence of the MLE with high probability. This is even more important in a simulation since it is required that the MLE exists for every iteration.

\cite{Dett:Bied:Robu:2003} derive maximin $D$-optimal design for the Michaeles-Menten mean function. A maximin design is defined as the design that maximizes the minimum of the $\Psi$-efficiency over an interval of $\boldsymbol{\theta}$. Maximin designs are considered to be more robust than the FLOD at an initial guess. To initialize both the $D$-and $c_{\boldsymbol{\theta}}$- optimal designs for the Michaeles-Menten model the following maximin optimal design with $\theta_{2}\in[100,500]$, given in \cite{Dett:Bied:Robu:2003}, is used
$\xi^{init} = \{(177.83,1/2),(2000,1/2)\}$. This initial design has an efficiency of 0.998 and 0.911 with respect to the $D$ and $c_{\boldsymbol{\theta}}$ criteria, respectively.

\cite{Dett:Lope:Rodr:Maxi:2006} find maximin $D$-optimal designs for exponential decay models. For the Danfaer experiment given in their paper they find the maximin $D$-optimal design for $\theta_{2}\in[0.0126,0.0156]$ to be $\xi^{init} = \{(70.43,1/2),(500,1/2)\}$. This is used to initialize the $D$-and $c_{\boldsymbol{\theta}}$- optimal designs for the exponential decay examples. This initial design has an efficiency of approximately 1 and 0.907 with respect to the $D$ and $c_{\boldsymbol{\theta}}$ criteria, respectively.

\cite{Atki:Chal:Herz:Opti:1993} find Bayesian optimal designs for the estimated parameter of the compartmental model for the Fresen (1984) data set. These designs do depend on the underlying parameters; however, they are more robust than local optimal designs since the parameters are assumed to have a prior distribution and designs are found by integrating over the prior. To initialize the compartmental models the following designs are used; $\xi_{D}^{init} = \{(0.2288,1/3),(1.4170,1/3),(18.4513,1/3)\}$ and $\xi_{c_{\boldsymbol{\theta}}}^{init} = \{(0.1829,1/3),(2.4639,1/3),(8.8542,1/3)\}$, This initial design has has an efficiency of approximately 0.988 and 0.719 with respect to the $D$ and $c_{\boldsymbol{\theta}}$ criteria, respectively. 

In a small number of cases (less than 0.5\%) the Newton-Raphson algorithm to find the MLE failed to converge for the Cauchy simulations. These cases were omitted from the simulation study.

Initial designs with equal weights have been selected for simplicity. In the simulation 5 observations will be placed on each point in the initial design. This means that $m(1)=10$, the sample size for the initial design, for the Michaeles-Menten and decay models; $m(1)=15$ for the compartmental model. 

The design space for each of the real world experiments was continuous. In an effort to be as practical as possible the design space is continuous in the simulations study. When the support is continuous it is not possible to ensure sufficient repeats at every point in the design. To accommodate this the approach described in Remark \ref{rem:repeat} was used to obtain the optimal RSD. 

\section{Information Efficiency}

Figure \ref{fig:MID} plots the $D$-efficiencies with respect to the Fisher information in the sample of the optimal RSD, ${\rm{RM}}_{\hat{\xi}_{RSD}}^{D}$ (solid line), and the AOD, ${\rm{RM}}_{\hat{\xi}_{AOD}}^{D}$ (dashed line), relative to the FLOD, represented by the dotted line at 1, for the Michaeles-Menten, Decay and Compartmental model (left to right) and the Cauchy, Exponential Power and $q$-Gaussian error distributions (top to bottom). Figure \ref{fig:MIC} is the same as \ref{fig:MID} except the relative efficiencies are defined in terms of the $c_{\boldsymbol{\theta}}$-optimal criterion. The figures have the same scale as Figures \ref{fig:OID} and \ref{fig:OIC}. The main point to be extracted from these figures is that there is very little difference between the Fisher information in the sample from the three designs. Particularly, the RSD and AOD are very similar with respect to this measure of information.

\begin{sidewaysfigure}
\setlength{\tempwidth}{.32\linewidth}
\settoheight{\tempheight}{\includegraphics[width=\tempwidth]{CauchyMMVD.eps}}
\centering
\hspace{\baselineskip}
\columnname{{\normalfont\scriptsize Cauchy}}\hfil
\columnname{{\normalfont\scriptsize Exponential Power}}\hfil
\columnname{{\normalfont\scriptsize q-Gaussian}}\\
\rowname{{\normalfont\scriptsize Michaeles-Menten}}
\begin{subfigure}[b]{\tempwidth}
       \centering
       \includegraphics[width=\tempwidth]{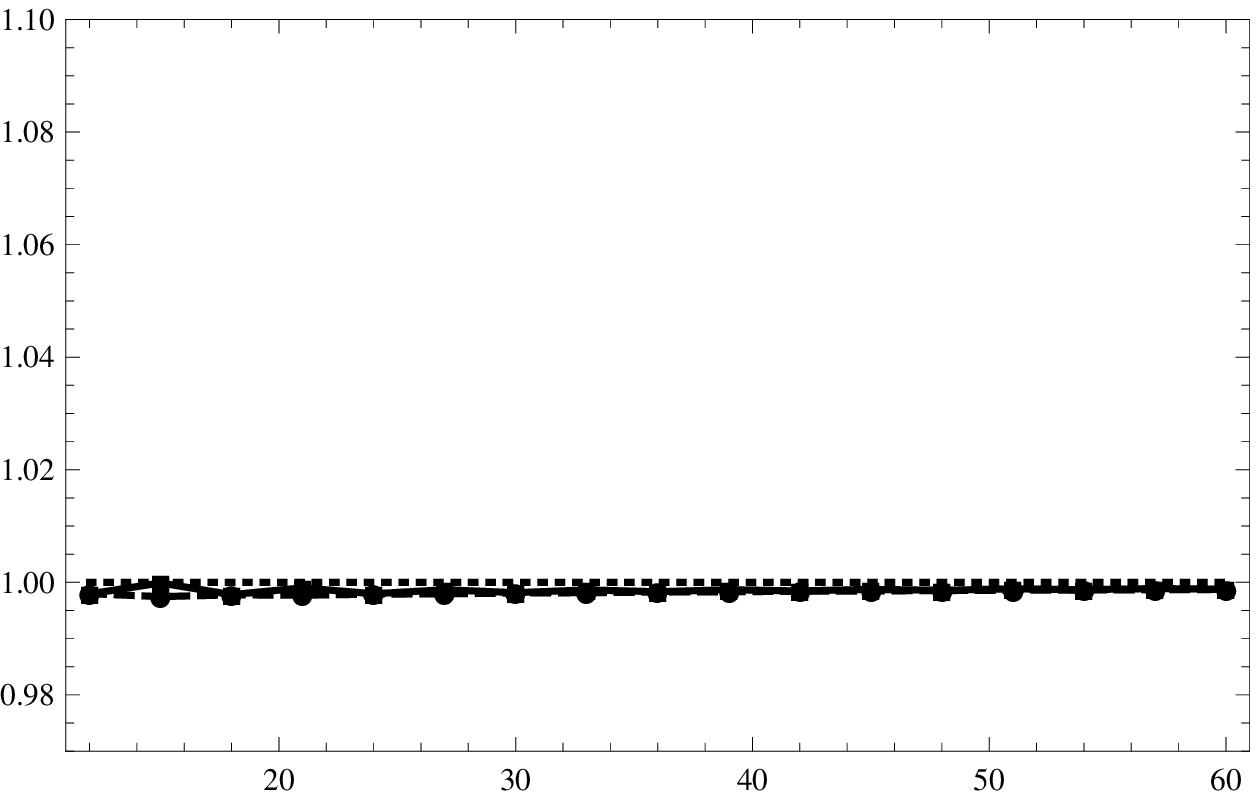}
\end{subfigure}
\begin{subfigure}[b]{\tempwidth}
       \centering
       \includegraphics[width=\tempwidth]{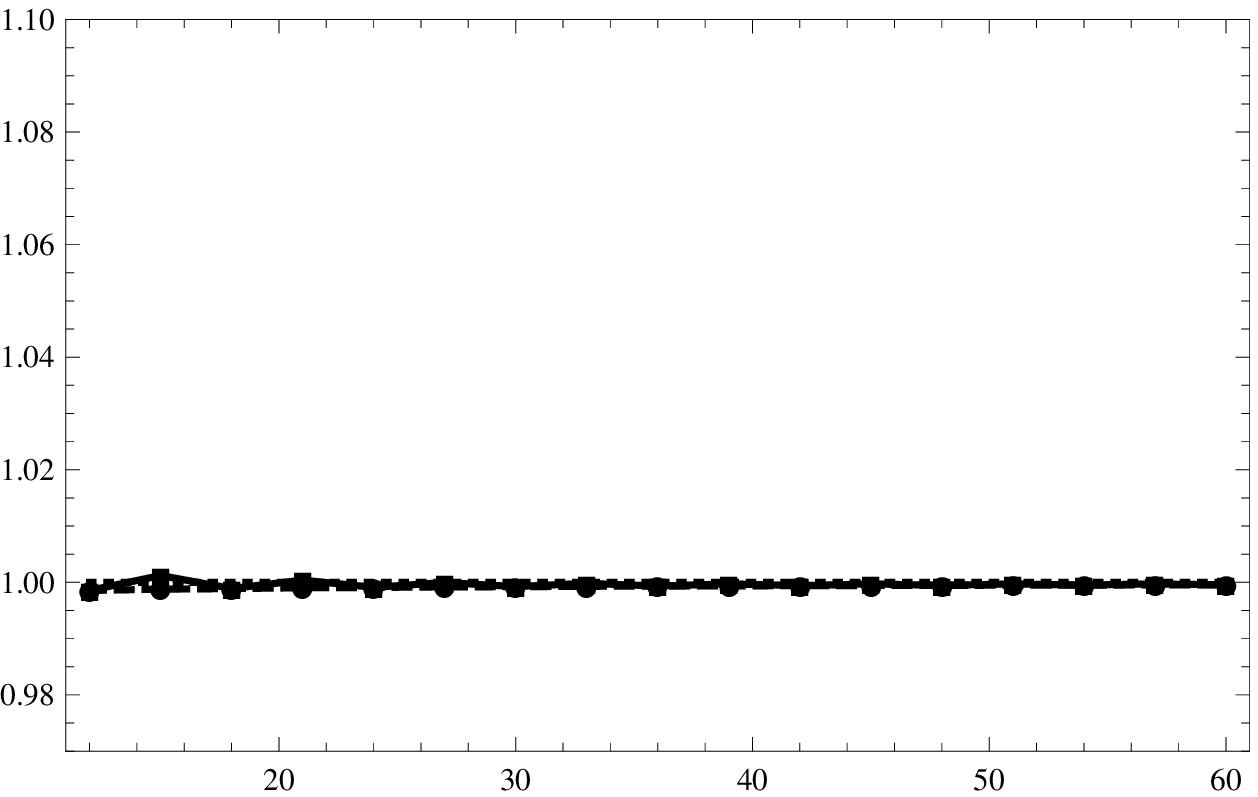}
\end{subfigure}\begin{subfigure}[b]{\tempwidth}
       \centering
       \includegraphics[width=\tempwidth]{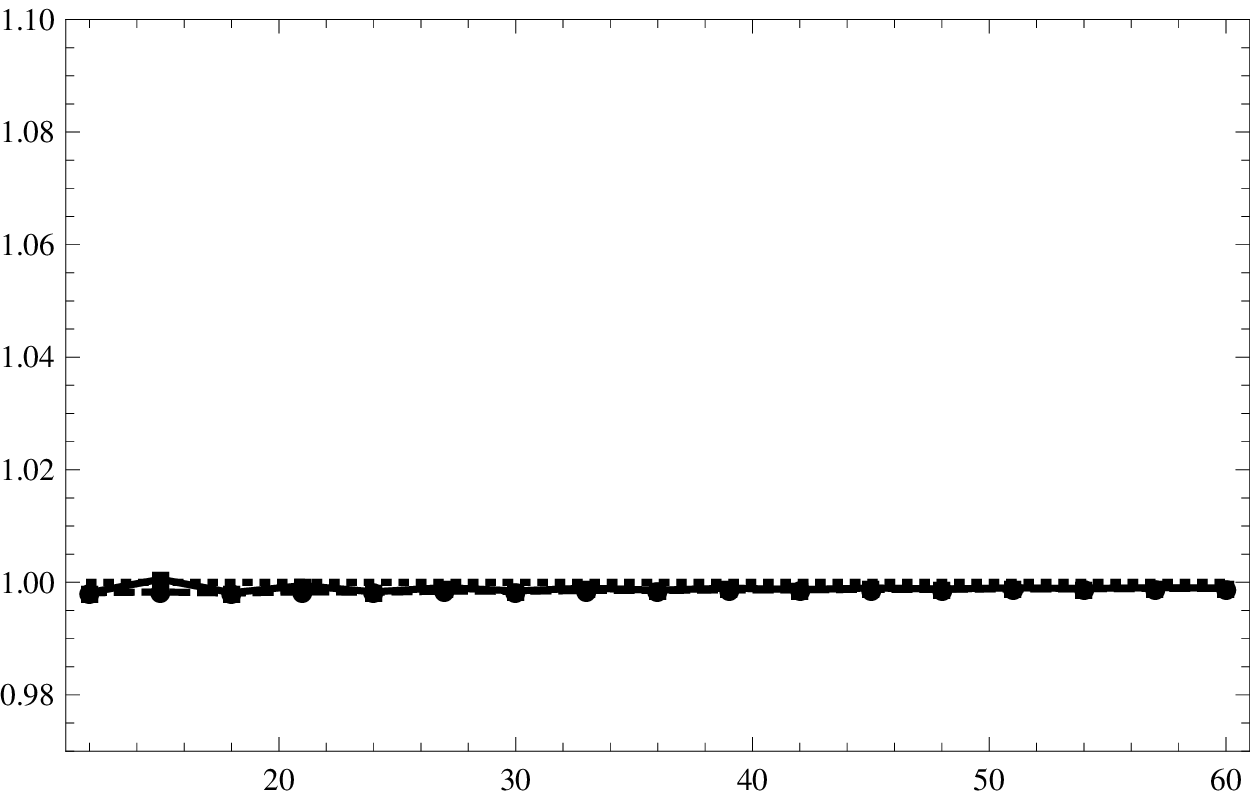}
\end{subfigure}\\
\rowname{{\normalfont\scriptsize Decay}}
\begin{subfigure}[b]{\tempwidth}
       \centering
       \includegraphics[width=\tempwidth]{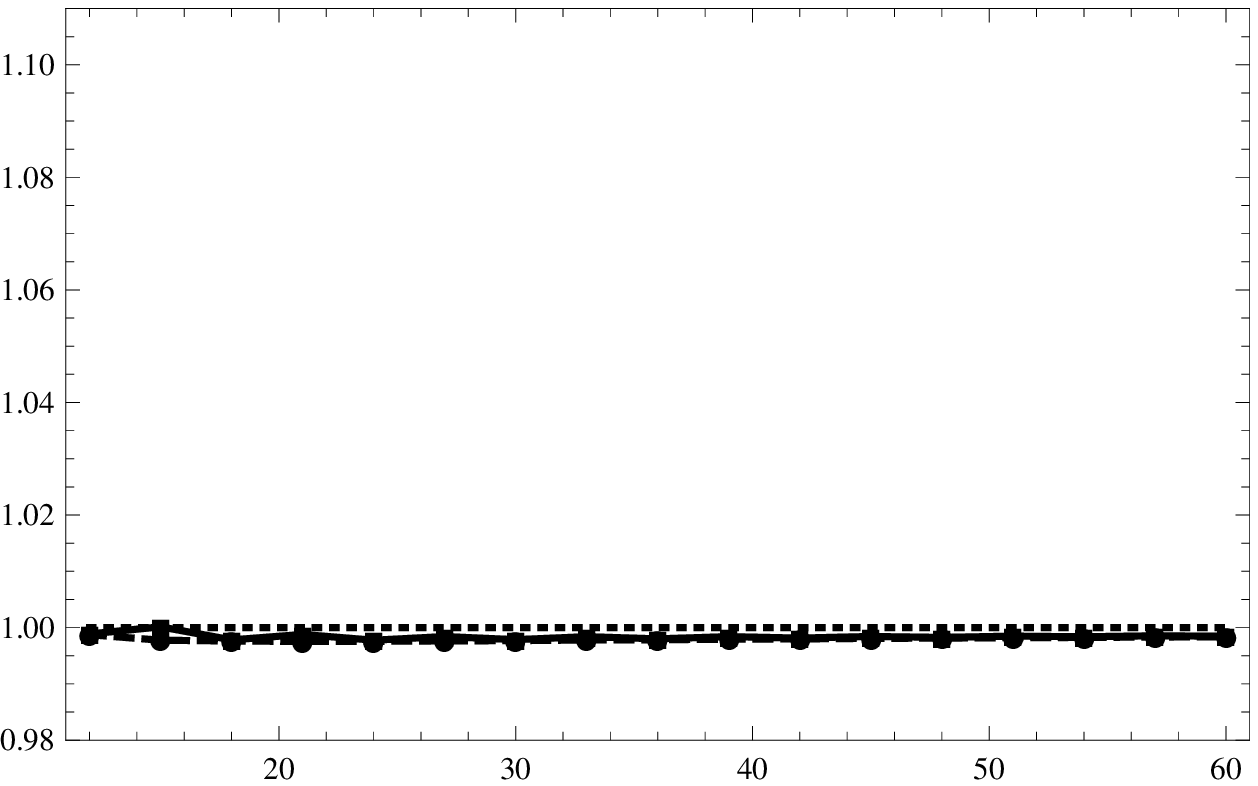}
\end{subfigure}
\begin{subfigure}[b]{\tempwidth}
       \centering
       \includegraphics[width=\tempwidth]{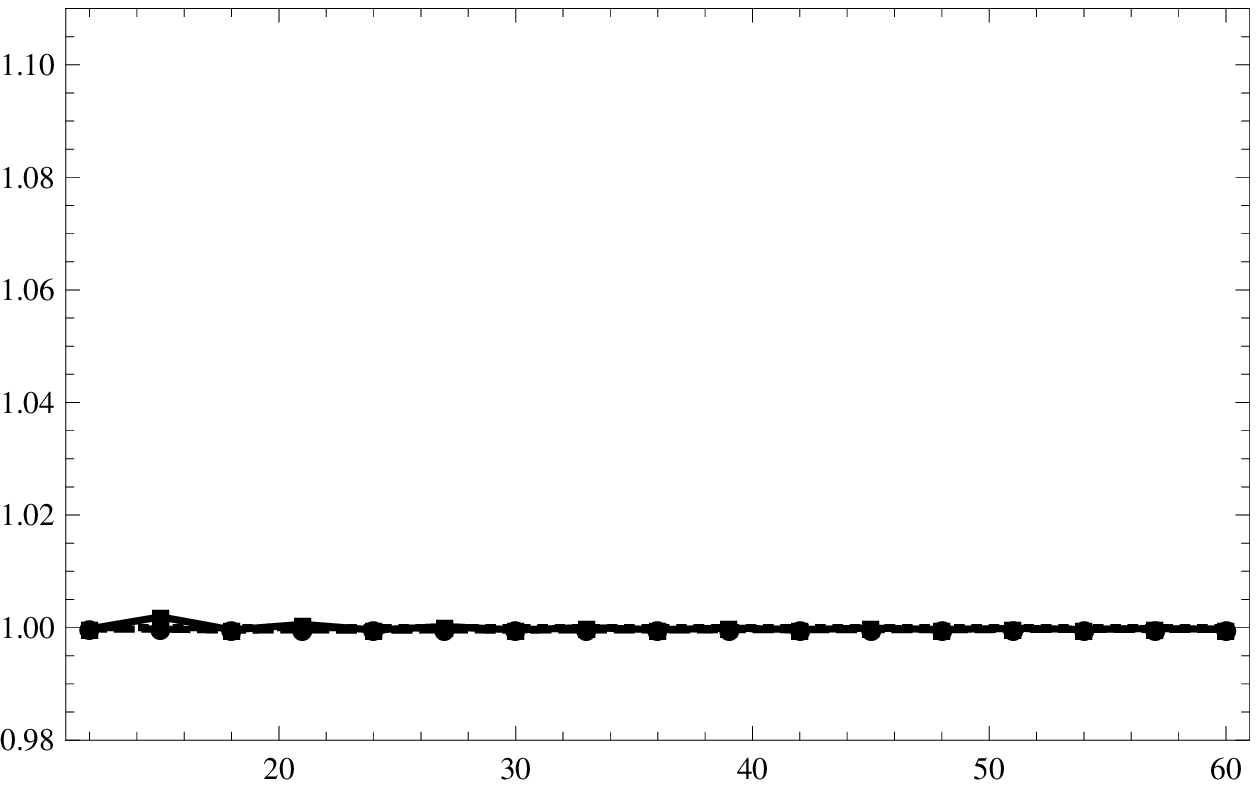}
\end{subfigure}\begin{subfigure}[b]{\tempwidth}
       \centering
       \includegraphics[width=\tempwidth]{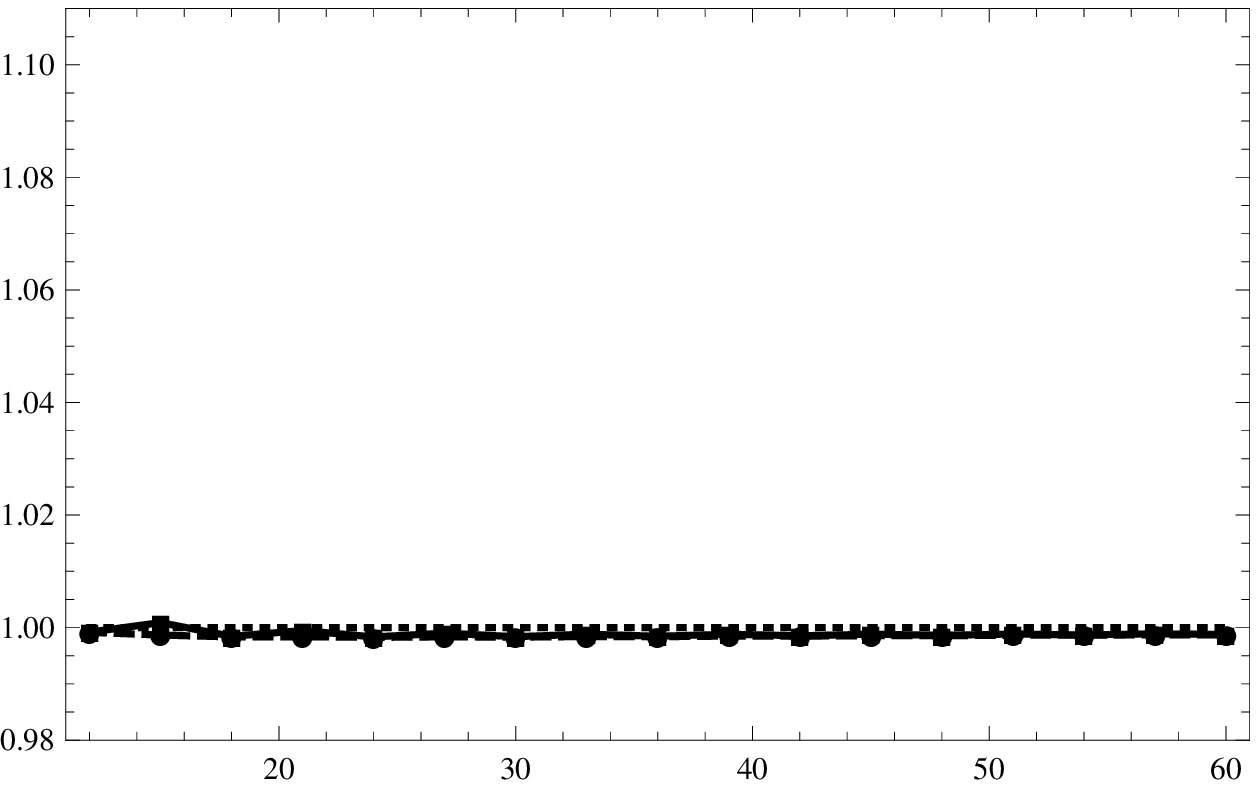}
\end{subfigure}\\
\rowname{{\normalfont\scriptsize Compartmental}}
\begin{subfigure}[b]{\tempwidth}
       \centering
       \includegraphics[width=\tempwidth]{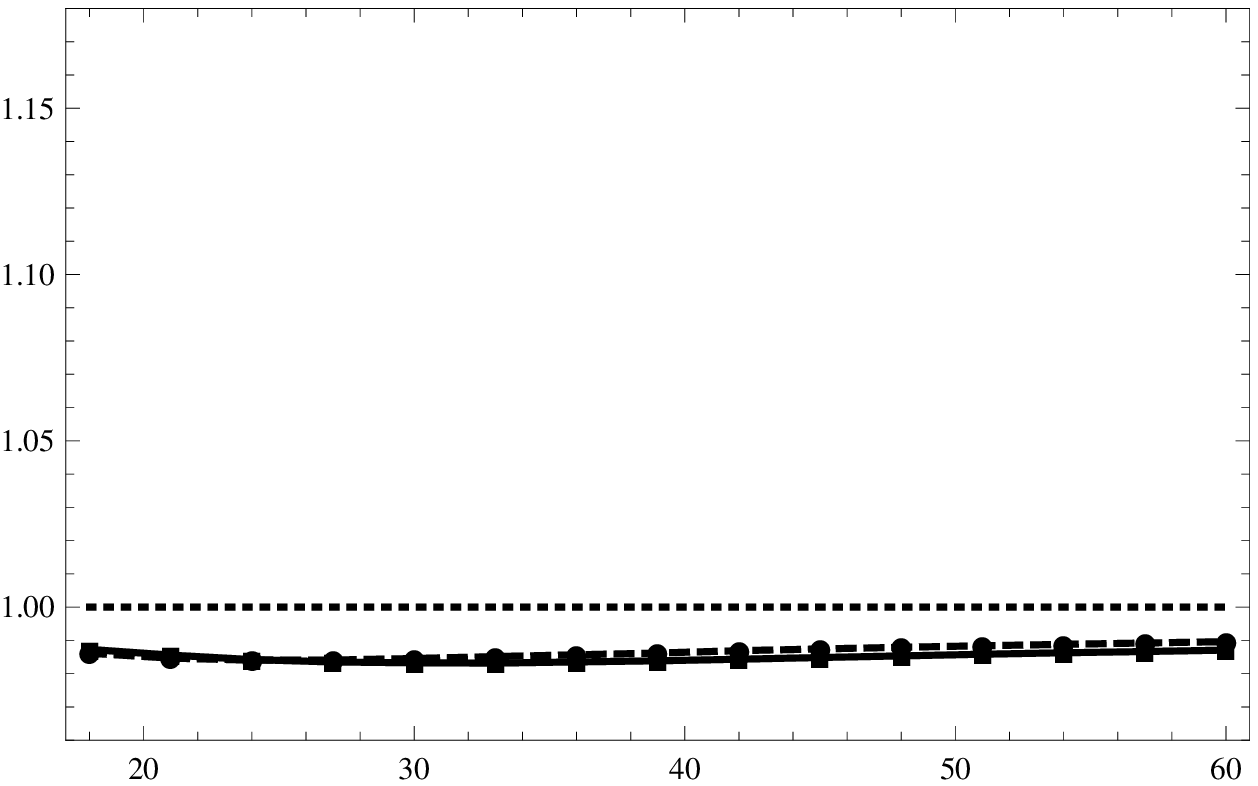}
\end{subfigure}
\begin{subfigure}[b]{\tempwidth}
       \centering
       \includegraphics[width=\tempwidth]{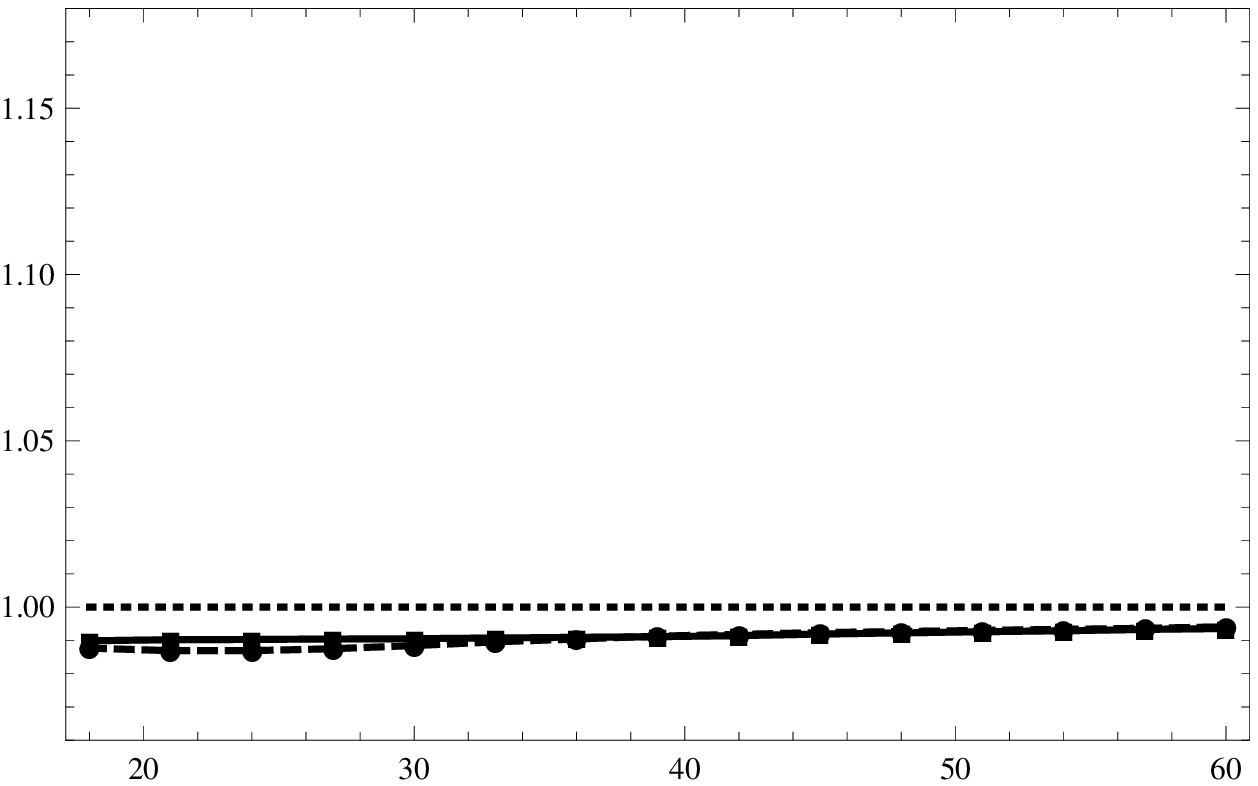}
\end{subfigure}\begin{subfigure}[b]{\tempwidth}
       \centering
       \includegraphics[width=\tempwidth]{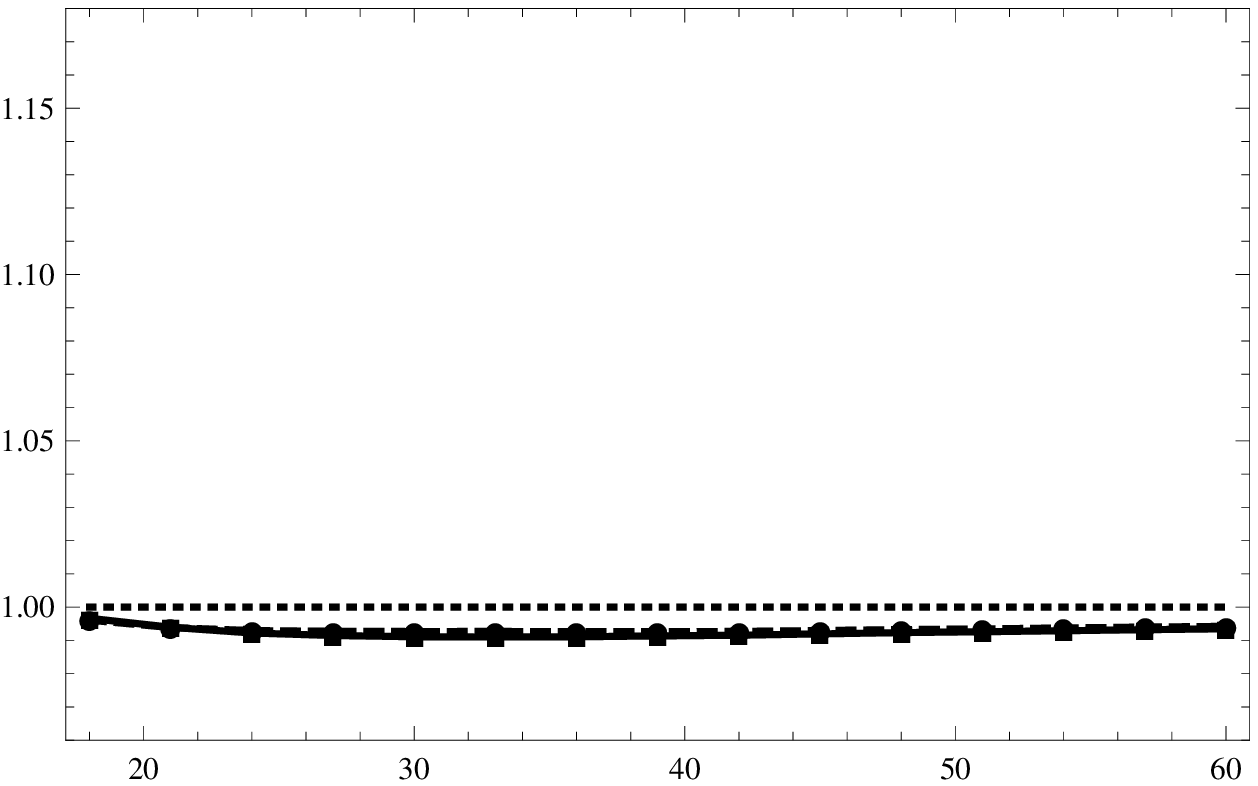}
\end{subfigure}\\
\caption{$D$-efficiencies with respect to the Fisher information in the sample of the optimal RSD, ${\rm{RM}}_{\hat{\xi}_{RSD}}^{D}$ (solid line), and the AOD, ${\rm{RM}}_{\hat{\xi}_{AOD}}^{D}$ (dashed line), relative to the FLOD, represented by the dotted line at 1, for the Michaeles-Menten, Decay and Compartmental model (top to bottom) and the Cauchy, Exponential Power and $q$-Gaussian error distributions (left to right).}\label{fig:MID}
\end{sidewaysfigure}
\begin{sidewaysfigure}
\setlength{\tempwidth}{.32\linewidth}
\settoheight{\tempheight}{\includegraphics[width=\tempwidth]{CauchyMMVD.eps}}
\centering
\hspace{\baselineskip}
\columnname{{\normalfont\scriptsize Cauchy}}\hfil
\columnname{{\normalfont\scriptsize Exponential Power}}\hfil
\columnname{{\normalfont\scriptsize q-Gaussian}}\\
\rowname{{\normalfont\scriptsize Michaeles-Menten}}
\begin{subfigure}[b]{\tempwidth}
       \centering
       \includegraphics[width=\tempwidth]{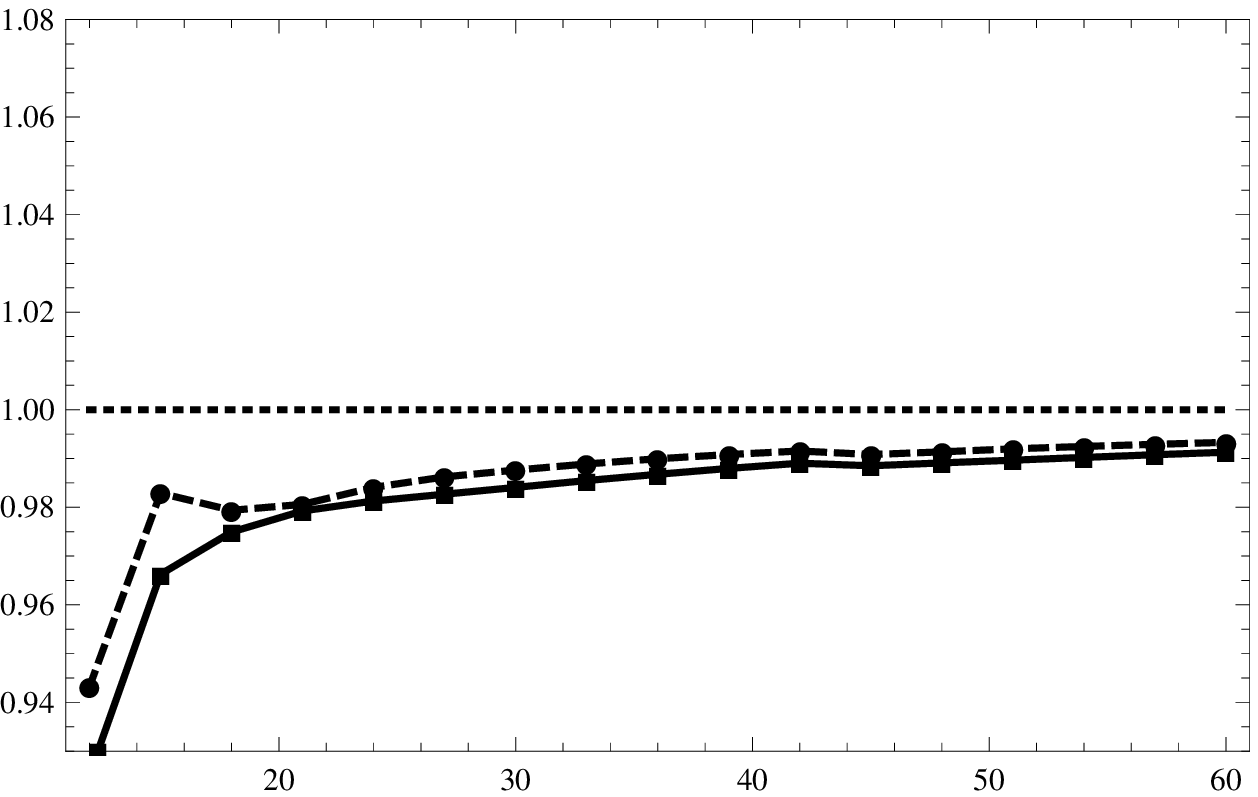}
\end{subfigure}
\begin{subfigure}[b]{\tempwidth}
       \centering
       \includegraphics[width=\tempwidth]{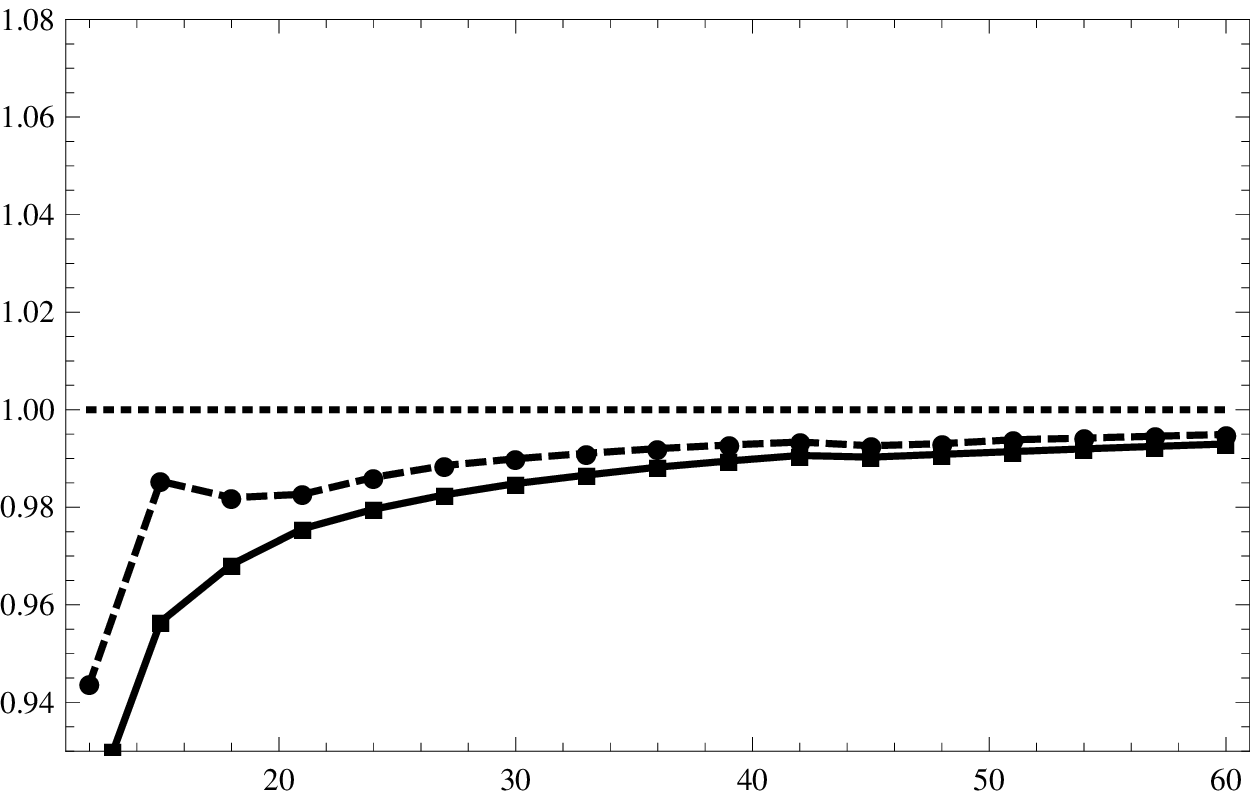}
\end{subfigure}\begin{subfigure}[b]{\tempwidth}
       \centering
       \includegraphics[width=\tempwidth]{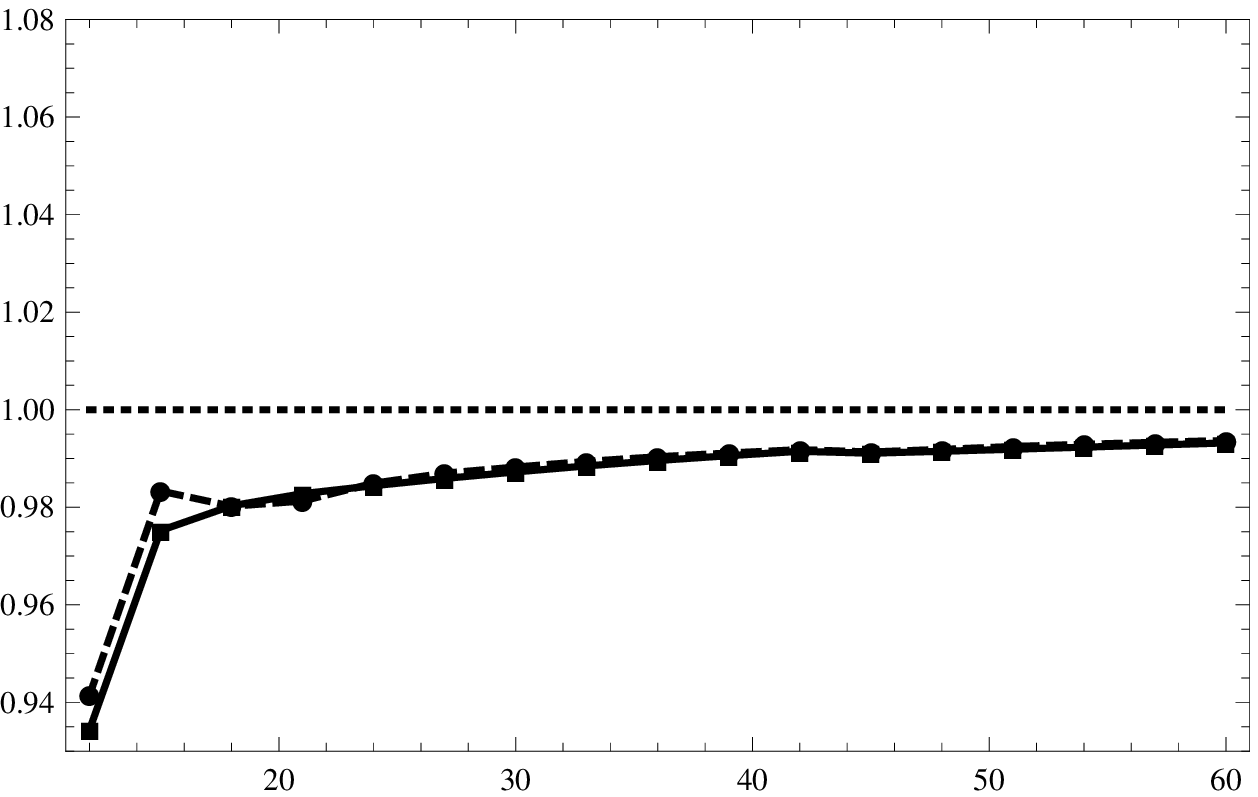}
\end{subfigure}\\
\rowname{{\normalfont\scriptsize Decay}}
\begin{subfigure}[b]{\tempwidth}
       \centering
       \includegraphics[width=\tempwidth]{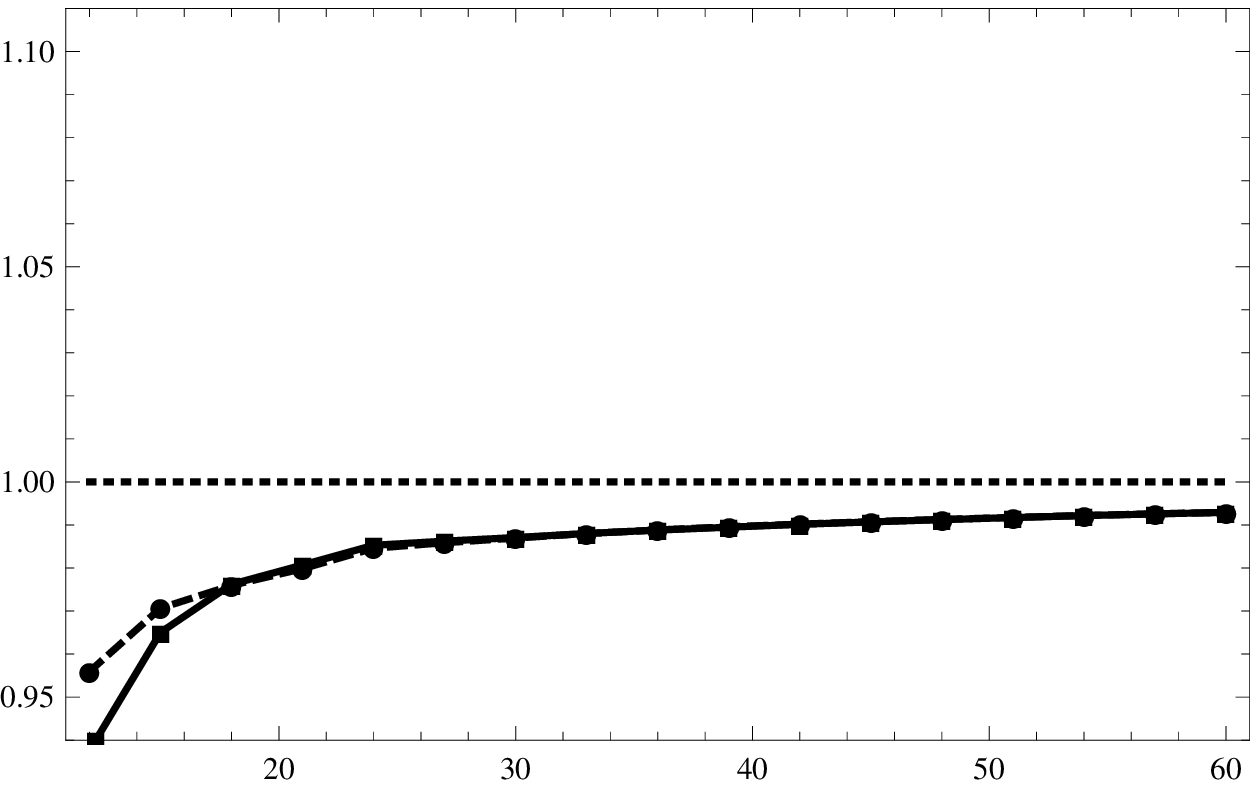}
\end{subfigure}
\begin{subfigure}[b]{\tempwidth}
       \centering
       \includegraphics[width=\tempwidth]{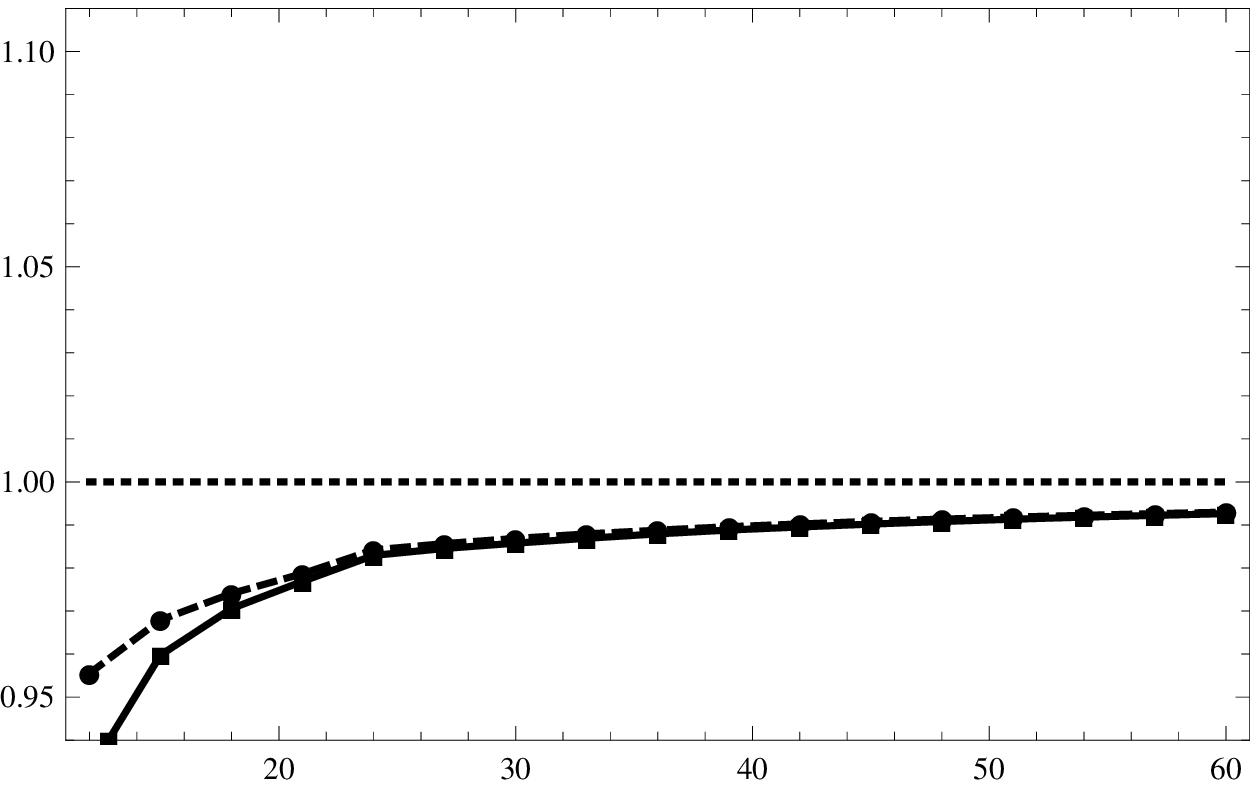}
\end{subfigure}\begin{subfigure}[b]{\tempwidth}
       \centering
       \includegraphics[width=\tempwidth]{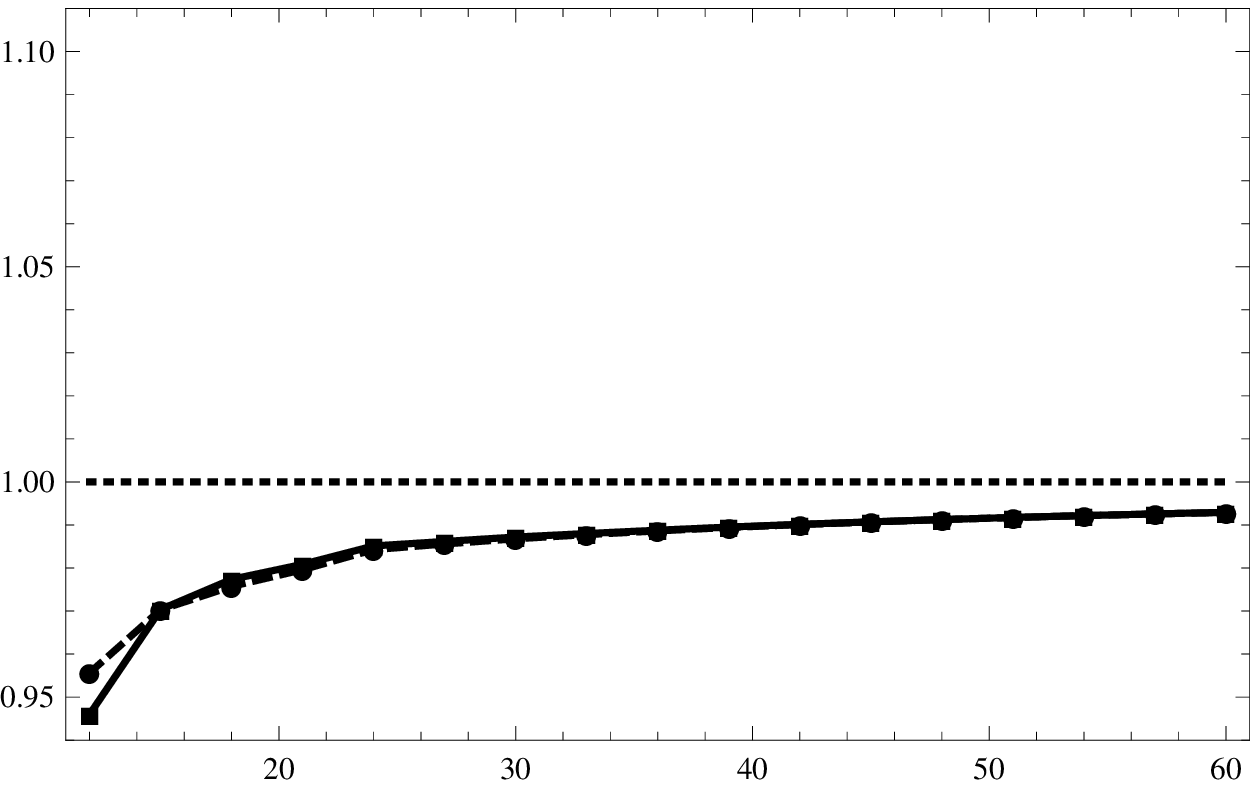}
\end{subfigure}\\
\rowname{{\normalfont\scriptsize Compartmental}}
\begin{subfigure}[b]{\tempwidth}
       \centering
       \includegraphics[width=\tempwidth]{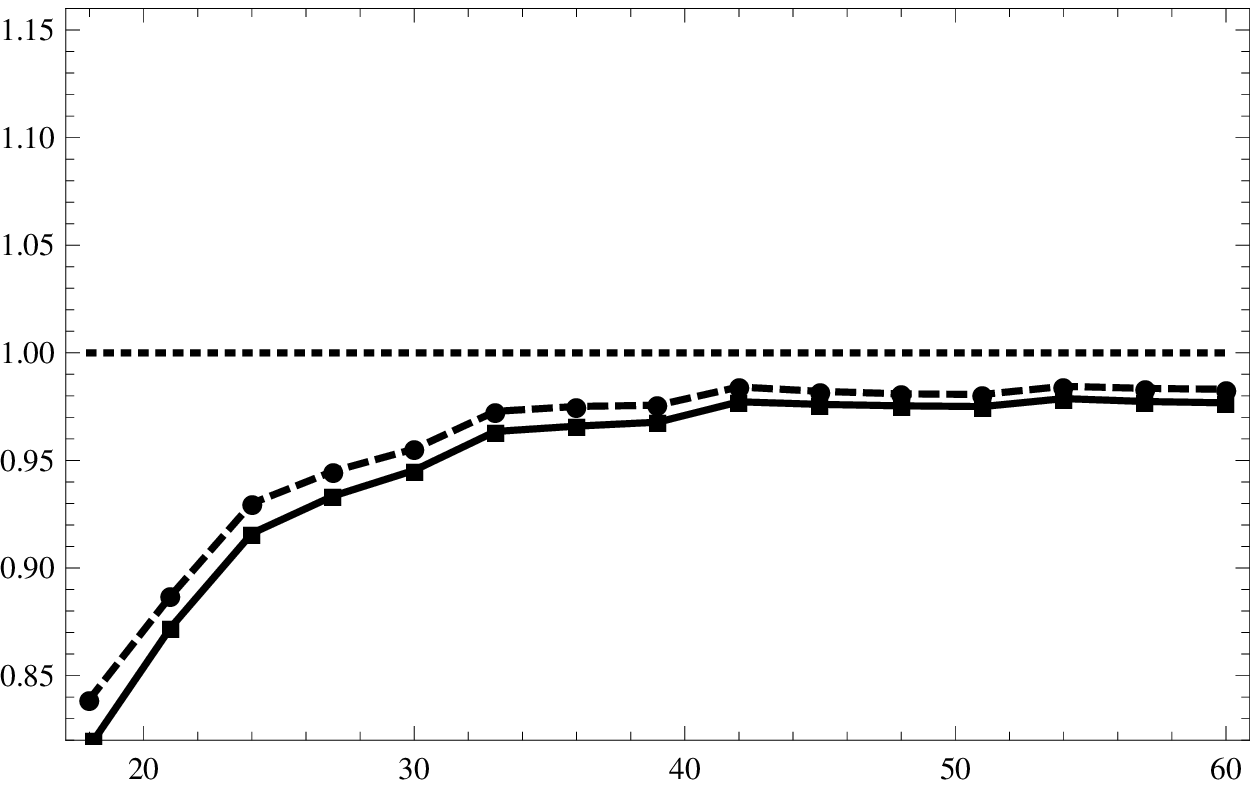}
\end{subfigure}
\begin{subfigure}[b]{\tempwidth}
       \centering
       \includegraphics[width=\tempwidth]{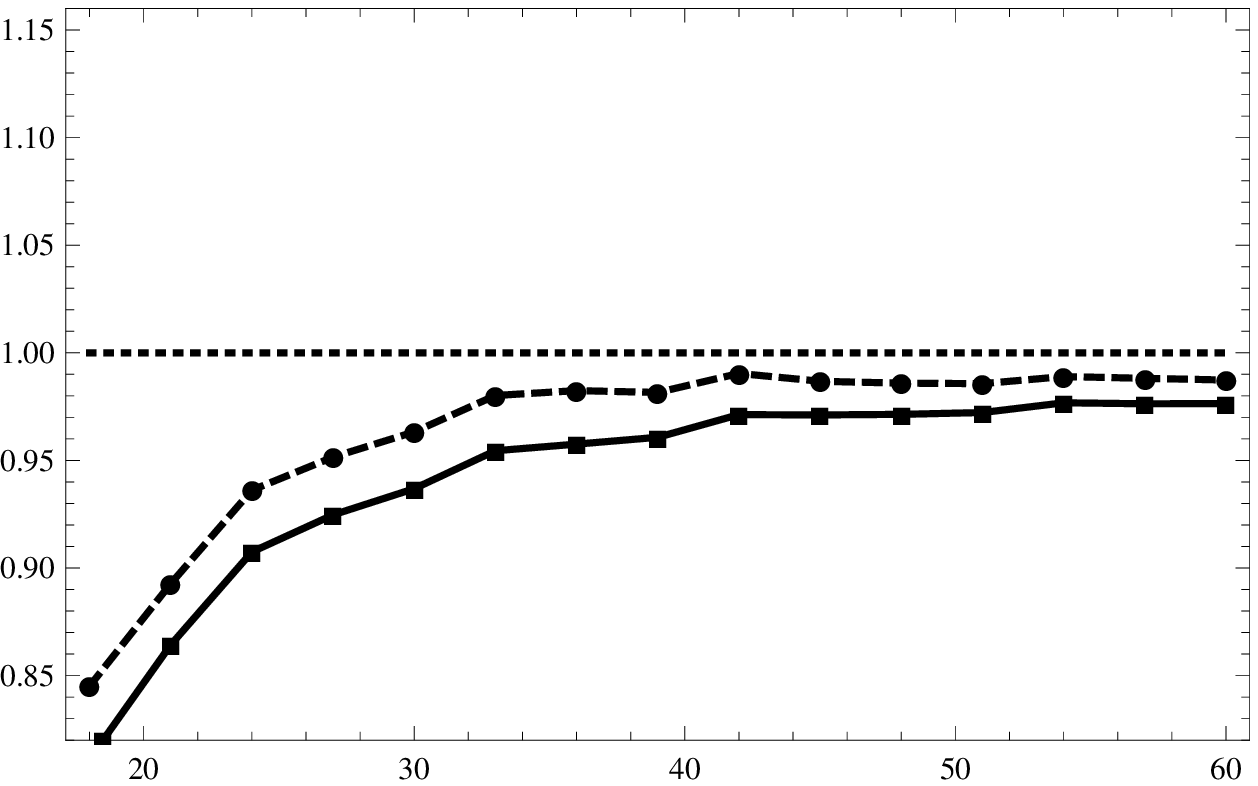}
\end{subfigure}\begin{subfigure}[b]{\tempwidth}
       \centering
       \includegraphics[width=\tempwidth]{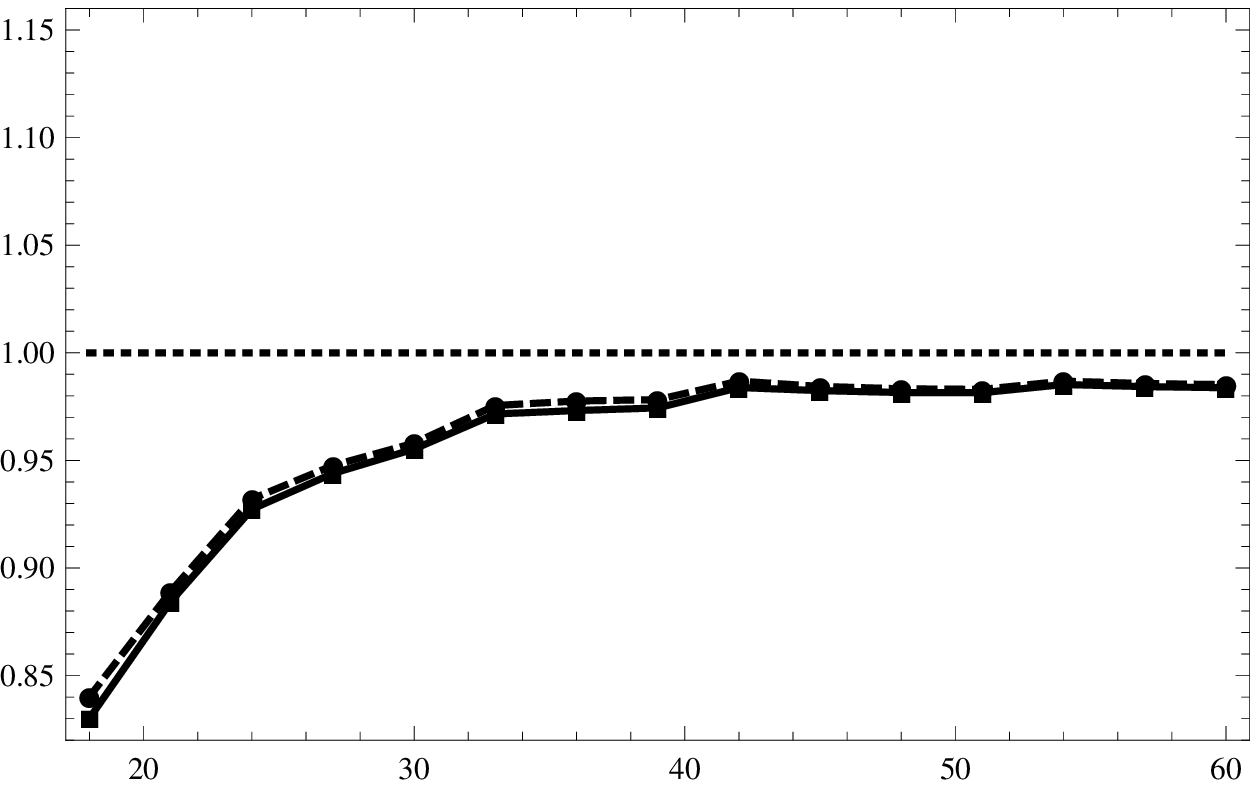}
\end{subfigure}\\
\caption{$c_{\boldsymbol{\theta}}$-efficiencies with respect to the Fisher information in the sample of the optimal RSD, ${\rm{RM}}_{\hat{\xi}_{RSD}}^{c_{\boldsymbol{\theta}}}$ (solid line), and the AOD, ${\rm{RM}}_{\hat{\xi}_{AOD}}^{c_{\boldsymbol{\theta}}}$ (dashed line), relative to the FLOD, represented by the dotted line at 1, for the Michaeles-Menten, Decay and Compartmental model (top to bottom) and the Cauchy, Exponential Power and $q$-Gaussian error distributions (left to right).}\label{fig:MIC}
\end{sidewaysfigure}

\end{supplement}

\bibliographystyle{imsart-nameyear.bst}
\bibliography{bibtex_entries}

\end{document}